\RequirePackage{rotating}
\documentclass[fleqn,usenatbib]{mnras}

\usepackage{newtxtext,newtxmath}

\usepackage[T1]{fontenc}

\DeclareRobustCommand{\VAN}[3]{#2}
\let\VANthebibliography\thebibliography
\def\thebibliography{\DeclareRobustCommand{\VAN}[3]{##3}\VANthebibliography}

\usepackage{graphicx}	
\usepackage{adjustbox}
\usepackage{amsmath}	

\usepackage{newtxmath}
\usepackage{amssymb}	
\usepackage{rotating}
\usepackage{lscape}
\usepackage{ulem}

\usepackage{longtable} 
\usepackage{rotating}

\title[Characterisation of the young stellar cluster Mon~R2]{Time series photometry and multi-wavelength characterisation of the young stellar cluster Mon~R2}

\author[S. Orcajo et al.]{
Santiago Orcajo$^{1,2}$\thanks{E-mail: santi@fcaglp.unlp.edu.ar},
Lucas A. Cieza$^{3}$\thanks{E-mail: lucas.cieza@mail.udp.cl},
Roberto Gamen$^{2,1}$\thanks{E-mail: rgamen@fcaglp.unlp.edu.ar}
\\
$^{1}$Facultad de Ciencias Astron\'omicas y Geof\'isicas, Universidad Nacional de La Plata, Paseo del Bosque S/N, 1900 La Plata, Argentina\\
$^{2}$Instituto de Astrof\'isica de La Plata, CCT La Plata, CONICET--UNLP, Argentina\\
$^{3}$N\'ucleo de Astronom\'ia, Facultad de Ingenier\'ia y Ciencias, Universidad Diego Portales, Av Ej\'ercito 441, Santiago, Chile\\
}

\date{Accepted XXX. Received YYY; in original form ZZZ}

\pubyear{2021}

\begin{document}
\label{firstpage}
\pagerange{\pageref{firstpage}--\pageref{lastpage}}
\maketitle

\begin{abstract}
Using the Las Cumbres Observatory Global Telescope Network (LCOGT), we have obtained multi-epoch photometry of the young cluster Mon~R2. We have monitored over 6000~sources with $i$-band between 13 and 23~mag within a $ 26'x26'$ field of view. For each star, we collected $\sim1,500$ photometric points covering a temporal window of 23~days. Based on these data, we have measured rotation-modulated of 136 stars and identified around 90 additional variables, including 14 eclipsing binary candidates. Moreover, we found 298 other variables with photometric high-scatter. In addition, we have obtained $r$-band and H${\alpha}$ narrow-band photometry of the cluster with LCOGT and low-resolution optical spectroscopy of 229 stars with GMOS-Gemini. We used the \textit{Gaia} data from the periodic stars and objects with H$\alpha$ or IR-excesses, which are mostly low-mass pre-main sequence stars ($<1$~M$_{\sun}$) in the cluster to estimate the distance ($825 \pm 51$~pc) and the mean proper motions ($\mu_{\alpha}cos(\delta)=-2.75$~mas~yr$^{-1}$ and $\mu_{\delta}=1.15$~mas~yr$^{-1}$) of its members. This allows us to use the \textit{Gaia} data to identify additional Mon~R2 member candidates. We also used Pan-STARRS photometry from our LCOGT sources to construct a more precise H-R diagram, from which we estimate the mean age of the cluster and identify other possible members including eleven spectroscopy brown dwarf with M7 to M9 GMOS spectral types. Finally, we combined our membership lists with \textit{Spitzer} infrared photometry to investigate the incidence of stars with discs and the effect these have on stellar rotation.
\end{abstract}

\begin{keywords}
stars: fundamental parameters -- 
stars: low-mass -- 
stars: rotation -- 
stars: variables: general --
open clusters and associations: individual: Mon~R2 --
brown dwarfs
\end{keywords}


\section{Introduction}

Open clusters (OCs) are essential targets for the study of galactic structure and composition, as well as of star formation in a galaxy, including the Milky Way  \citep[see  e.g.][]{1982ApJS...49..425J,2019MNRAS.488.2158M,2020A&A...640A.127Z}.
The members of a young OC are usually assumed to have the same origin, age and composition, although there may be some substructures of star-formation within a given cluster \citep{2003ARA&A..41...57L}. In our nearby neighbourhood ($<$1~kpc) there are at least 655 such OCs \citep{2019JKAS...52..145S}. 

The young OCs contain PMS stars, molecular gas and HII regions.
The dust associated with the molecular gas in young OCs interferes with the estimates of stellar distance, age and mass. 
However, the combination of data sets from observatories such as \textit{Spitzer}, \textit{Gaia}, and Pan-STARRS provide us with a more precise and in-depth view which allow to obtain these main stellar parameters with greater precision \citep[e.g.][]{2020A&A...635A..45C,2021A&A...649A...5F}. 
In turn, a more complete census of the stellar members permits to  better define the structure and age of a young OC and provides observational constrains to test theories of star-formation and stellar structure \citep[e.g.][]{1987ARA&A..25...23S,2003ARA&A..41...57L}. 

In addition to age and mass, angular momentum is one of the fundamental parameters in stellar astrophysics  \citep{2004AJ....127.1029R,2014prpl.conf..433B} . Several studies have shown that, during the PMS stage, stars drain angular momentum through their circumstellar discs \citep[e.g.][]{2001AJ....121.1676R,2020AJ....159..273R, 2008MNRAS.384..675I,2017A&A...599A..23V,2019MNRAS.487.2937O}.
Therefore, measuring the rotation period in OCs  of stars with and without a disc is key to study the evolution of angular momentum as a function of age and stellar mass. Currently, such studies can be performed through the combination of photometric rotation periods and infrared (IR) observations (used for disc identification).  

In order to contribute to the understanding of OCs in our Solar neighbourhood, we have obtained time-series photometry and conducted a multi-wavelength characterisation of the stellar population in the OC Mon~R2. 
The paper has a broad scope and several main goals, including: identifying new stellar and substellar members,  refining the distance, age, and proper motion of the cluster, identifying stars with circumstellar disks, and investigating their effect on stellar rotation.  In addition to this main goals, we use the time-series photometry   to identify eclipsing binaries and several types of non-periodic variables in the field.

The Mon~R2 cluster is composed of a prominent reflection nebulae,  dark nebulae, molecular outflows, and HII regions \citep{2008hsf1.book..899C} and located in the third quadrant of the Galaxy at an estimated distance of $830\pm50$~pc and with an estimated age of 1-10~Myr \citep{1976AJ.....81..840H,1997AJ....114..198C}.
The structure and composition from this complex region have been studied in the X-ray, IR, and radio regimes \citep[e.g.][]{2011A&A...535A..16L,2017A&A...607A..22R,2019MNRAS.483..407S}.
B-type stars illuminate the HII regions and the IR shell at the centre of the OC \citep{2001ASPC..251..254K}. 
The dense molecular regions show arc-like structures surrounding the HII regions composed of three clumps, and an infrared cluster is located at the far-side of OC \citep{2000ApJ...538..738C}. \citet{2008hsf1.book..899C} described this cluster in detail, concluding that Mon~R2 is very young and complex star-formation site. 

Also, \citet{1997AJ....114..198C} studied the central region photometrically and 34 stars spectroscopically. They obtained a lower limit of $\sim475$ cluster members with a limiting  $K$-band magnitude of 17.4~mag.
\citet{Andersen_2006} found that $27\pm9$~per~cent of the stars with estimated masses between 0.1 to 1.0~M$_{\sun}$ have a near–infrared (NIR) excess indicative of a protoplanetary disc.
However, the low-mass (sub)stellar population in Mon~R2 still remains to be studied and characterised in detail. 

Here, we present a multi-technique and multi-wavelength characterization of Mon~R2.  
Firstly, we used Las Cumbres Observatory Global Telescope Network (LCOGT) to obtain time-series photometry to identify periodic and variables sources in $I$-band. 
The $I$-band was chosen as a compromise between the S/N and the amplitude of the expected variations due to stellar spots and to minimize the contribution from either accretion or disk emission to the measured stellar flux.
In particular the red colours of M-type stars imply that the stars are significantly brighter in the $i$-band with respect to the $r$-band. In low-mass stars, the rotation-modulated variability is due to cool (stellar activity) and hot (accretion-shocked) spots. The cool spots can cover up to the $\sim$40~\% of the stellar surface and usually dominate the periodic variability \citep{1994AJ....108.1906H,2001AJ....121.1676R}.  Additional photometry was obtained in $R$-band and H$\alpha$ narrow-band to select sources with  H$\alpha$ emission. The $(R - $H$\alpha)$ colour allows us to identify PMS stars and discard non-member stars, i.e. giant field stars. The  $(R - $H$\alpha)$ colour of many PMS stars, in particular accreting classical T~Tauri stars (CTTSs), is larger than the colour of a main-sequence (MS) stars of the same spectral type. The H$\alpha$ emission in non-accreting weak-lined T~Tauri stars (WTTSs) is weak (from chromospheric activity) and typically not detectable through narrow-band photometry. The recombination H$\alpha$ line is a power full accretion indicator because it traces the ionised hydrogen produced by the UV radiation, which is caused by the magnetospheric accretion shock in young stellar objects \citep[][]{2012A&A...548A..56R}. Furthermore, the giant stars have smaller $(R - $H$\alpha)$ colours than MS stars of the same $R-I$ colours for spectral types later than $\sim$K3. So, it is possible to discriminate between background giants and PMS stars with a $(R - $H$\alpha)$ vs. $(R - I)$ colour-colour diagram \citep{2004A&A...417..557L}. Furthermore, since the H$\alpha$ line in CTTSs is mostly due to disc accretion, the $(R - $H$\alpha)$ colour is also a useful disc indicator.  
On the other hand, observational works \citep[i.e.][]{2010A&A...510A..72F} found that the fraction of stars with mass accretion decreases fast with time, going from $\sim$60~per cent at 1.5-2~Myr down to $\sim$2~per cent at 10~Myr, this fraction is systematically lower than the fraction of stars showing near-to-mid infrared excess. So, the NIR \textit{Spitzer} data is typically a more a robust disc indicator than the H$\alpha$ photometry; however, we find that some of our sources lack \textit{Spitzer} colours but do have $(R - $H$\alpha)$ colour data.

Then, we studied the LCOGT sources with astrometric data ($\varpi$, $\mu_{\alpha} \cos{\delta}$ and $\mu_{\delta}$) from \textit{Gaia} Data Release 2 (GDR2) and Early Gaia Data Release 3 (EGDR3) to define specific \textit{Gaia} membership criteria for the cluster. 
Also, using mid-IR photometry we identified objects with and without IR excess from a disc.
The rotation periods of Mon~R2 members were correlated with disc indicators to investigate disc regulation.
We also used the Pan-STARRS 1 (PS1) Data Release 1 and 2 (DR1 and DR2 respectively) photometric data to construct the H-R diagram of the OC and calculate its average age with isochrones from theoretical models.   
Finally, we used the low-resolution spectroscopy with the GMOS instrument on Gemini-South Observatory (GS) to classify low-mass stars and brown dwarfs.

This paper is organised as follow. 
In Section~\ref{Sec:Observations},  we describe our own observations: the photometry and spectroscopy data from Las Cumbres Observatory and Gemini South Observatory, respectively.  We also describe the different  public data sets used for our analysis. 
In Section~\ref{sec:Analysis},  we present the analysis of the LCOGT photometry and the GMOS spectroscopy data, with the study of light curves of the variable sources and the spectral characterisation of the targets. Then, we combine the data sets to identify new (sub)members and characterise the distance, extinction, age, disc frequency, and rotation period distribution of the  cluster population.    
Finally, in Section~\ref{sec:Conclusion}, we present a summary of our results and main conclusions.

\section{Observations and public data sets}
\label{Sec:Observations}

\subsection{Observations}

\subsubsection{Photometry with LCOGT}
\label{sec:TSP}

LCOGT has twenty three telescopes in seven sites around the world working together, allowing to monitor a given field during several weeks or months with an almost continuous time coverage.  
We used the SCICAM-SINISTRO instrument at the 1.0-m telescopes, which consist of a $4Kx4K$~CCD camera and provides a field of view (FOV) of about $26.5x26.5$~arcmin$^2$, with a scale of 0.389~arcsec~pixel$^{-1}$.  
Our data were obtained from the South African Astronomical Observatory (CPT), Siding Spring Observatory (COJ), McDonald Observatory (ELP), and Cerro Tololo (LSC) \citep[][]{2013PASP..125.1031B}.

We obtained the images with the filters $I$-band, $R$-band and H$\alpha$-narrow band for different purposes.

We used the $I$-band filter to obtain time-series photometry to identify variable and periodic sources.
We obtained 2400 images with the $I$-band in a field centred at RA(J2000.0)=06$^{h}$07$^{m}$53$^{s}$.57 and DEC(J2000.0)=-06º21'59''.4 over a period of 23 days, from 2016 December 1 to 23.
Overall, we monitored a total of 6843 sources in a field, with $I$-band apparent magnitudes ranging from  $\sim~14$  to $\sim24$~mag. We took sets of 10 images with 180 s of integration per image, with an observational overhead of approximately one and a half hours. Finally, an observational cadence of 30 minutes between each set of 10 images was achieved.

\begin{figure*}
	\includegraphics[width=0.9\textwidth]{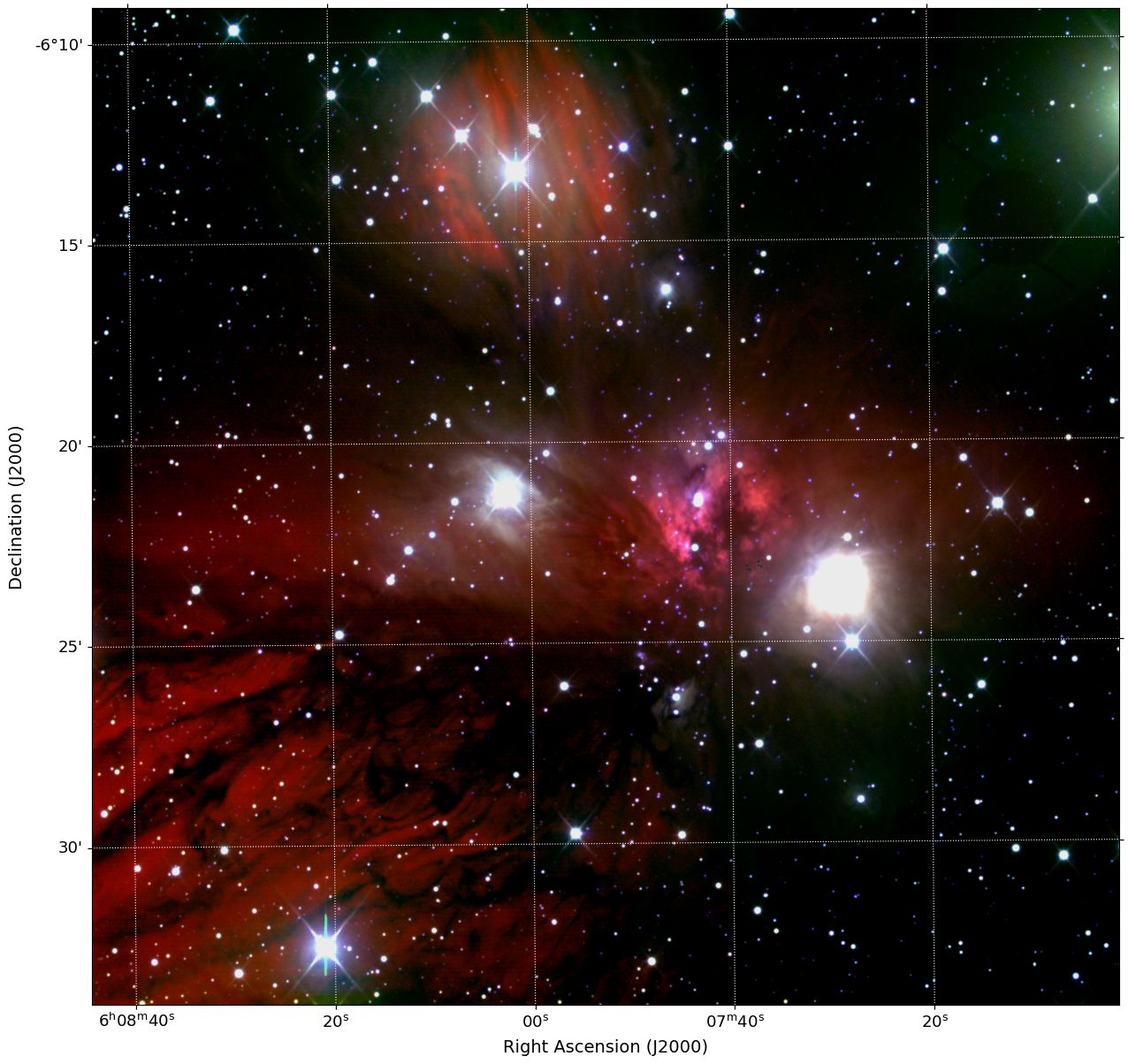}
	\caption{Image composite of the young stellar cluster Mon~R2, based on $I$-band (blue), $R$-band (green) and H$\alpha$ (red) data taken with LCOGT Sinistro 1-m telescopes.}
\label{MonR2}
\end{figure*}

We also obtained $R$-band and H$\alpha$ narrow-band photometry of Mon~R2 in order to construct colour-magnitude and colour-colour diagrams.
For H$\alpha$ and $R$-band data, we obtained 18 and 20 exposures, respectively, 600 and 300~s each, in the second half of December 2016. We detect 915 sources in H$\alpha$ and 2355 objects in $R$-band, for a total of 844 targets with $(R - $H$\alpha)$ colours. 
A colour image of the Mon~R2 Cluster based our $I$-band, $R$-band, and H$\alpha$ data taken with LCOGT is shown in Fig.~\ref{MonR2}.

\subsubsection{Spectroscopy with GS}

The determination of stellar mass and age in Mon~R2 by photometric colour indexes is difficult due to the intrinsic differential extinction in the field.
However, from the spectral types we can obtain the effective temperature and intrinsic photospheric colours of the targets using calibration tables. The latter allows us to correct the photometry for extinction, enabling a more accurate determination of the mass and age of the targets.
We observed Mon~R2 with the Gemini Multi-Object Spectrograph (GMOS) in the multi-object mode from GS. We divided the photometry field in 16 sub-fields according the FOV of GMOS. We observed the cluster in three different programs over three years (programs ID GS-2017B-Q-39, GS-2018B-Q-204 and GS-2019B-Q-121) covering 12 sub-field (see Table \ref{subfieldtb} and Fig. \ref{subfieldim}). For these three programs, we used the R150 grating, which delivers optical spectra between 450 and 1100~nm with a resolving power R$\approx$630. Our GMOS targets have $I$-band magnitudes in the 15--22 range, and were observed with integration times ranging from  5 to 11.5~min, to reach typical Signal to Noise ratios (S/N) of 50.

Data were processed, in the standard way (CCD processing, aperture tracing, aperture extraction, wavelength calibration and sky substraction), using the GMOS-Gemini package within \textsc{IRAF} as in \citet{2019MNRAS.487.2937O}. 
For each mask we obtained 3 sets of images consisting of one flat-image (called “flat”), one CuAr lamp-image (“arc”) and one science-image (``sci''). Each set was processed individually and manually. The flats were normalised and corrected for overscan and gaps between CCD’s. The slits were recognised in the flat from the gradient method and then this information was applied to the arc and sci images. Pixel-to-wavelength transformations were found using arc image calibration. Each slit was analysed individually and the results were checked until the optimal transformation was found. Sci images were corrected for overscan, flat, bad pixels and pixel-to-wavelength transformation. Spectral extractions were interactive and spectrum was separated into a single file.
The spectrum is the result of the combination of three reduced spectra, achieving an improvement of $\sim \sqrt{3}$ in the S/N. Also, we obtained telluric spectra which were used to minimise the atmospheric contamination.
We were able to obtain 297 spectra in the 12 masks employed. 

\begin{table}
\caption{Details of each GMOS mask employed. In successive columns we show the number of sub-field (according the Fig.~\ref{subfieldim}), program ID, integration times (exptime), and the number of spectra obtained (n).}
\label{subfieldtb}
\centering
\begin{tabular}{c l l c c}
\hline
sub-field & program ID & exptime (s) & n \\
\hline
01 &  GS-2019B-Q-121 & 420 & 20 \\
02 &  GS-2019B-Q-121 & 420 & 20 \\
03 &  GS-2017B-Q-39  & 330 & 19 \\
04 &  GS-2018B-Q-204 & 435 & 21 \\
05 &  GS-2017B-Q-39  & 300 & 26 \\
06 &  GS-2018B-Q-204 & 435 & 25 \\
07 &  GS-2018B-Q-204 & 435 & 31 \\
08 &  GS-2018B-Q-204 & 690 & 26 \\
09 &  GS-2018B-Q-204 & 680 & 25 \\
10 &  GS-2018B-Q-204 & 435 & 25 \\
11 &  GS-2018B-Q-204 & 690 & 24 \\
12 &  GS-2017B-Q-39  & 330 & 30 \\
\hline
\end{tabular}
\end{table}

\subsection{Public data sets}

In our study of Mon~R2, in addition to our own data, we also used different public multi-wavelength and multi-technique data sets, which we describe below.

\subsubsection{Pan-STARRS}

The Panoramic Survey Telescope and Rapid Response System (Pan-STARRS) is an innovative wide-field astronomical imaging and data processing facility developed at the University of Hawaii’s Institute for Astronomy \citep{2002SPIE.4836..154K,2010SPIE.7733E..0EK,2018AAS...23110201C}.
\citet{2018AAS...23110201C} provides an overview of the fully implemented Pan-STARRS System. The first telescope of the PanSTARRS Observatory is the Pan-STARRS Telescope \#1, (Pan-STARRS1 or informally PS1). The PS1 survey used a 1.8-m telescope and a 1.4 Gigapixel camera to image the sky in five broadband filters ($g$, $r$, $i$, $z$, $y$).  
The PS1 survey covers the entire Northern sky and extends to the South to declination $-30\degr$, contains wide range of magnitudes with millimagnitude accuracy \citep{2012ApJ...750...99T,2020ApJS..251....6M}. 
We used the data from DR2 and added a small fraction of sources with DR1 (about 15~per cent). We downloaded the data from website\footnote{\url{https://catalogs.mast.stsci.edu/panstarrs/}}.  DR1 occurred on December 19, 2016 and DR2 on January 28, 2019. DR2 shows improvements over DR1, but both were necessary to cover all our LCOGT sources (some of these sources only had complete data, or the necessary flags, in DR1, but not in DR2)
\citep{2020MNRAS.493..267S}. We matched the PS1 sources in the $g$, $r$, $i$, $z$ and $y$-band filters with the sources obtained in the LCOGT field.We use the mean PSF magnitude from each filter with its associated uncertainty\footnote{Some photometric values in the PS1 catalogue do not have an associated error.} Sources with missing photometric errors in the catalogue were discarded. In subsec. \ref{sec:PS1} we explain how we select the data.

\subsubsection{\textit{Gaia}}

The \textit{Gaia} EDR3 represents a major and the newest release for the \textit{Gaia} mission. The data were collected during 34 months started in 2014 and contain high-precision parallaxes and proper motions. The parallaxes and proper motions were derived from \textit{Gaia} observations alone. These data releases cover sources brighter than $G=21$~mag. The EDR3 presents an increment of 30~per~cent in parallaxes precision and  doubled the precision of proper motions respect to the previous release (GDR2) \citep{2018A&A...616A...1G,2020yCat.1350....0G,2021AJ....161..147B,2021A&A...649A...4L}. Firstly, we used the GDR2 data and then, we worked with EDR3 and checked the improvements. We matched the \textit{Gaia} sources downloaded from the EDR3 archive page\footnote{\url{{https://gea.esac.esa.int/archive/}}} with  sources obtained from the LCOGT field. We used the \textit{GAIA} data to determine the proper motions and distances from Mon R2. In subsec. \ref{sec:Gaia} we explain how the data were collected.

\subsubsection{\textit{Spitzer}}
\label{sec:Spz}
With the goal of studying the membership of sources to the cluster and the incidence of circumstellar material on stellar rotation, we used \textit{Spitzer}-IRAC colours, which are particularly good disc indicators (see subsec. \ref{sub:disc}). 
To build a catalogue with the \textit{Spitzer}-IRAC photometry data for our sample, we first obtained the data from the \textit{Spitzer} Enhanced Imaging Products (SEIP) Source List on a box with 1800~arcsec centred at RA$=91\degr.94417$ and DEC$=-6\degr.38306$ (Equinox J2000) from the IPAC Infrared Science Archive (IRSA)\footnote{ 
\url{https://irsa.ipac.caltech.edu/Missions/spitzer.html}}. 

We followed the recommendations of the 'Spitzer Enhanced Imaging Products Explanatory Supplement\footnote{\url{http://irsa.ipac.caltech.edu/data/SPITZER/Enhanced/SEIP/docs/seip\_explanatory\_supplement\_v3.pdf}}' (subsection 1.1.2) to obtain the most reliable fluxes. However, the second recommendation, adopting \textsc{fluxflag}=0 to exclude sources that may be affected by nearby saturated sources or nearby extended sources. This eliminated most sources located in a circular area of radius $\sim8.5$~arcmin from the cluster centre. For this reason, this recommendation was not followed. On the other hand, we chose the data with \textsc{fluxtype}=1 and used the 3.8~arcsec diameter aperture with its associated uncertainties for the IRAC channels, this makes a cut-off at SRN$\geq$3.

Furthermore, IRAC provides the photometric data in flux, which were transformed to magnitudes using the equations acquired from the same supplement in table 4.6. For this transformations we used the zero points 280.9~Jy, 179.7~Jy, 115.0~Jy, 64.9~Jy for fluxes $I1=[3.6]$, $I2=[4.5]$, $I3=[5.8]$ and $I4=[8.0]$ respectively. 

On the other hand, we found two works with \textit{Spitzer} data not included in the IRAC database.
\citet[][hereafter G9]{2009ApJS..184...18G} presented an uniform mid-infrared imaging and photometric survey of 36 young, nearby, star-forming clusters and groups, including Mon~R2. They used IRAC data and calibrated with \textit{Spitzer} Science Center Standard Basic Calibrated (SSC BCD) pipeline. \citet[][hereafter K12]{2012AJ....144...31K} identified protostars from nine star-forming molecular clouds within 1~kpc and Mon~R2 is one of them. They extracted photometry from \textit{Spitzer} surveys using point-spread function (PSF) fitting photometry techniques. 

We cross-matched the data in the three samples mentioned above finding a good agreement among fluxes of coincident sources (within errors). Then, we constructed an unique table prioritising data according their sources, as follows: the \textit{Spitzer} SEIP data, G9, and K12 data. 
Thus, the \textit{Spitzer} SEIP data were complemented with 48 sources from G9 and 1 from K12.  
The final catalogue contains 983 targets with both $3.6~\mu$m and $8.0~\mu$m data and is shown in Appendix Table~\ref{tab:Spitzer}. For completeness the $4.5~\mu$m and $5.8~\mu$m fluxes are also included,  even though they are not used in the analysis. In addition, we note that the \textit{Spitzer} SEIP, G9, and K12 data sets virtually show a 100~per cent agreement in terms of disc identification.

\section{Results and analysis}

\subsection{LCOGT imaging data}
\label{sec:Analysis}
LCOGT delivers the images processed with the \textsc{BANZAI} pipeline \citep[][]{2018SPIE10707E..0KM} as \textsc{FITS} multi-extension files, together with a catalogue of sources detected Source Extraction and Photometry and the bad pixel mask\footnote{\url{https://lco.global/documentation/data/BANZAIpipeline/}}. This photometry was not deep enough ($I_\mathrm{max} \sim $ 18.5~mag) to detect all objects of interest; therefore, so we decided to apply our own photometry and object search using the the tasks provided by the software \textsc{IRAF}, \textsc{DAOFIND} and \textsc{APPHOT}.
Saturation occurred near $I \sim12$~mag and the completeness limit was $I \sim 22.61$~mag. Where $I$ is the I-instrumental corrected for an approximation of ZeroPoint obtained with the PS1 matched data in I-band. $I$ is a correction for filter change, but it is not an absolute photometry

\begin{figure}
\includegraphics[width=1\columnwidth]{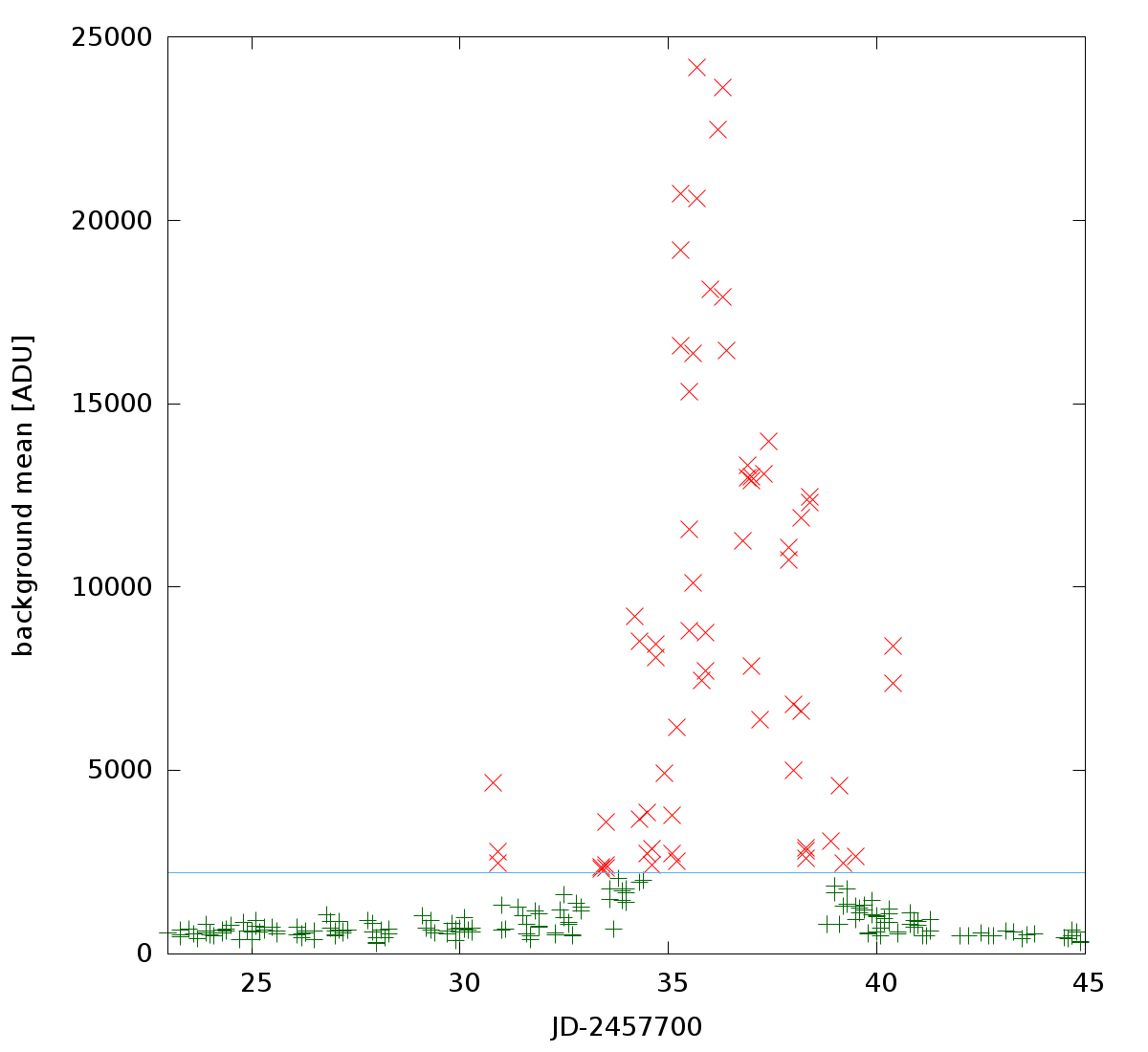}
	\caption{Mean statistic in two background sectors 
	from each $I$-band master images. Red X simbols depict the poor-quality images (high background), while the green + signs the good-quality images, i.e. mean $<2200$~ADU (see text).
	}
\label{statics}
\end{figure}

From the 2400 images obtained in $I$-band, only 1560 were useful. Some poor-quality images showed very low S/N, poor tracking or focus etc. In order to identify poor-quality images, we studied and analysed statistically two background sectors on master images created from the median of ten consecutive individual images with very few bright sources.
We considered that if the mean of background sectors is great than 2200~ADU ( 3~$\sigma$ above the mean value of the images with the best background), the background is too bright and the S/N is too low to be useful and thus we discard all the individual images in the master image.
In Fig.~\ref{statics}, we depict the statistic study, a gap of almost four days in the time-series due to bad images was unluckily obtained.
To obtain a sample containing even the faintest objects, we created a super image combining all 1560 good frames in $I$-band and applied the \textsc{DAOFIND} package to obtain their coordinates. Finally, with the coordinates obtained, we applied aperture photometry to the 1560 individual images with the \textsc{PHOT} task of the \textsc{APPHOT} package.
Finally, we cross-matched the LCOGT sources with others catalogues (\textit{Gaia} EDR3, PS1, \textit{Spitzer}), which drastically reduced the number of spurious sources.  

\subsubsection{Differential photometry and lightcurves}

\begin{figure}
\centering
    \includegraphics[width=0.8\columnwidth]{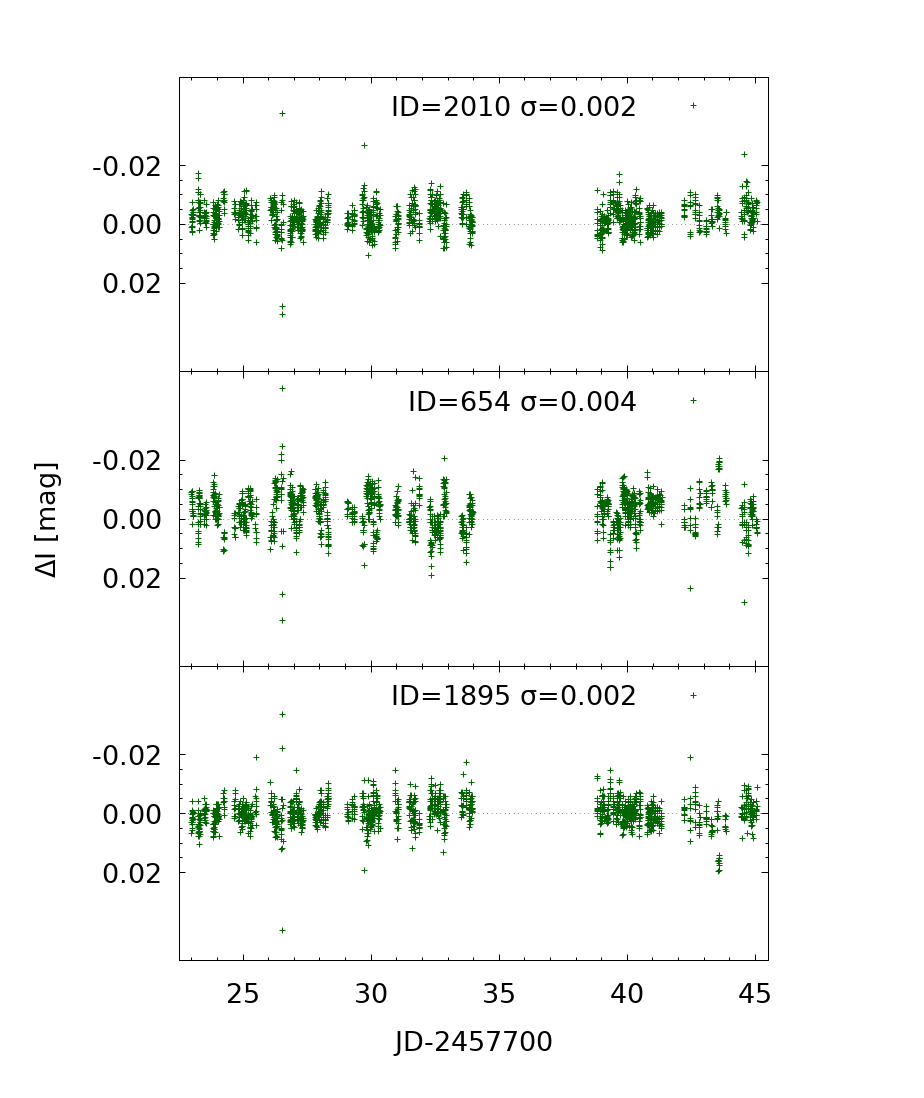}
	\caption{Differential magnitudes ($\Delta i$) vs. Julian Day for three non-variable stars with the lowest photometric standard deviation. The magnitudes in $i_{PS1}$-band are: ID 2010 $= 14.616 \pm 0.004$, ID 654 $= 13.924 \pm 0.011$ and ID 1895 $= 14.726 \pm 0.001$.$\Delta mag$ is calculated between a star and the average of the three stars. The reference star has $i_{PS1}=14.442$ mag and the $\sigma$ symbol is the amplitude of light curve.  
}
\label{nonvariablesdif}
\end{figure}

We calculated the differential photometry of the 6843 sources respect to a mean reference star, which was constructed by the averaging the fluxes of the three stars with the lowest photometric variations and resulted in a magnitude of $14.422$ mag in the $i$-band from PS1.
These reference stars were selected because they have extremely consistent differential photometry between them, indicating that they are intrinsically non-variable. Light curves of these non-variables stars are shown in Fig.~\ref{nonvariablesdif}. Each light curve was generated with the 1560 $I$-band images obtained in 23 consecutive days. The gap in the time-series is due to the discarding of bad images (see Sec.~\ref{sec:Analysis}).

We reached $rms < 0.001$~mag, which allows us to identify variable stars with amplitudes as low as 0.01~mag. 
From this well-sampled and well-calibrated time-series data, we are able to find photometric periods between 0.2 to 16 days. In bottom Fig,~\ref{power} we show the mean photometric error per magnitude in the $i$-band, observing that periodic stars have errors below 20~per~cent, indeed, 90~per~cent remain below 5~per~cent of the instrumental error.

\subsubsection{Periodograms}
\label{periodogram}

We used the Lomb--Scargle (LS) algorithm, described by \citet{1982ApJ...263..835S} and \citet{1986ApJ...302..757H} and implemented in the Interactive Data Language (\textsc{IDL}) software, to  search for periodic signals in the 6843 sources detected. 
The LS algorithm is a variation of the Discrete Fourier Transform, in which a time series is decomposed into a linear combination of sinusoidal functions. 
The maximum power in the resulting periodogram indicates the most likely period, and the height of the maximum power is associated to a false alarm probability.

For all the stars, we searched for periodic signals with periods between 0.02 and 25~days and identified the peak of the periodogram as the most likely period.

\begin{figure}
\includegraphics[width=1\columnwidth]{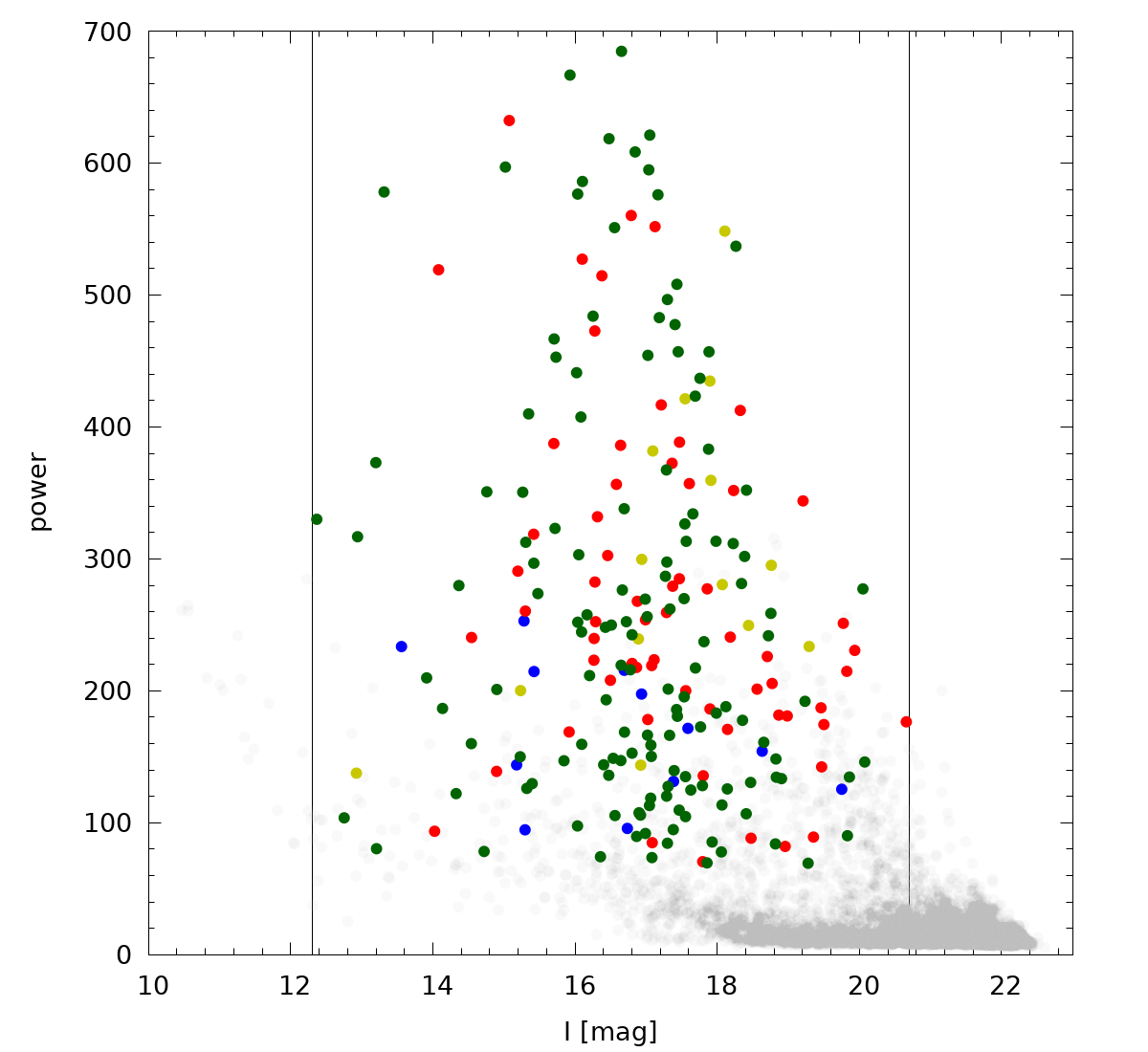}
\includegraphics[width=1\columnwidth]{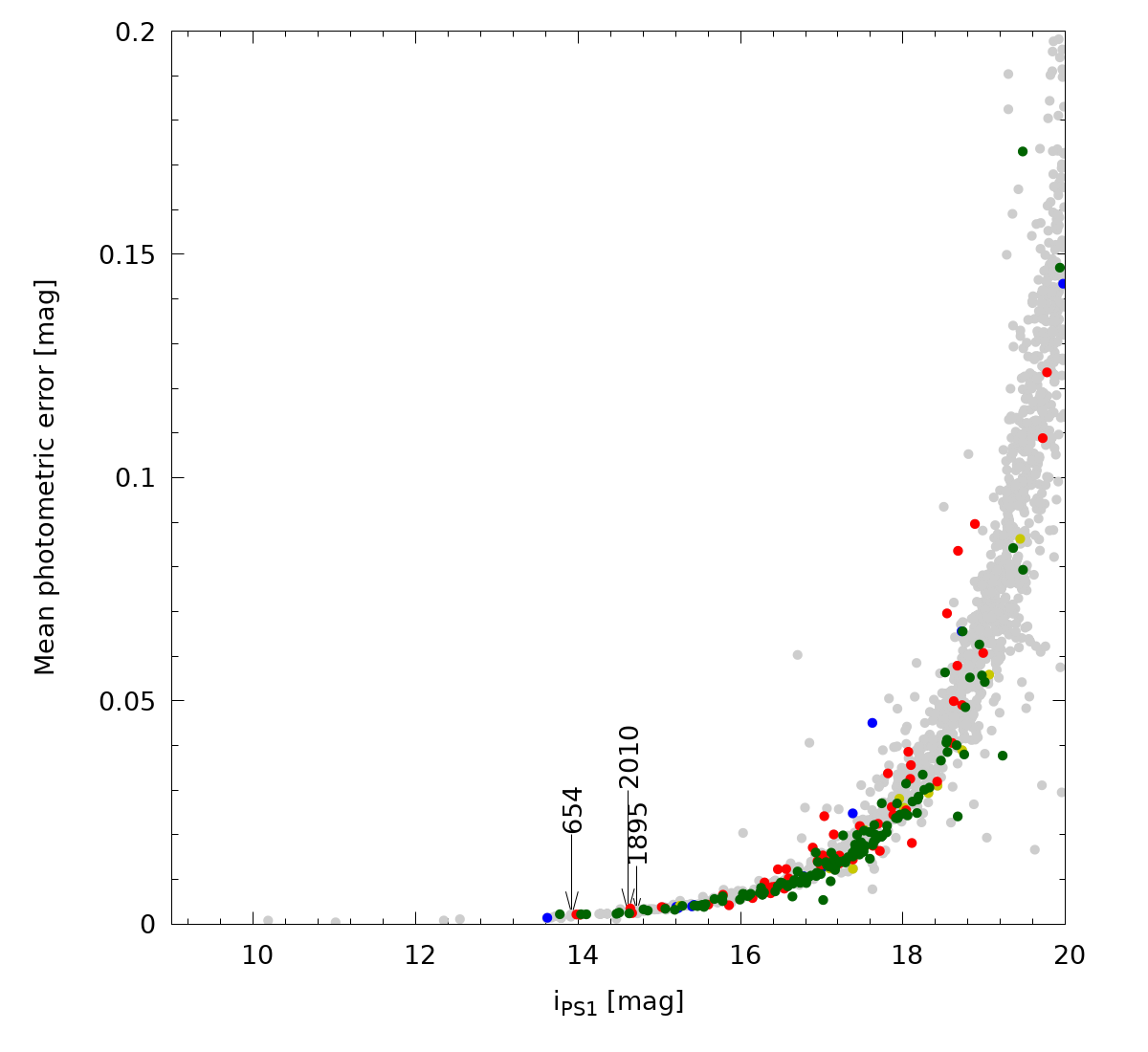}
	\caption{\textbf{Top:} Periodogram power vs. $I$-corrected magnitudes. All sources are shown in grey; the rotation-modulated variable stars, in green; possible rotation-modulated stars, in blue; potential binary stars, in dark-yellow; and non-periodic/miscellaneous variables, in red. The black lines depict the $I_c$ magnitude limits where we found the variables stars (outside this limits, the stars tend to be saturated or too faint to obtain accurate photometry). \textbf{Bottom:} Mean photometric error vs. PS1 magnitude in the $i$-band with the same colours reference. Ninety~per~cent of the periodic stars discovered in this work have a mean photometric error of less than 5~per~cent, while the remainder do not exceed 20~per~cent error. In addition, we show the position of the 3 reference stars marked with an arrow indicating each ID.}
\label{power}
\end{figure}

In order to identify good-quality periods in the light curves, we calculated the False Alarm Probabilities (FAP) using the code \textsc{PracticalLombScargle} from package \textsc{astropy.timeseries}\footnote{http://www.astropy.org} \citep[][]{2018ApJS..236...16V} which uses the default \textsc{baluev} method (similar results were obtained with the other available methods). 
The FAP indicates the probability of reaching a given maximum in a power spectrum due to Gaussian noise.  
We find that, depending on the stellar brightness, a peak power of 12 to 20 corresponds to a $FAP\approx1$~per cent.  However, this result is very optimistic given the assumption of purely Gaussian noise. In practice, additional sources of noise, such as those due to the changes in air-mass, produce spurious signals with peak powers up to 50 and periods close to 1~day (see top Fig.~\ref{power}). 
Therefore, based on visual inspections and the distribution of observed peak powers seen in top Fig.~\ref{power} (top), we adopt a conservative threshold of 70 for the peak power to identify periodic stars. However, some stars with a period of about 1-day passed visual inspection, although they should be taken with caution.

We further investigated the periodicities in the light curves, with peak powers greater than 70, through the NASA Exoplanets periodogram tool\footnote{\url{https://exoplanetarchive.ipac.caltech.edu/cgi-bin/Pgram/nph-pgram}}, which also utilises the LS algorithm, but allows for an interactive inspection of multiple peaks in the periodogram of each light curve. 
Finally, we inspected the light curves by eye and check the spatial positions of the variables candidates. 
We discarded saturated stars ($I_{c} < 12.3$~mag) and stars close to diffraction spikes and the edges of the detectors. Most of these rejected stars had spurious periods of 0.5, 0.99, and 1~days. 

\subsubsection{Classification of variables}
\label{periodic}

From the results of the periodograms and the visual inspections discussed in  \ref{periodogram}, we classified the variable sources with repeated photometric patterns onto the following categories: 

\begin{itemize}
    \item rotation-modulated: LC with robust periodic signal, e.g. Peak Power $>$~70 and convincing LC from visual inspection (136 sources).
    \item possible: LC with marginal periodic signal, e.g. Peak Power $\sim$~70 and convincing LC from visual inspection (12).
    \item binary: Peak Power $>$~70 and LC with one or more abrupt drops in magnitudes, mimicking eclipses (14).
    \item other: Peak Power $>$~70, but LC with irregular or large periodic variability but uncertain origin (64).
\end{itemize}

\begin{figure*}
	\includegraphics[width=2\columnwidth]{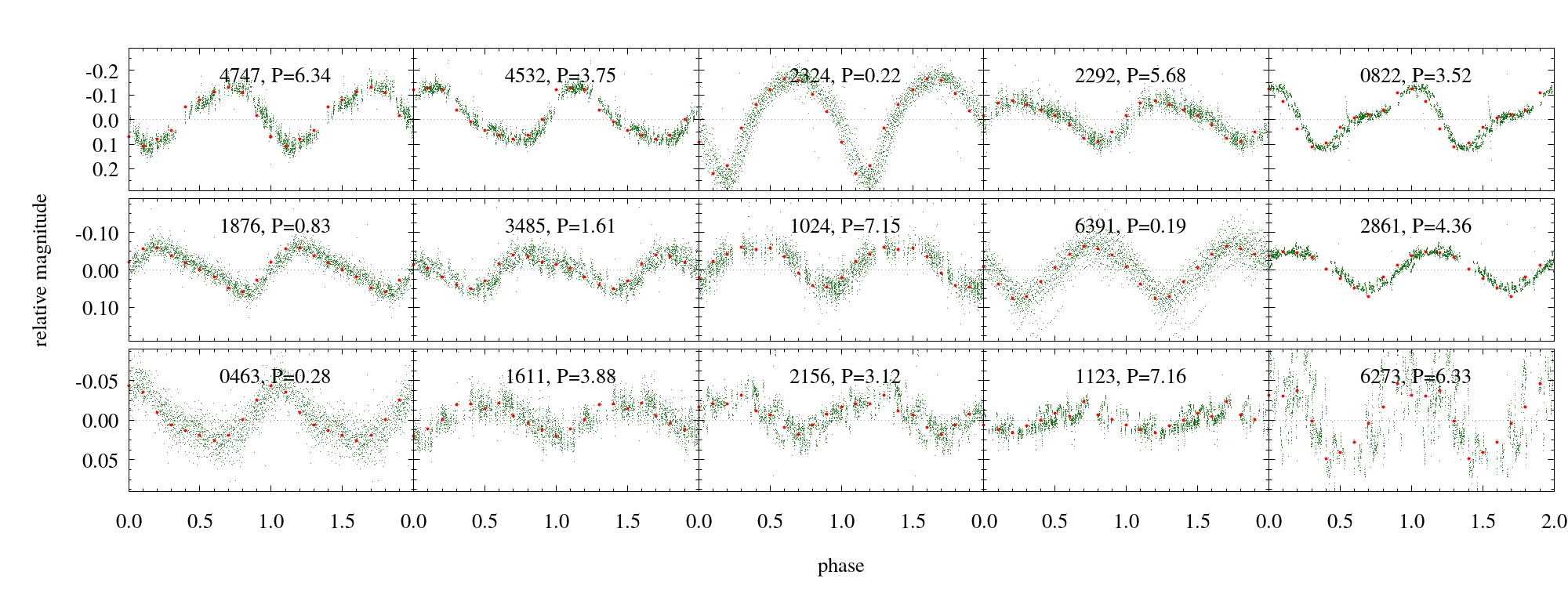}
	\caption{Phased light curves with different amplitudes. Top row: amplitude $\geq$ 0.1 mag. Centre row: between 0.05 to 0.1 mag. Bottom row: amplitude $\leq$ 0.05 mag.	The light curves in each row are sorted by periodogram power. The green dots are the $I$-band data from LCOGT. The red circles are the averages of the differential magnitude every tenth of a phase. The thickness of each light curve is an approximate measure of the sample error.}
\label{panelPrin}
\end{figure*}

We are particularly interested in stars in this FOV that have a robust periodic signal because their periodicity is likely to be modulated by stellar rotation and low-mass PMS stars with photometric rotation periods have a high probability of being  members of the Mon~R2 cluster. We called these sources ``rotation-modulated variables''.
We separated the rotation-modulated variables in three groups according to their amplitude in greater than 0.1 mag, 0.05 to 0.1 mag , and lower than 0.05 mag. In Fig.~\ref{panelPrin} we show five light curves of each group per line and the rest of the rotation-modulated variables LC (grouped as above) and possible periodic in the Appendix~\ref{sec:other} in Fig.~\ref{Amp101}, Fig.~\ref{Amp102} and Fig.~\ref{Amp103}, Fig.~\ref{Amp5101}, Fig.~\ref{Amp5102} and Fig.~\ref{Amp5103}, Fig.~\ref{amp5} and Fig.~\ref{quiza} respectively. 

Finally, we show in Appendix the binary system candidates with one abrupt drop in magnitude in Fig.~\ref{binaries1ec}, and more abrupt drops in magnitudes (mimicking eclipses), in Fig.~\ref{binariesmulti} and the miscellaneous variables Fig.~\ref{other}.

\subsection{GMOS spectroscopic data}
\label{sec:Espectros}

\subsubsection{Spectral classification}

\begin{figure}
	\includegraphics[width=1\columnwidth]{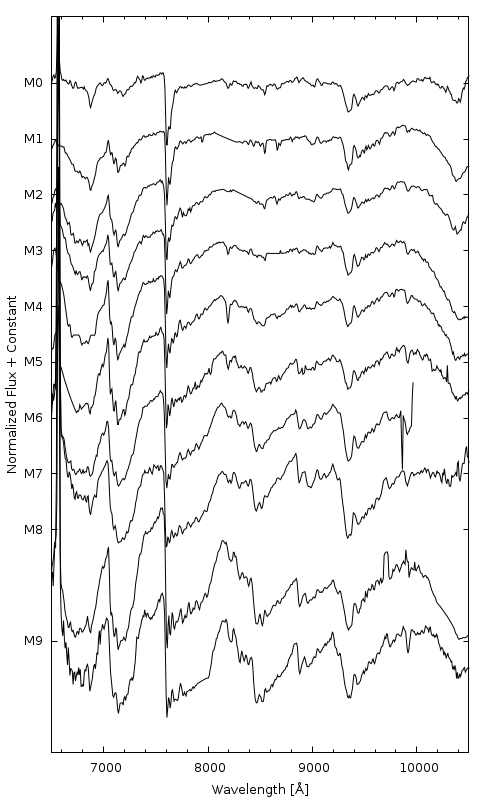}
	\caption{Spectral sequence from M0 to M9 generated with our GMOS spectra. We normalised the spectra with order 5 or 7 polynomial fits. The prominent emission line at 6560~\AA\, is H${\alpha}$. The sequence differences are most evident in the spectral range between $\sim 6500$ to $\sim 8500$~\AA\ (due to TiO and VO bands).}
\label{sequenceM}
\end{figure}

M-type stars present molecular absorption bands mainly from titanium oxide (TiO) and vanadium oxide (VO). These bands are deeper in latter sub-spectral type. For all M-type stars the TiO band is prominent, but the VO band dominates from M5 to later types.
As in \citet{2019MNRAS.487.2937O}, we created our own internal spectral sequence of GMOS data (see Fig.~\ref{sequenceM}) by comparing with published templates \citep{1991ApJS...77..417K,1994AJ....108.1437H,2015hsa8.conf..441A}. 
This  allows for a more direct and self-consistent comparison of all our GMOS spectra.   
For each of our spectra, we performed a visual comparison against the GMOS spectral sequence to  determine the best match. 

From the 12 GMOS fields, we obtained 297 spectra, 229 of which  had enough signal to noise for spectra classification.    
Of these 229 objects,  140 have spectra matching M-type stars, where 58 have  M2 or earlier types, and 82 have  M3 to M9 spectral types.  The rest are likely to be K-type or earlier types, and were not classified further as the GMOS spectra lack the required resolution.
For our spectral classification, we use the reference spectra obtained in the study of the cluster NGC2264 \citep[][]{2019MNRAS.487.2937O}, which were taken with the same instrument. For this reason, we estimate that our spectra have an uncertainty of 1 spectral subtype based on comparisons with results from the literature.

We find that 178 of the  229 targets with useful GMOS spectra have \textit{Spitzer} data, and 86 are variables. 
In Table ~\ref{espectros}, we summarise the number of stars classified using our GMOS data. The sample could not be enlarged as we did not find any useful spectral classifications in the literature.

\begin{table}
\caption{Number of stars according to their spectral type determined in this work. In the ``nonM'' tag we included spectral types earlier than M0, most of them are likely to be K-types. 
}
\label{espectros}
\centering
\begin{tabular}{c c}
\hline
sub-type & number\\
\hline
nonM & 89 \\
M0 & 28  \\
M1 & 17 \\
M2 & 13 \\
M3 & 26 \\
M4 & 15 \\
M5 & 21 \\
M6 & 9 \\
M7 & 5 \\
M8 & 3 \\
M9 & 3 \\

\hline
Total M & 140 \\
\hline
\end{tabular}
\end{table}

\subsection{Membership}
\label{member}

To assess whether the sources in our sample are (sub)stellar members of the Mon~R2 cluster we adopted a variety of criteria based on photometric variability, disc indicators, position in H-R diagram, proper motions and parallaxes.  The different criteria are discussed below.   

\subsubsection{Rotation-modulated variability}

The light curves of the variable stars classified in \ref{periodic} as rotation-modulated variability were visually inspected and it is notable that most of them show sinusoidal shapes. These light curves are typical of very young and active late stars (typically M-types) which present cool or hot stellar spots that cover a significant fraction of their surfaces, thus producing high-amplitude (of a few $\%$ or more), rotation-modulated variations. Cool spot associated with magnetic activity usually  dominate the light curves of non-accreting weak-line T Tauri stars, while hot spots from magnetospheric accretion might dominate the light curves of accreting classical T Tauri stars \citep{1994AJ....108.1906H}.

Field stars can also present periodic signals, but for stars in the temperature range we are working with, the amplitudes are usually less than 1~per~cent of magnitude \citep[][]{2014ApJS..211...24M}.
Considering that we can approximate the amplitude as $Amp=\sqrt{2}*rms$, the rotation-modulated stars in our sample have values above 1~per~cent. Approximately 80~per~cent have amplitudes between 0.01 and 0.1~magnitude, which are high for the amplitudes expected in field stars.

\subsubsection{High-scatter variables}
\label{sub:npVari}

\begin{figure}
	\includegraphics[width=1\columnwidth]{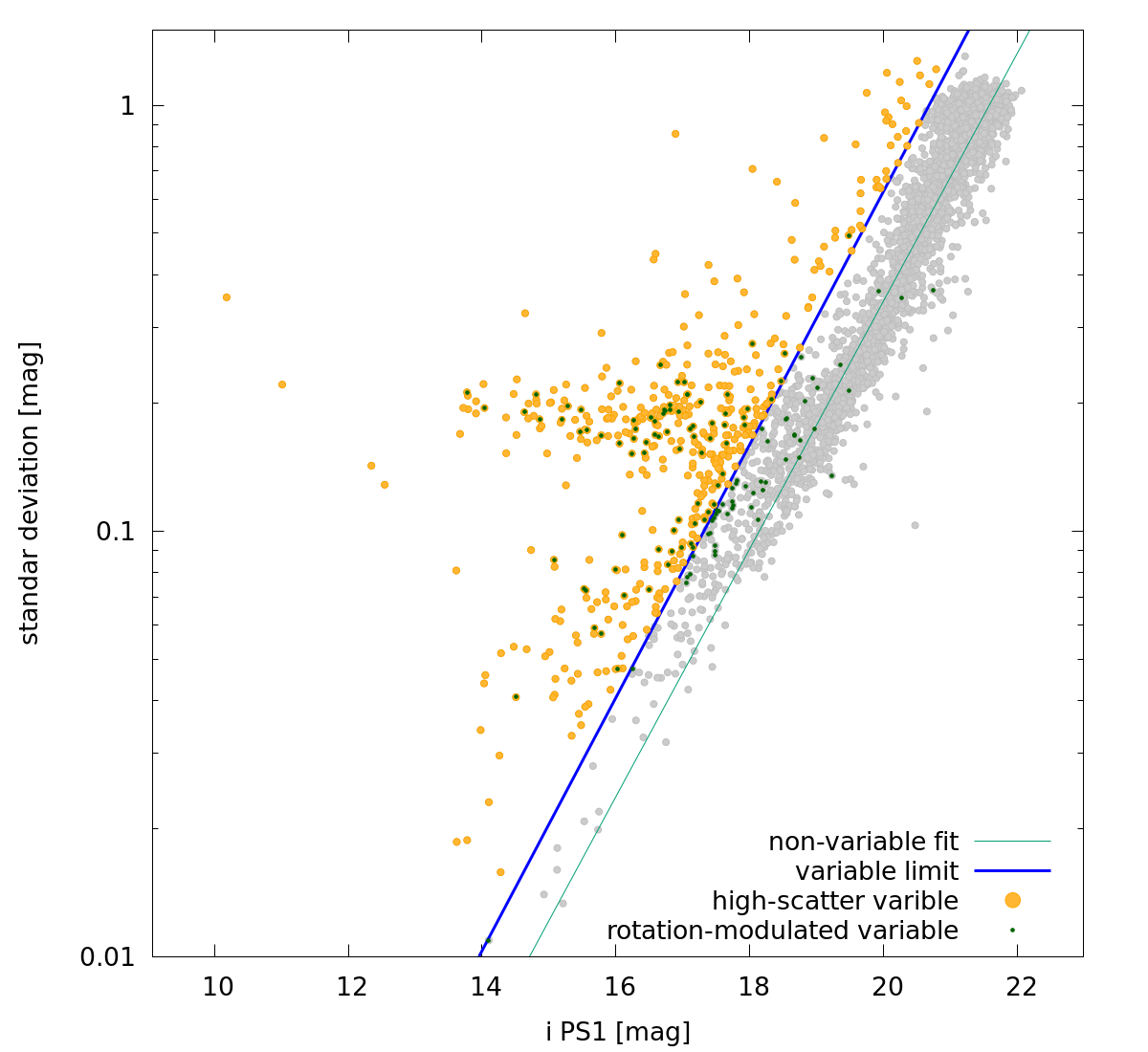}
	\caption{$i$-band (PS1) vs standard deviation in the mean differential photometry in I-band. The green line is a fit ($f(x)=10^{(0.2912*x-6.2837)}$) for the overdensity of objects. Above 3~$\sigma$ of this fit (blue line) are non-periodic variable sources, some of them already located from the periodogram analysis. Most variable sources have a characteristic standard deviation of 0.2~mag. From this figure, we identified a total 444 high-scatter variables (yellow circles).}
\label{nube}
\end{figure}

Using only sources with periodic variability as a membership criterion introduces a bias against accreting stars. Young Stellar Objects (YSOs) may also have non-periodic light curves (see Appendix~\ref{sec:other}), due to variable accretion. One way to find possible non-periodic variables is from the scatter of the light curve with respect to the mean magnitude value obtained. If this scatter or standard deviation is larger than the expected photometric error for a given magnitude, then the probability that a source is intrinsically variable is high. In Fig.~\ref{nube} we show this standard deviation of the instrumental differential magnitude in the $I$-band versus the magnitude in the $i$-band of PS1. In Fig.~\ref{nube} we notice an overdensity of objects, mainly field stars, following a straight line. We make a linear fit to this overdensity and determine a limit above the 3~$\sigma$ of the fit. We note that most of the viable stars have a standard deviation of around 0.2~mag, which we interpret as the characteristic variability amplitude of the T Tauri stars in the cluster \citep[][]{2015A&A...581A..66V}. Then, we define a new membership criterion called ``high-scatter variables'' (``V\_hs'' in Tables~\ref{MemberCrit} and~\ref{MS}), as the sources above the ``variable limit'' line in Fig.~\ref{nube}. Using this procedure, we identify  444 high-scatter variables. Of these 444 objects,  81 are also rotation-modulated variables based on the first criterion defined in section~\ref{periodic} and 65 objects are other types of variables identified by the periodogram analysis.

\subsubsection{Disc identification}
\label{sub:disc}

The presence of IR excess from a protoplanetary disc can also be considered a clear indicator of young and membership to a young cluster as Mon~R2 because protoplanetary discs dissipate very quickly ($<$ 1--10~Myr) and are not present in older stars in the field.
The warm dust ($T\sim100-300$~K) in protoplanetary discs is typically optically thick and produces a characteristic mid-IR excess \citep{2011ARA&A..49...67W}. 
Thus, the \textit{Spitzer}-IRAC [3.6]-[8.0] colour index can be used  as a robust disc indicator. 
These wavelengths are close to the Rayleigh-Jeans regime of the stellar photospheres and thus the photospheric colour shows little dependence on spectral types.
The colour boundary between stars with and without a disc usually occur in [3.6]-[8.0] colours between  0.5 and 0.7 \citep{2007ApJ...671..605C,2019MNRAS.487.2937O}.
We found only 4 M-type stars with [3.6]-[8.0] colours in the 0.5-0.7 mag range and hence adopt [3.6]-[8.0] $>$ 0.5 as disc identification criterion. 
We note that the disc identification criterion is only weakly affected by extinction as the median A$_{V}$ of 2.09~mag determined for Mon~R2 (see below in \ref{subsub:Av}) corresponds to a reddening of just 0.07~mag in the [3.6]-[8.0] colour \citep{2009ApJS..181..321E}.

\begin{figure}
	\includegraphics[width=1\columnwidth]{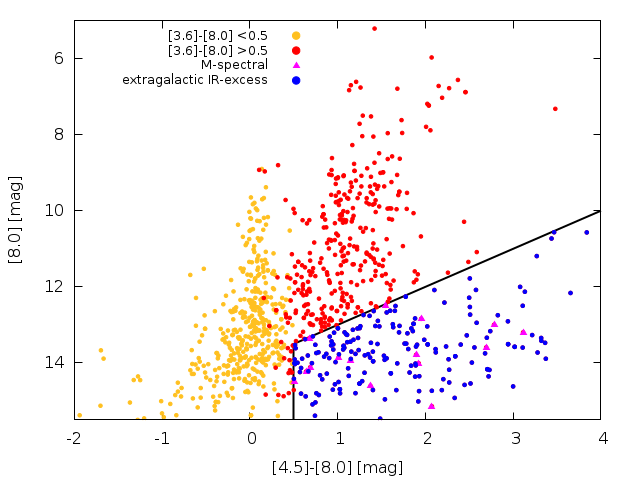}
	\caption{Diagram color-magnitude with \textit{Spitzer} data indicating the location of stars with and  without a disc and the region occupied by extragalactic sources \citep{2006ApJ...644..307H}.  Note that some Spectroscopically confirmed M-type stars with protoplanetary discs (red triangles) fall in the same region of the diagram as galaxies.}
\label{colorMagSpt}
\end{figure}

While the [3.6]-[8.0] colour is very useful to distinguish between stars with and without a disc, we note that extragalactic sources can have similar colours as stars with discs. 
In order to identify potential extragalactic objects, we adopt the criteria from \citet{2006ApJ...644..307H}, 
according to which objects with [4.5]-[8.0] $>$ 0.5 and [8.0] magnitudes fainter than 14  - ([4.5]-[8.0]) are consider potential background galaxies (see Fig.~\ref{colorMagSpt}). 
However,  we find that some spectroscopically confirmed M-type stars with IR excess fall in this region of the [8.0] vs [4.5]-[8.0] diagram, implying that the  \textit{Spitzer} colours  alone are not enough to distinguish between these two types of sources. 
On the other hand, the \textit{Spitzer} data obtained nominally correspond to galactic objects (\textsc{irac\_obstype}=0). However, we have performed a visual inspection of the possibly extragalactic objects, discarding 10 extended sources and about 50 sources that do not appear in the $I$-band super image. The remaining sources have a high probability of being background or cluster sources.

\begin{figure}
	\includegraphics[width=1.\columnwidth]{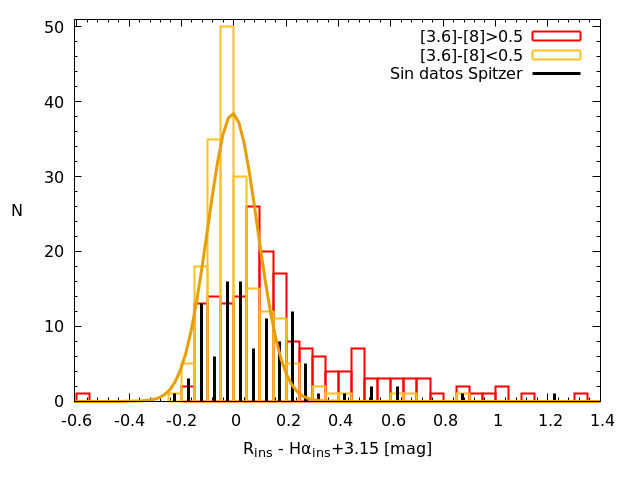}
		\caption{ $(R-$H$\alpha)+3.15$ (instrumental magnitude) histogram with and without \textit{Spitzer} data. Two different distributions are observed. For objects with $[3.6]-[8.0]<0.5$ the distribution is well-defined, with a peak centred at -0.025 and to be cut off near the value 0.25, while the distribution with $[3.6]-[8.0]>0.5$ has a peak centred at 0.05 but has a tail at values above 0.25. Objects without \textit{Spitzer} data beyond 0.25 have a high probability of having emission in the H$\alpha$ line due to the presence of a circumstellar disc.}
\label{Hacolormag}
\end{figure}

PMS stars surrounded by protoplanetary discs can also be identified by the  presence of strong H$\alpha$ emission indicating accretion. Therefore, we also used our LCOGT photometry in $R$, $I$ broad bands and the H$\alpha$ narrow band to identified additional targets with accretion discs.  
We find a total of 838 sources with high quality H$\alpha$ and $R$ photometry (S/N $>$ 10). From this photometry, we find that most stars have instrumental $(R - $H$\alpha)$ colours of -3.15  (see Fig. \ref{Hacolormag} centred in 3.15). In Fig.~\ref{Hacolormag}, two distributions are clearly visible for objects with $[3.8]-[8.0]<0.5$ (180 sources) and $[3.8]-[8.0]>0.5$ (197 sources). The cut-off of the $[3.8]-[8.0]<0.5$ objects around the value 0.25 is also clear. While those with $[3.8]-[8.0]>0.5$ extend beyond that value. The black peaks are the objects that do not have enough \textit{Spitzer} data (106 sources) to define the presence of a disc. In addition, the few sources with $[3.8]-[8.0]<0.5$ beyond 0.25 have a high probability of having a disc due to their high emission in H$\alpha$. They could be discs with dust cavities (known as transition discs), which is why \textit{Spitzer} does not detect it \citep[figs. 1 and 2][]{2014prpl.conf..497E}. Therefore, we have adopted $(R-$H$\alpha)+3.15>0.25$ as a disc identification criterion.

Using this criterion, we added 20 sources with $(R-$H$\alpha)+3.15>0.25$ to the list of member candidates. 
This criterion is very restrictive, and we seek not to contaminate our sample of stars with disc with objects without disc. \citet[][]{2011MNRAS.415..103B} shows an overlap between Weak T Tauri stars and stars without H$\alpha$ line. With this criterion we only detect two stars without IR-excess and with H$\alpha$ emission line. 

\subsubsection{\textit{Gaia} distances and proper motions}
\label{sec:Gaia}
We also studied membership to Mon~R2 using the  5-parameters solutions ($\alpha$, $\delta$, $\mu_{\alpha} \cos\delta$, $\mu_\delta$, and $\varpi$) in the \textit{Gaia} EDR3 database \citep[][]{2021A&A...649A...1G,2021A&A...649A...5F}.
Based on position matches within 1~arcsec,  we found $\sim$1600 coincidences with our 
LCOGT photometric sample with parallax and proper motions data. Regarding parallaxes, \citet{2021A&A...649A...4L} explains their limitations and the different approaches to calculate distances.

\begin{figure}

    \includegraphics[width=1.\columnwidth]{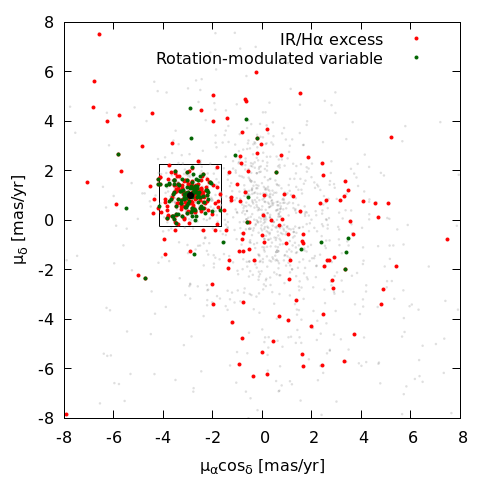}
	\caption{The proper motions from \textit{Gaia} EDR3 data of our LCOGT sources are shown in grey (transparent) dots. The rotation-modulated variable stars are shown as green circles, while the objects with IR or H${\alpha}$ excess are shown as red circles. A clear conglomeration of periodic and IR or H${\alpha}$ excess sources is seen around $\mu_{\alpha}cos(\delta)=-2.75$~mas~yr$^{-1}$ and $\mu_{\delta}=1.15$~mas~yr$^{-1}$. We define a square $2.5~masyr^{-1}$ ($\sim2.5\sigma$) on a side around this kinematic centre as an additional criterion to identify member candidates.   
	}
\label{pmotion}
\end{figure}

\begin{figure*}
	\includegraphics[width=2\columnwidth]{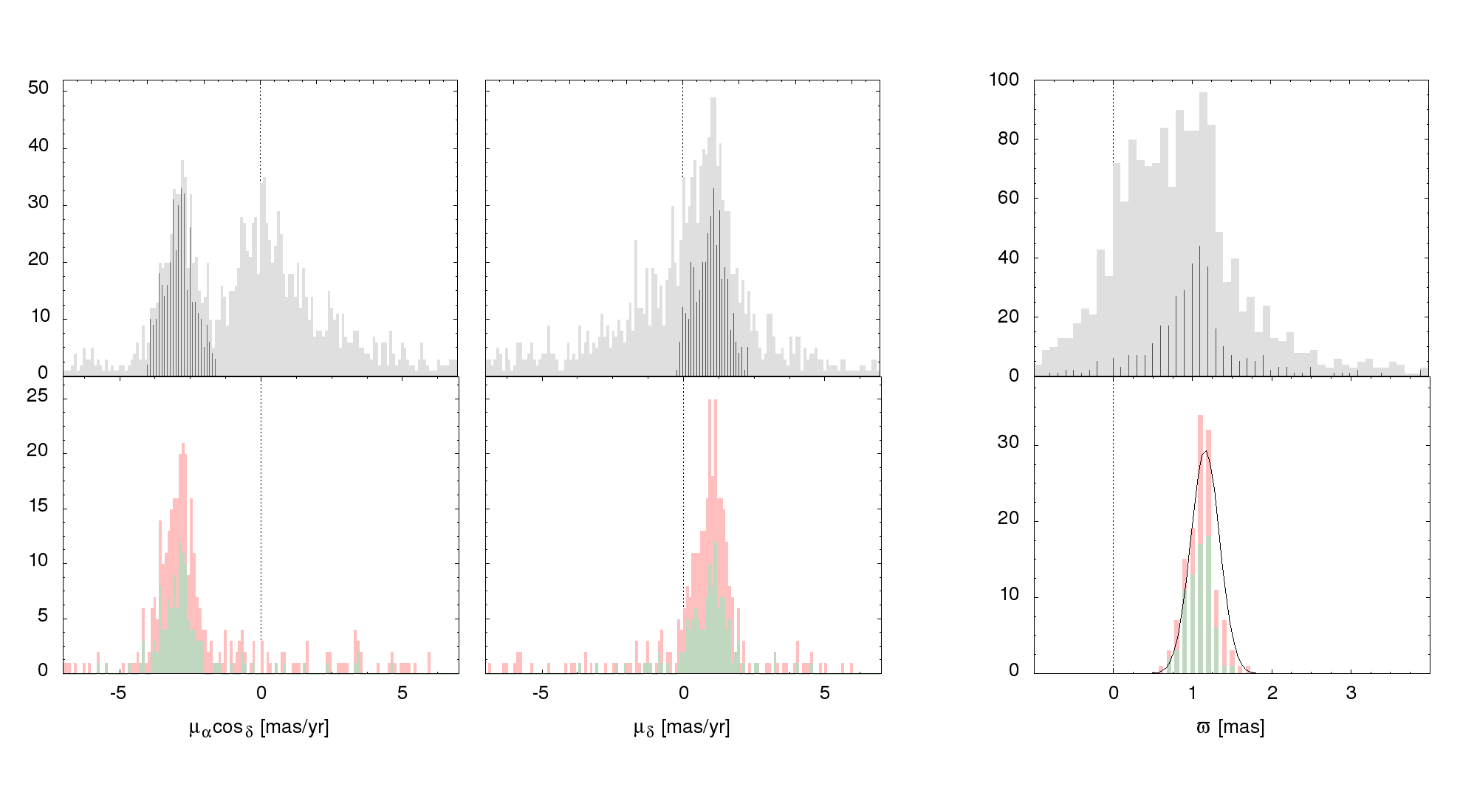}
	    \caption{Member candidates from \textit{Gaia}. \textbf{Top:} histograms of the 1600 LCOGT objects with 5-parameter data from EDR3. The left panel corresponds to $\mu_{\alpha}cos(\delta)$, where the objects with $\mu_{\alpha}cos(\delta)$ =-2.75$\pm$1.25~mas~yr$^{-1}$ are shown in dark-grey. The centre panel corresponds to $\mu_{\delta}$, objects with $\mu_{\delta}$ =1.15$\pm$ 1.25~mas~yr$^{-1}$ shown in dark grey. 
	    The right panel corresponds to the parallax data, where the dark-grey fills are the $\varpi$ of objects restricted by the proper motion limits indicated above. The dark-grey objects correspond to objects inside the box in Fig.~\ref{pmotion} and therefore cuts have been applied in both proper motion dimensions.} \textbf{Bottom:} The same data as in the  top, but limited to the rotation-modulated variables (green) and non-periodic sources with IR or H${\alpha}$ excess (red stacked bars). The right panel shows the $\varpi$ with the additional restriction of  $\varpi$ errors < 20~per~cent.
	    
    \label{pmHisto}
\end{figure*}

The first two parameters, $\alpha$ and $\delta$, are defined by the limits of  the field obtained with LCOGT. 
The other three parameters were constrained by statistical considerations. In Fig.~\ref{pmotion}, we plot the $\mu_{\alpha}cos(\delta)$ vs. $\mu_{\delta}$ plane of our LCOGT sources with EDR3 data. 
Clearly, a clump is appreciated around ($-2.75$, $1.15$)~mas~yr$^{-1}$. Real cluster members might have different proper motions due to dynamical interaction with other members during or after formation. Therefore, objects with proper motions outside the box can not necessarily be excluded as cluster members.
Since there is no unique way to define a `membership region' around this kinematic centre, we try to define the limits from the observed distributions.  
In Fig.~\ref{pmHisto},  we show the distributions of proper motions. 
We find a well-defined bimodal distribution in $\mu_{\alpha}cos(\delta)$, with a peak around $\mu_{\alpha}cos(\delta)=-2.75$~mas~yr$^{-1}$ in addition to the peak at $\mu_{\alpha}cos(\delta)=  0.0$~mas~yr$^{-1}$ expected from very distant sources.  
We took advantage of the bimodality to set the half-width of the distributions to $1.25$~mas~yr$^{-1}$ (equivalent to 2.5~$\sigma$) so that the upper limit in $\mu_{\alpha}cos(\delta)$ coincides the with the observed minimum between the bimodal distributions. 
Since the $\mu_{\delta}$ distribution does not present this bimodality, we simply adopt the same half-width of $1.25$~mas~yr$^{-1}$. Hence, we define a square membership region with a kinematic centre at $(-2.75;1.15)$~mas~yr$^{-1}$ and a size of $2.5$~mas~yr$^{-1}$ on each side.  

To verify that the adopted ranges in proper motions are appropriate to identify new cluster member candidates, we inspect the distributions of proper motions of the rotation-modulated variable sources and/or sources with IR or H$\alpha$ excess. 
We find that their proper motions are very consistent with the adopted ranges (see Fig. \ref{pmHisto} bottom left and centre panel).
We emphasise that excellent agreement between independent membership indicators (periodicity, IR/H$\alpha$ excess,  and kinematics) reinforces each one of the criteria adopted to select cluster member candidates.   

In Fig \ref{pmHisto} we show the parallax's distribution in top right panel of the whole sample in light grey with the first selected sample in proper motion superimposed in dark grey. Even in the latter sample, there is a high dispersion that even contains negative parallaxes. 
Using rotation-modulated sources and/or sources with IR or H$\alpha$ excess and selecting sources with errors in parallax smaller than 20~per~cent, as shown in the right-bottom panel of Fig \ref{pmHisto}, we adopt limits for the candidate member parallaxes. Thus, fitting a Gaussian to this final distribution, we adopt limits from 0.888 to 1.42~mas for the candidate member parallaxes. This values correspond to the 1.5~$\sigma$ limits of the distribution.
We find that 206 sources fulfil the  adopted five-parameters requirements from \textit{Gaia}, 146 of them are also rotation-modulated variables and/or present IR or H$\alpha$ excess.

\subsubsection{PS1 colour-magnitude diagram}
\label{sec:PS1}
Another membership criterion is the selection of photometric sources which fall over, i.e. redder and brighter, the isochrone representing the Zero Age Main Sequence (ZAMS) of the Mon~R2 cluster. To perform this task,  we generated the $i-y$ vs. $i$ diagram using the PS1 data and placed the corresponding ZAMS (considering the distance modulus and mean reddening of the cluster). 
We selected the photometric data in both $i$ and $y$ bands from DR2 and DR1 (prioritising DR2). We cross-matched 3078 LCOGT targets with $i$ and $y$ data from PS1 with well-defined errors in DR2 and DR1. These sources have $10.6<i<23.1$~mag and $10.7<y<20.9$~mag. We analysed different combinations of bands for the colour-magnitude diagram but considered $i-y$ vs. $i$ to be the least affected by reddening (see subsection~\ref{subsub:Av}).

\begin{figure}
	\includegraphics[trim=0.cm 0cm 0cm 0cm, clip=true, width=1\columnwidth]{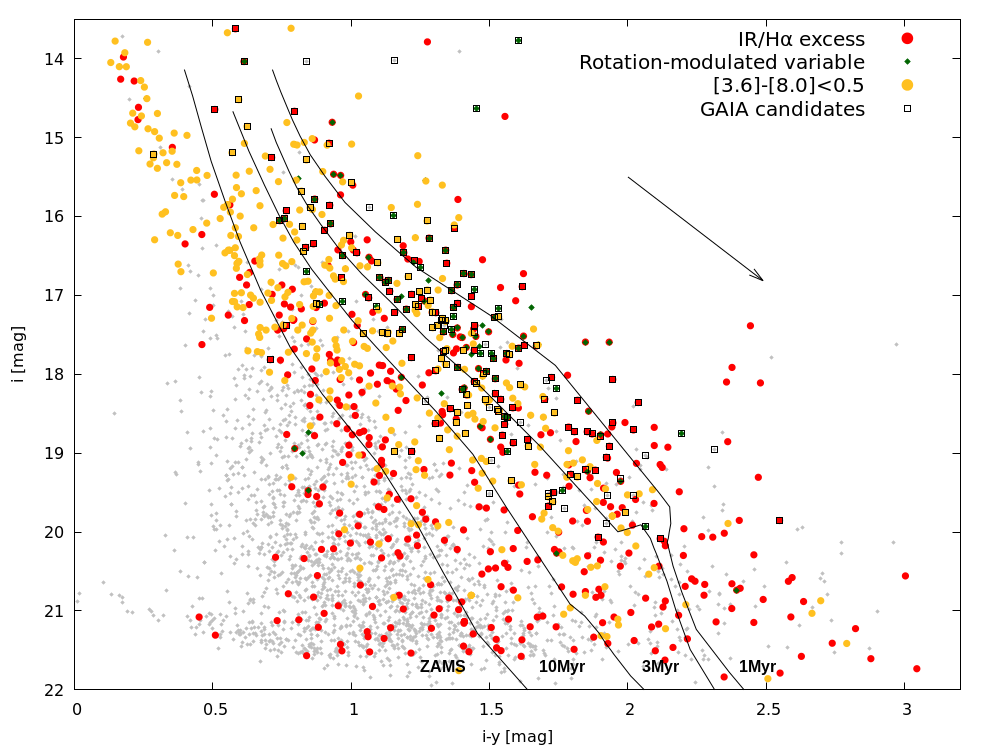}
	\caption{Colour-magnitude diagram of sources in the Mon~R2 field with PS1 photometry (grey dots). We overplotted four lines depicting the 1, 3, 10~Myr and ZAMS isochrones (see text). Note how \textit{Gaia} candidates and periodic variables seem to concentrate around the 3~Myr line.}
\label{HR}
\end{figure}

We used the isochrones from the Lyon University web from Phoenix\footnote{\url{https://phoenix.ens-lyon.fr/Grids/BT-Settl/}}. 
We used the 2015 BT-Settl model with $[M/H]=0.0$ and AB photometry consistent with PS1.
The BT-Settl model includes low-mass stars down to the hydrogen-burning limit and consistently couples atmosphere and interior structure. BT-Settl present $T_\mathrm{eff}$=15000 to 1500~K and age of pre-sequence, starting from 1 Myr \citep{2015A&A...577A..42B,2016sf2a.conf..223A}. 
We selected four isochrones, 1, 3, 10~Myr and the ZAMS, and we shifted them by $m-M=9.58$ (see \ref{sec:distance}), $A_i$=1.315~mag, and E($i-y$)=0.488~mag (see \ref{subsub:Av}).
In Fig.~\ref{HR} we show the $i-y$ vs. $i$ diagram with all data collected and isochrones. As all samples selected with a single membership criterion, the member candidates located above the ZAMS should be taken with caution.  In particular, this sample could be contaminated by binary objects \citep{1998MNRAS.300..977H} and field stars. The uncertainty in this criterion could be as high as $\approx$~50~per cent (see Appendix~\ref{Ap:zams}). Therefore, we caution that this criterion should not be considered, by itself, as a reliable membership indicator and that  the degree of contamination increases as the objects get closer to the ZAMS. 
We find that 1267 PS1 objects are located above the ZAMS isochrone. 
We also show the placements of \textit{Gaia} member candidates, periodic variables, and stars with disc indicators. Their positions clearly indicate the youth of the cluster. 

In Table~\ref{MS}, we show the Mon~R2 member candidates distinguishing the different flags, i.e. rotation-modulated and high-scatter variables, star with IR or photometric H${\alpha}$ excess, and \textit{Gaia} 5-parameter and a summary of the membership criteria is shown in Table~\ref{MemberCrit}. Also, we provide the period of photometrical variations, the spectral type and the presence of H$\alpha$ in emission.

Finally, we obtained 1566 candidate members with at least one flags and 23 objects have 5 flag simultaneously (rotation-modulated, high-scatter, Disc, ZamsFlag and GaiaFlag = 1). In addition, only one star with variability not fulfilling the membership criteria, although this may be due to insufficient information i.e. \textit{Spitzer}, PS1 and/or \textit{Gaia} data (source with ID=4579).

 We emphasise that all sources that meet some membership criteria are member candidates, and note that no single criterion, by itself, guarantees membership. The membership criteria should be used with caution, keeping in mind the balance between completeness and reliability. The larger the number of simultaneous membership criteria used, the larger the reliability will be, but the lower the completeness.

There are few studies of membership in this cluster. \citet[][]{1997AJ....114..198C} studied Mon~R2 in the near-IR ($J$, $H$,  and $K$-band). They identified 115 member candidates with an $A_{V}$ less than 11.3~mag and 309 objects brighter than 14.5~mag in $K$-band in a field of view of 15´ x 15´ centred approximately at the same place as our FOV. They also  estimated a total stellar population of at least 475 stars.  This number is broadly consistent with our results, in which we identify 1267 objects  above the ZAMS and estimate a contamination up to 50~per cent, for an estimated membership of 600 in a slightly larger field of view (26.5´x 26.5´). However, we note that the member candidates from \citet[][]{1997AJ....114..198C} might contain several deeply embedded objects that are not detected in our optical observations.

\begin{table}
\caption{Summary of membership criteria.}
\centering
        \begin{tabular}{| p{1.5cm} | p{5cm} | p{0.5cm} |}
    \hline
        Flags & Description & n \\ 
    \hline
    \hline
    ZamsFlag  & Objects are located above the
    ZAMS isochrone.   & 1267  \\
    \hline
    GaiaFlag  & Objects with 5-parameters of membership from \textit{Gaia}: $\alpha$ and $\delta$, are defined by the limits of  the field obtained with LCOGT. Square $2.5~masyr^{-1}$ ($\sim2.5\sigma$) on a side around kinematic centre in $\mu_{\alpha}cos(\delta)=-2.75$~mas~yr$^{-1}$ and $\mu_{\delta}=1.15$~mas~yr$^{-1}$. And $\varpi$ between 0.888 to 1.42~mas. & 206     \\
    \hline
    V\_type   &  Value 1. Objects showing periodic variability consistent with rotation modulation  in this work. & 136          \\
    \hline
    V\_hs   &  objects showing high photometric scatter for given magnitude in Fig.~\ref{nube}. & 444         \\
    \hline
    Disc      & Objects with IR or H$\alpha$ excess consistent with disc presence & 515          \\
        \hline
    \end{tabular}
    \label{MemberCrit}
\end{table}

\vspace{0.5cm}

\subsection{Fundamental parameters of Mon~R2} 

After presenting the members candidates, we now discuss the fundamental parameters of the Mon~R2 cluster. However, these tasks were performed somewhat iteratively. For example, the distance modulus and A$_V$ obtained affect member candidates selected based on their position in PS1 colour-magnitude diagram.

\subsubsection{Distance}
\label{sec:distance}

In literature, the distance to Mon~R2 has been calculated mainly studying its molecular clouds and brightest stars. 
\citet{1968AJ.....73..233R} studied the brightest stars in colour magnitude diagrams and obtained a distance of $d=830\pm50$~pc. 
\citet{1976AJ.....81..840H} improved the study and found essentially the same distance. 
\citet{2011A&A...535A..16L} derived a distance of $d=905\pm37$~pc creating a 
dust column density map with 2MASS data. 
\citet{2014ApJ...786...29S} used the PS1 photometry data and obtained to distances, $d=830\pm83$~pc for clouds near the core and $d=1040\pm104$~pc for the “Crossbones” toward the northeastern edge. 
\citet{2016ApJ...826..201D} studied the cluster with Very Long Baseline Interferometry and derived a parallax $ \varpi = 1.12 \pm 0.05$~mas, that corresponds to a distance $d = 893^{+42}_{-40}$~pc.  
One of the most recent papers regarding the distance is \citet{2019ApJ...870...32K}. They used \textit{Gaia} DR2 to estimate the distance of 28 Open Clusters and associations with ages from 1 to 5~Myr. For Mon~R2, they found a distance of $d=948_{-38}^{+42}$~pc. 
 
We decided to employ the geometric and photogeometric distances calculated by \citet[][hereafter BJ21]{2021AJ....161..147B}, for almost 1.5 billion stars. 
We obtained 1605 matches with distances in BJ21. 
We made our own distance determination using the distances calculated by BJ21 for the sources with \textit{GaiaFlag}=1 and $\sigma_{\varpi}/\varpi<20$~per~cent and removing a few clear outliers in distance. 
We obtained $d=834\pm79$~pc using the mean photogeometric distance with 100 sources with a distance distribution peak at $d=825$~pc.
We adopt this latter distance, which in turn corresponds to a distance modulus of 9.58~mag. 

Finally, we note that the results of GDR2 are different from those of EDR3. Also, we studied these results without taking into account variability and excess IR and the distances of EDR3 are still lower than those of GDR2 (see Appendix~\ref{sec:distEdr3}).\\

\subsubsection{$A_{V}$}
\label{subsub:Av}

\begin{figure}
	\includegraphics[width=1\columnwidth]{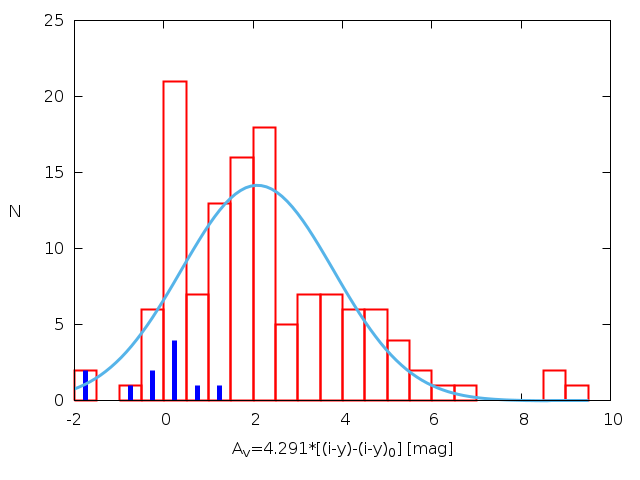}
	\caption{\textbf{Top:} $A_{V}$ histogram in ($i-y$) colour. In blue boxes are the spectroscopic brown dwarf with $A_{V}$ $\sim$ 0.0~mag. Also, we noted that three out of five objects with extreme values are variable stars. The blue spikes represent spectroscopic brown dwarf stars.}
\label{Av}
\end{figure}

Mon~R2 is a HII region with dark nebula zones and regions of active star formation.  
To characterize the A$_{V}$  in the region, we studied the distribution of PS1 colour excesses, E($r-i$), E($r-z$), E($i-y$), E($i-z$) and E($z-y$), in stars with known spectral types. We used the extinction curves from \citet[table 3]{2019ApJ...877..116W} (with $R_{V}=3.16\pm0.15$~mag) and  
employed the intrinsic colours determined by \citet[table 4 \& 5]{2018ApJS..234....1B} for M0--T9 dwarfs (also using PS1 photometry,  see Appendix~\ref{ColourExc}). 
We note that some stars resulted with (unphysical) negative E($r-i$) and E($r-z$) values. This is likely to be due to a combination of factors, including the uncertainties in the photometry and spectral types and, in some highly accreting sources, the contribution of the H$\alpha$ emission to the $r$-band flux of late M-type stars.
Also, since the $A_{V}$ values obtained from colours using adjacent filters, like ($z-y$) and ($i-z$),  are very sensitive to error in photometry, we decided to use E($i-y$) to estimate $A_{V}$.
In Fig. \ref{Av}, we show the results of $A_{V}$ using the ($i-y$) PS1 colour. We find a mean $A_{V}$ of $2.09 \pm 1.7$~mag with a few outliers outside the   -1 to 7~mag $A_{V}$ range.
 The distribution of extinction is the result of the differential absorption produced by the molecular cloud in which the cluster is embedded, which has significant structure. 
 
Finally, the $A_V$ reached corresponds to a stellar extinction obtained from visible stars with spectral type found (spectral type M). Mon~R2 contains areas with star-forming activity, so there are very dense clouds of up to more than 30 mag \citep[i.e.][]{1997AJ....114..198C,Andersen_2006,2019MNRAS.483..407S}.

\subsubsection{$Age$}
\label{subsub:Age}

According to the features studied in Mon~R2, we observe that it is a young stellar cluster with an heterogeneous population in terms of age. Indeed, \citet[][]{1976AJ.....81..840H} found evidence for the existence of at least two stellar groups with different ages. The first would reach an age between 6 and 10~Myrs and the second between 10$^{4}$ and 10$^{5}$~yrs. The first estimate is given by the presence of B stars without emission lines, with the B1 stars located on the main sequence. The second estimate is derived from the presence of maser emission in the central region, indicating ongoing star-formation. This idea is supported by \citet[][]{1992PhDT.........2X} and \citet[][]{1994ApJ...430..252X} who studied the shell structure and outflows. \citet[][]{1997AJ....114..198C} made an HR diagram from photometric (near-IR) and spectroscopic data and estimated a characteristic age of~$\leq$3~Myrs.

To estimate the mean age of Mon~R2, we plotted the member candidates and different isochrones on the colour-magnitude diagram (CMD) and calculate the number of members between isochrones. 
The age of the cluster was determined as the range with the highest density of members. 

We used the isochrones described in subsection \ref{sec:PS1} in $i$ vs $i-y$ CMD, corrected by the distance and mean extinction obtained in the previous sections, in 1~Myr steps from 1 to 10~Myr. Older stars were grouped in two additional bins, 10-50 Myr and 50 to 100 Myr. We extracted the 248 objects that are above the ZAMS in the H-R diagram and meet  at least two  of the membership criteria  (\textit{Gaia} candidates, rotation-modulated variables, high-scatter variables, and stars with discs) and found that  $\sim$50~per~cent of these stars have ages  $<$~2~Myr, and than $\sim$60$\%$ of the member candidates have ages between $\sim$1 and 3 Myr. We, therefore,  adopt a mean age of 2~Myr with a characteristic dispersion of $\sim 1$~Myr, corresponding to the size of the bins used. 
These results are in line with those of \citet[][]{1997AJ....114..198C}.

\subsection{Brown dwarf and brown dwarf candidates}
\label{subsec:BD}

\begin{figure}
	\includegraphics[width=1\columnwidth]{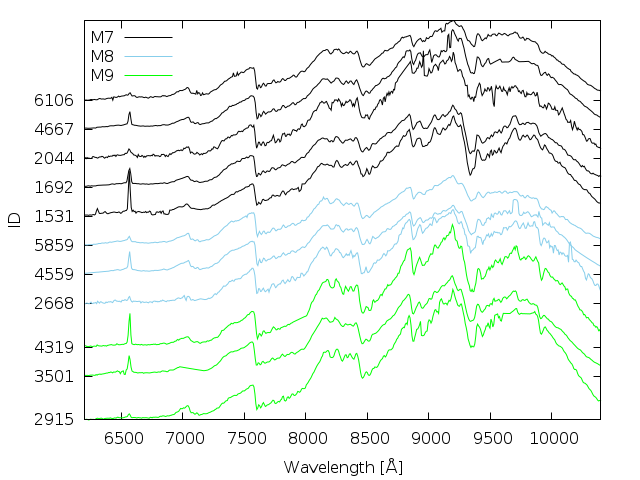}
	\caption{M7, M8 and M9 normalized spectras of BD stars. The M7 stars are in black line, M8 in light-blue line and M9 in green line.}
\label{BDspec}
\end{figure}

\begin{figure}
	\includegraphics[width=1\columnwidth]{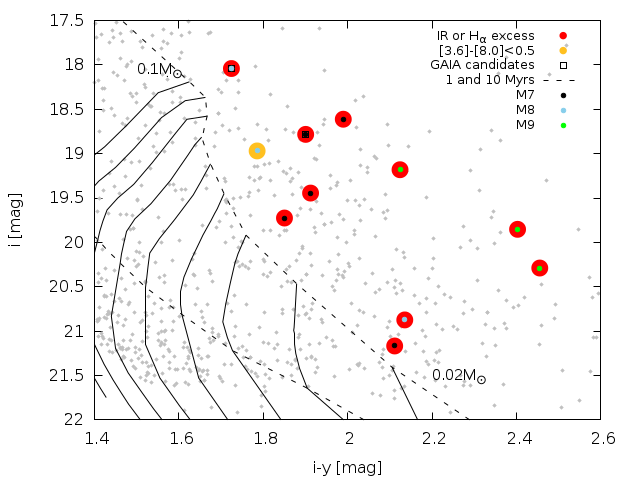}
	\caption{($i-y$) vs $i$ CMD with 1 and 10~Myr isochrones and evolutionary tracks from 0.02 to 0.1~M$_{\odot}$. The position of the spectroscopically confirmed brown dwarfs are indicated by coloured circles.In this area we find BDs candidates but we cannot confirm them as there is a contamination of low-mass stars with high reddening. }
\label{iyi}
\end{figure}

In Section \ref{sec:Espectros}, we found 11 objects with spectral types between M7 to M9. 
Most of these objects clearly show H${\alpha}$ emission in their spectra (see Fig. \ref{BDspec}). 
According to their A$_V$ (blue spikes in fig. \ref{Av}) and position in the 
$i-y$ vs $i$ colour-magnitude diagram (fig. \ref{iyi}), these M7 to M9-type objects are consistent with very young ($<$~1~Myr) brown dwarfs (BD) with little to no extinction.  The youth and low extinction of these objects are likely to be selection effects because more extincted or older BD would be too faint for our GMOS observations.  
The eleven spectroscopically confirmed BD are listed in Table \ref{tab:BDspec}, while the photometric BD candidates are listed in  Table \ref{tab:BDcand} in Appendix \ref{sec:BDcand}.
These photometric BD candidates are objects that occupy the same region in the  $i-y$ vs $i$ CMD (fig. \ref{iyi}) as the spectroscopically confirmed BD. However,  this list should be taken with a lot of caution as it is likely to be contaminated by reddened low-mass stars.

\begin{table*}

\begin{adjustbox}{max width=\textwidth}

\centering

\begin{tabular}{|l|l|l|l|l|c|l|l|l|c|l|c|c|c|c|c|c|c|}
    \hline
     Id & RA & DEC & I & rms & Idif & V\_type & P(days) & Power & V\_hs & H$\alpha$  & SpT & Disc & ZamsFlag & GaiaFlag & Gaia\_id & Spz\_id & PS1\_id \\
    \hline 
   (1) & (2) & (3) & (4) & (5) & (6) & (7) & (8) & (9) &  (10) &  (11) & (12) & (13) & (14) & (15) & (16) & (17) & (18) \\
     \hline
6818.0 & 6:8:7.1856 & -6:33:59.796 & 19.0309 & 0.00975613 & 2.02544 & 0 & 0 & 73 & 1 & - & - & 0 & 1 & 0 & 3018444780168540288 & J060807.18-063359.3 & 100120920299320610 \\
6725.0 & 6:7:50.2488 & -6:33:58.788 & 22.3372 & 0.162477 & 5.33534 & 0  & 0  & 10 & 0  & -  & - & - & 1 & 0 & 3018444569715178112 & J060750.25-063358.3 & 100120919593990938 \\
6781.0  & 6:8:18.3696 & -6:33:57.06  & 22.5164 & 0.26959 & 5.5192 & 0 & 0 & 17 & 0 & - & - & - & 1 & 0 & - & J060818.36-063356.5 & 100120920766011381 \\
6711.0 & 6:7:57.3912 & -6:33:56.736 & 21.6147 & 0.0929442 & 4.61286 & 0 & 0 & 16 & 0 & - & - & - & 1 & 0 & 3018444638434525696 & J060757.39-063356.4 & 100120919890031543 \\
6674.0 & 6:7:7.7976  & -6:33:56.556 & 20.2315 & 0.0267705 & 3.22733 & 0 & 0 & 36  & 1 & - & - & 1 & 1 & 0 & 3018446768737432576 & J060707.81-063355.7 & 100120917825831736 \\
6718.0 & 6:8:6.2544  & -6:33:56.232 & 19.1179 & 0.0108045 & 2.11367 & 1  & 3.515 & 168  & 0 & - & - & 0  & 1  & 1 & 3018444745809316864 & J060806.25-063355.8 & 100120920260641786 \\
6719.0 & 6:8:20.8008 & -6:33:54.468 & 23.7159 & 0.930684 & 6.72568 & 0 & 0 & 10 & 0 & - & -  & - & 1 & 0  & -  & J060820.78-063353.8 & 100120920865812352 \\
6661.0 & 6:7:22.6368 & -6:33:52.128 & 19.9262 & 0.0223622 & 2.92198 & 0 & 0 & 54 & 1 & - & - & 1 & 1 & 0 & 3018446940535925760 & J060722.64-063351.6 & 100120918443553172 \\
6683.0 & 6:8:2.8656  & -6:33:50.256 & 22.8973 & 0.339716 & 5.905 & 0 & 0 & 12 & 0 & - & - & - & 1 & 0 & -  & J060802.86-063349.8 & 100120920119593785 \\
        \hline
      \end{tabular}
 \end{adjustbox}

   \caption{Some lines of the table of candidate members. The full table is electronic. References: (1)~Internal identification number. (2)~Right ascension and (3)~declination in Equinox J2000. (4)~Photometric instrumental $I$-band. (5)~Instrumental rms $I$-band. (6)~Differential magnitude. (7)~Variability type: 1=rotation-modulated, 2=possible periodic, 3=binnary and 4=other variability. (8)~Periodiod in days. (9)~Peak power of periodogram. (10)~high-scatter variables (1=yes, 0=no). (11)~H$\alpha$ spectral line in emission (Ha), in absorption (Ha-) and "HaCa" is H$\alpha$ and CaII emission line presence. (12)~Spectral Type: M1 to M9, ``nonM'' for earlier spectral types and ``N'' for unidentified spectra. (13)~Excess in IR or H$\alpha$ (1=excess, 0=no excess). (14)~Object above zams in $i-y$ colour (1=yes, 0=no). (15)~\textit{Gaia} candidate according us parameters (1=yes, 0=no).  (16)~\textit{Gaia} source identifications. (17)~\textit{Spitzer} source identifications. (18)~\textit{Pan-STARRS} source identifications. The ``-'' symbol represent no sufficient data.}
    \label{MS} 
\end{table*}

\vspace{0.5cm}

\subsection{Rotation period distributions}

Together with mass and metallicity, angular momentum is one of the fundamental properties determining  stellar evolution.  Therefore,  the evolution of stellar rotation and angular momentum in young stellar clusters remains an active area of study in the field \citep[][]{2014prpl.conf..433B}. As PMS are still contracting, they are expected to spin up as they approach the main sequence, unless they are able to efficiently drain angular momentum through  star-disc interactions, a process known as disc-braking \citep[][]{2002ApJ...566L..29H}. Indeed, previous studies have shown that  PMS stars hosting protoplanetary discs tend to rotate slower than their disc-less counterparts \citep[][]{2002A&A...396..513H,2007ApJ...671..605C,2019MNRAS.487.2937O}.

\begin{figure}
	\includegraphics[width=1.\columnwidth]{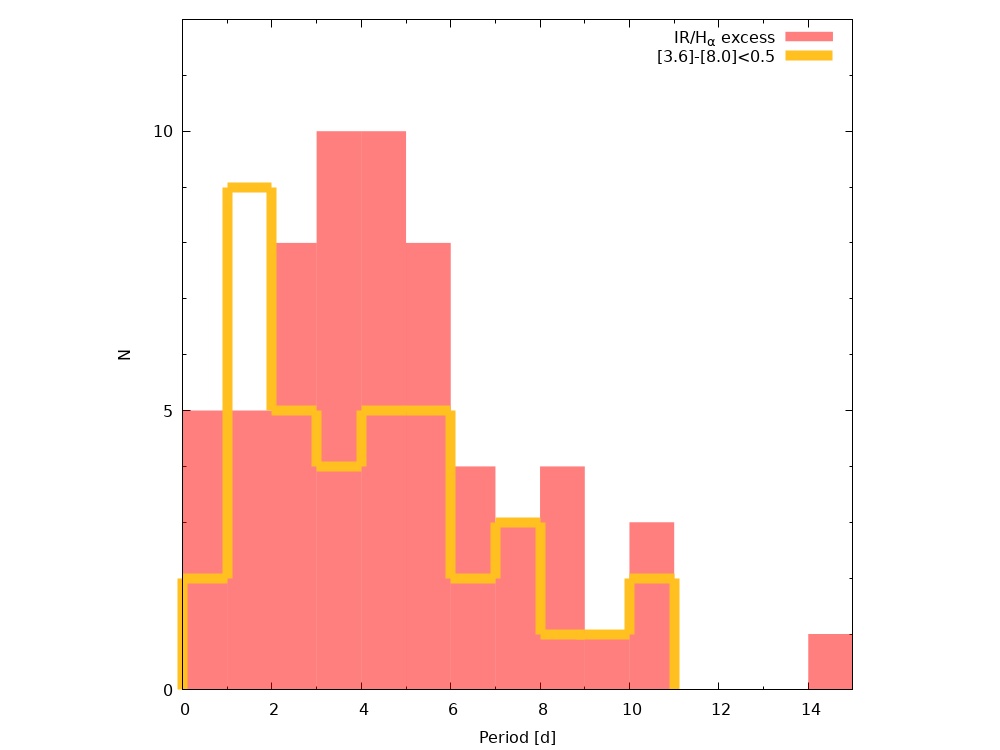}
		\caption{The period distributions of stars with (solid histogram) and without (empty histogram) a disc. K--S test p-value=0.5946, $statistic$=0.1492, D=0.2775, indicating that the null hypothesis cannot be discarded, but noting the trend that stars with a disc rotate more slowly than those that have lost it. }
\label{PerDisc}
\end{figure}

In the context of our study of the Mon~R2 cluster, we investigate disc braking  by selecting the rotation-modulated variable stars for which the presence of a protoplanetary disc can be established using the IR or H$\alpha$ data discussed in Sec. \ref{sub:disc}. Thus, we perform our analysis with 101 sources  (62 of them with a disc and 39 disc-less objects). We have removed 4 rotation-modulated sources from this sample due to the rotation periods being even faster than the break-up speed, so these periods ($p$<0.35~d) would be more consistent in binary stars \citep[][Chapter 2]{2009pfer.book.....M}.
In Fig.~\ref{PerDisc}, we show the rotational period distribution for stars with and without a disc. As expected, the stars with a disc (i.e., with IR and/or H${\alpha}$ excess) tend to rotate slower than those without it. The two--sample K--S test (\textsc{SciPy:KS}\_2samp) does not reject the null hypothesis ($p-value$=0.5946, $statistic$=0.1492, $D$=0.2775), probably because the sample size is modest. Although there is a tendency in the direction predicted by disc-braking (faster rotation for diskless stars), a further study is needed to increase the sample size, which could be done by spectroscopic estimations of the mass of the remaining periodic sources. Due to the differential absorption of the cluster, characterising the mass from photometry alone can lead to large uncertainties.

In this context, the distribution shown in Fig.~\ref{PerDisc} could also be explained by a strong mass-period dependence. Work such as \citet[][]{2020AJ....159..273R} studies the mass-period and the disc-period dependence in clusters of different ages. For clusters larger than 7~Myr, the evidence shows a strong mass--period dependence that increases with increasing age. This is expected when the circumstellar disks have dissipated and the stars increase their spin rate due to gravitational contraction until angular momentum stability is reached on the main sequence.

\begin{figure*}
	\includegraphics[width=2.\columnwidth]{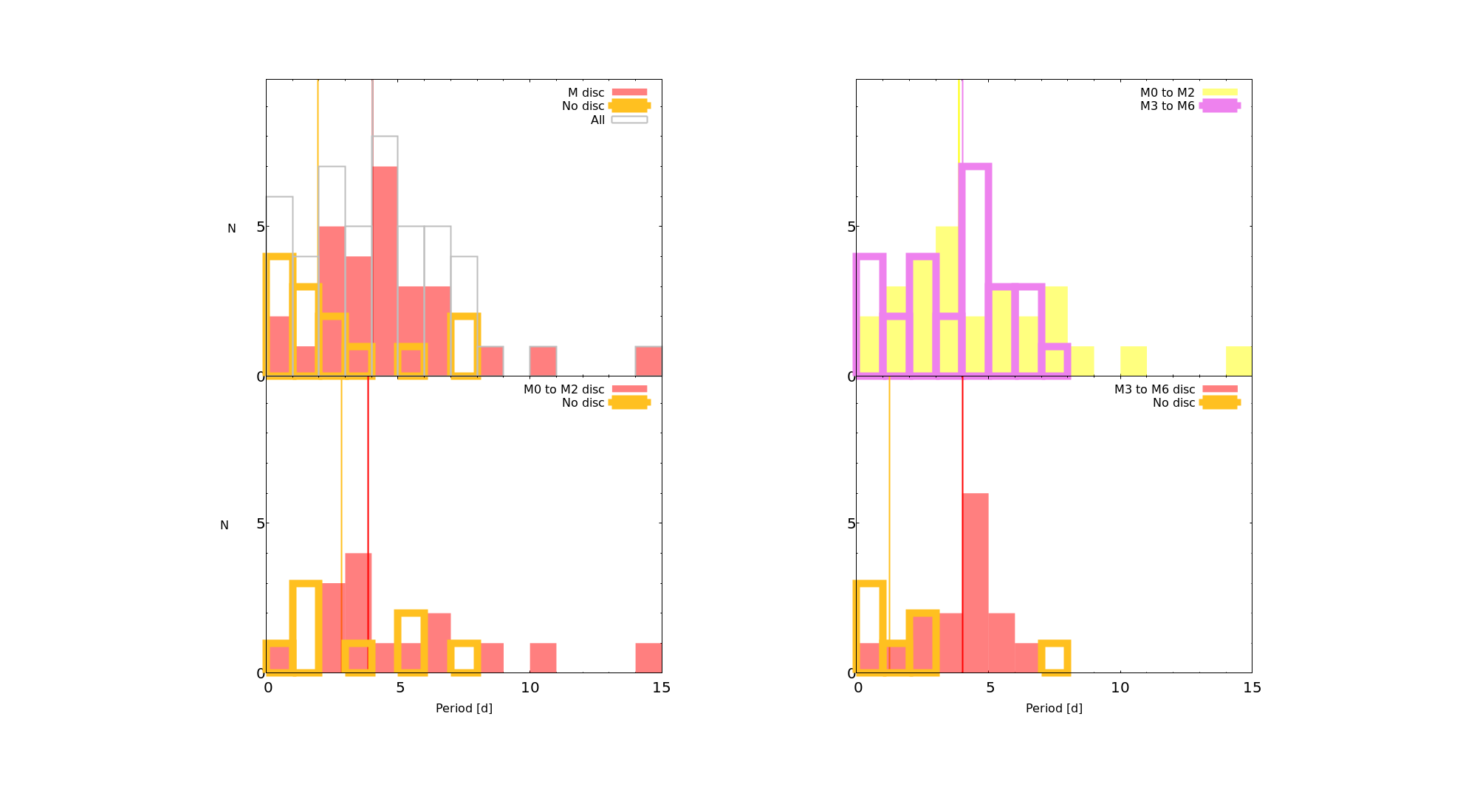}
		\caption{The period distributions for M-stars. \textbf{Top left:} period distribution for M stars with and without disc in red and goldenrod boxes respectively and all periodic M stars in grey line. \textbf{Top right:} period distribution for M0 to M2 and M3 to M6 stars in yellow and violet boxes respectively. The distribution for M-early is flatter and spreads towards slow rotators while the distribution in M-late is more concentrated towards fast rotators with a peak between 4 to 5 days. \textbf{Bottom:} period distribution for M0 to M2 (left) and M3 to M6 (right) stars with and without disc in red and goldenrod boxes respectively. In addition, in all histograms, we add the medians of each distribution respecting their colour.}
\label{MPer}
\end{figure*}

In our work we only spectrally identify M-type stars and in Fig.~\ref{MPer} we analysed the mass-period and disc-period dependence. In the top left panel we separated the M-stars with and without IR/H$\alpha$-excess and we still noticed the tendency to show the same as in Fig.~\ref{PerDisc}. On the other hand, if we separated this stellar group in two between M0 to M2 and M3 to M6 \citep[we chose this cut according to][]{2019MNRAS.487.2937O}, we see that there is a tendency for the M-late stars to rotate faster \citep[as expected, e.g.][]{2017A&A...599A..23V,2019MNRAS.487.2937O,2020AJ....159..273R} but even so, the IR excess selection still makes a difference, i.e. stars with an excess tend to rotate slower than stars without an excess in both stellar groups. However, these results are statistically poor (47 M stars), but tend to the expected ones. 
To broaden the sample of periodic stars according to their mass, we perform a photometric cut according to the colour of an M1 star. We show this analysis in Appendix~\ref{Ap:Period}. The results are inconclusive because the sample is not statistically significant.
On the other hand, for the BD we have not found a convincing periodical signal.

\subsubsection{Comparison to other clusters}

As discs dissipate in 2--3~Myr timescales \citep[][]{2011ARA&A..49...67W} and stars contract toward the main sequence and spin up, the distribution of rotation periods is expected to shift toward shorter periods.  In order to place Mon~R2 in the context of the angular momentum evolution of young stellar clusters,  we compare the rotation results obtained in Sec. \ref{periodogram} with other clusters with rotation periods in a similar mass range (of $< 0.5$~M$_{\sun}$ equivalent to spectral type M0 and later). To do this, we gleaned the literature and found the following four clusters:  The  2~Myr NGC~6530 \citep[][]{2013MNRAS.434..806B,2012ApJ...747...51H} and  Orion Nebula Clusters \citep{2009IAUS..258..103N,2014MNRAS.444.1157D}, the 3~Myr old NGC~2264 cluster  \citep{2009IAUS..258..103N,2004AJ....127.2228M,2005A&A...430.1005L,2013yCat..74240011A,2017A&A...599A..23V,2019MNRAS.487.2937O} and the 5~Myr Cepheus~OB3b cluster \citep{2013MNRAS.434..806B,2010MNRAS.403..545L}.
For the MonR2 sample, we used the sources with spectral types from M0 and later and added the sources with colours consistent with these spectral types (111 objects with $(i-y)$>0.9~mag colour reddened).

\begin{figure*}
	\includegraphics[width=2\columnwidth]{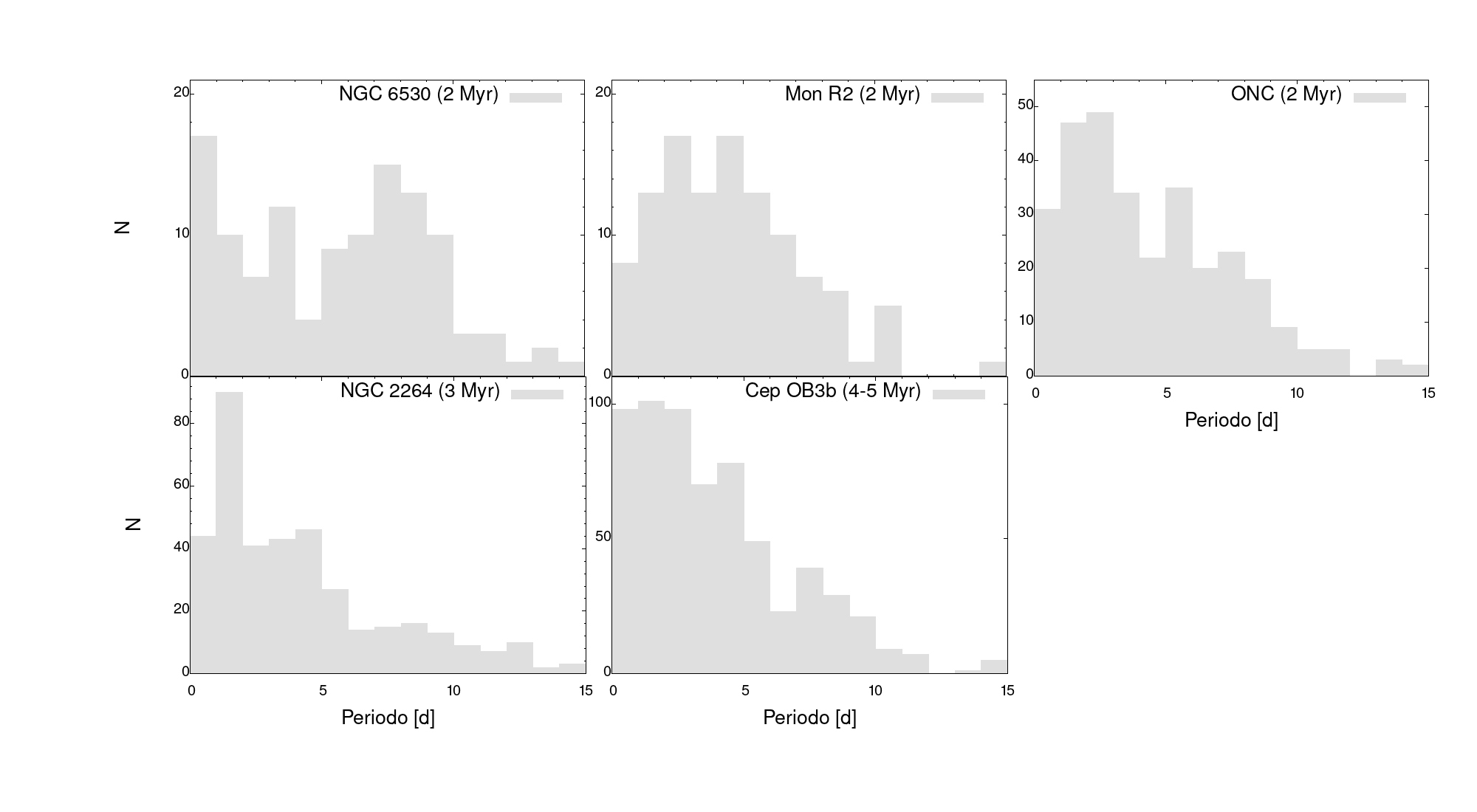}
		\caption{Period distribution of different clusters showing a progressive shift toward shorted periods with age. Samples are limited to masses less than 0.5~M$_{\odot}$, equivalent to spectral type M0.} 
\label{Cumulos}
\end{figure*} 

The rotation-period distribution collected for each cluster is shown in the Fig.~\ref{Cumulos}.
There, it can be noted that the period distribution of Mon~R2 members is intermediate between NGC 6530 the ONC cluster and is consistent with the age we calculated in Subsec.~\ref{subsub:Age}. Moreover, the figure illustrates how stellar members tend to rotate faster as they age, i.e. rotation distribution tend to narrow towards shorter periods, in agreement with the predictions of the disc-braking paradigm. Such distributions can be compared to Monte Carlo models to constrains the physical parameter of disc-braking. For example, \citet[][]{2015A&A...578A..89V} used the Monte Carlo simulations to reproduce the period distributions in young clusters assuming the hypothesis of disc-braking. Started with a bimodal distribution of periods for disc and disc-less stars and obtained results consistent with observational data in young clusters (i.e. NGC~2264, ONC). Mon~R2 has a period distribution statistically similar to that of ONC (see Tab.~\ref{tab:KScluster}), so it could be represented by Monte Carlo simulations. Therefore, the disc-braking paradigm would explain this distribution.

In Tab.~\ref{tab:KScluster} we show the values obtained with the K-S~test for two samples comparing the clusters with each other. Taking as a reference a significance level of~0.05, we find that NGC~6530 is statistically different from the rest of the clusters. 
On the other hand, the rest of the clusters seem to follow a timeline, as ONC is statistically equal to MonR2 and NGC2264 is equal to Cep~OB3b, this could also support, not only the relative age of the clusters already studied, but also the relative age of of Mon~R2.

Furthermore, we observe that the younger a cluster is, the more slow rotators it contains. On the other hand, the median disc lifetimes is between 2 and 3~Myrs \citep[][]{2011ARA&A..49...67W}, this coincides with the increase of fast rotators at older ages. As the disc is lost, the star contracts and this leads to an increase in its rotation rate. If this process occurs while the star is young, then we have a fast rotator.
Although we will always see a remnant of slow rotators without a disc.

\begin{table*}
    \centering
    \begin{tabular}{|l|l|l|l|l|l|l|l|l|l|l|}
    \hline
    & \multicolumn{2}{|c|}{NGC~6530} & \multicolumn{2}{|c|}{ONC}& \multicolumn{2}{|c|}{NGC~2264} & \multicolumn{2}{|c|}{Mon~R2} \\
    \hline
    \hline
    & D & $p-value$ & D & $p-value$ & D & $p-value$ & D & $p-value$ \\
    \hline
     ONC    & 0.14352 & 0.00101 & \\
     NGC~2264 &  0.14005 & 3.59388e-07 & 0.10231 & 0.01175 & \\
     Mon~R2 & 0.17744 & 6.71806e-05 & 0.14947 & 0.27211 & 0.14613 & 0.01737 & \\
     Cep~OB3b & 0.13336 & 2.13161e-09 & 0.09295 & 0.00030 & 0.08748 & 0.15364 & 0.13974 & 0.01498 & \\
     \hline
    \end{tabular}
    \caption{Statistical comparison of the cluster with the K-S test for two samples.}
    \label{tab:KScluster}
\end{table*}

\section{Summary and conclusions}
\label{sec:Conclusion}

In this paper, we studied and characterised the young open cluster Mon~R2 using a multi-wavelength and multi-technique approach. We used $I$-band time-series photometry to monitor the stars and to identify variability. We  also used optical spectroscopy to determine the spectral types in the low-mass regime. Moreover, we identified cluster member candidates using different criteria, including variability, proper motions from \textit{Gaia}, disc indicators (IR colors and H$\alpha$ emission)  and the position in the H-R diagram. From the position of member candidates in the H-R diagram and theoretical isochrones, we also estimated the age of the cluster. Finally, we compared distribution of rotation periods of objects with and without a disc as well as the overall period distribution seen in Mon~R2 to those observed in clusters of different ages. Here we summarise our main results: \\

\begin{enumerate}
    
    \item We used the LCOGT Network to construct light curves for 6843 sources, each one containing 1560 differential photometry points in $I$-band. Using a periodogram analysis, we found 226 periodic stars: 136 stars consistent rotation-modulated variability and rotation periods between 0.2 to 14~days, 12~possible periodic stars, 14~eclipsing binary candidates and 64~stars with other types of periodic variability. We also use the photometric scatter as a function of magnitude as an independent criterion to identify 444 ``high-scatter'' variables and likely member candidates. \\
    
    \item We spectroscopically classified 229 targets. We find that 140 objects are M-type, including 11  M7-M9 brown dwarfs,  while 89 objects have K and earlier spectral types.   \\
    
    \item Of the 986 LCOGT targets with \textit{Spitzer}-IRAC 3.6 and 8.0~$\mu$m photometry, 549 have IR colours consistent with IR-excesses above a stellar photosphere, indicating the presence of a protoplanetary disc.   
    Using narrow-band photometry, we identified 20 additional disc candidates with H${\alpha}$ excess.  \\
    
    \item We matched 1591 LCOGT sources with \textit{Gaia} EDR3 and establish the following membership criteria for Mon-R2:
        \begin{itemize}
            \item $\mu_{\alpha}cos(\delta)$ =-2.75$\pm$1.25~mas~yr$^{-1}$ 
            \item $\mu_{\delta}$ =1.15$\pm$ 1.25~mas~yr$^{-1}$
              \item 0.888$\leq \varpi \leq$1.42~mas
        \end{itemize} 
    
    \item Using sources \textit{GaiaFlag}=1 and have Gaia parallaxes with error smaller than 20~per cent, we have estimated a distance to the cluster of 825$\pm$51~pc.

    \item We matched  3078  LCOGT targets with the Pan-STARRS catalogue.
    Using the objects with known spectral types, we calculated an A$_{v}=2.09\pm1.7$~mag for the Mon-R2 cluster  using their $(i-y)$ colour excesses.  \\
    
  \item We  identified a total of 1566 cluster member candidates using the following (non-exclusive) criteria:  
        \begin{itemize}
            \item 1267 PS1 member candidates above ZAMS
            \item 206 \textit{Gaia} member candidates based on proper motions and distances
            \item 136 member candidates based on the rotation-modulated variability. 
            \item 444 member candidates based on high-scatter variability.
            \item 515 member candidates based on the presence of IR and/or H$\alpha$ excess. 
        \end{itemize}
        
  \item From the position of the H-R diagram of the member candidates and the comparison to evolutionary tracks, we estimated an age of $2\pm1$~Myr for the cluster and identified 248 brown dwarf candidates. \\
        
        \item We compared the distributions of rotation periods of stars with and without a disc signature (IR and/or H$\alpha$ excess) and find tendency of stars with a disc to rotate slower, in agreement with the disc braking paradigm.  We also compared the overall distribution of rotation periods in Mon~R2 to that observed in clusters of different ages and find a reasonable agreement with the expectations of the angular momentum evolution in PMS stars (a progressive shift toward shorter periods with age). 

\end{enumerate}

\section*{Acknowledgements}

We thank the anonymous referee whose comments and suggestions have helped to improve the paper significantly. We thank Tim Naylor for his recommendations on the use of his tau² code \url{https://www.astro.ex.ac.uk/people/timn/tau-squared/}.
S.O. and R.G. acknowledge support from grant PICT 2019-0344. L.A.C.  acknowledges support from ANID FONDECYT grant 1211656.
This work makes use of observations from the Las Cumbres Observatory global telescope network. And, this work is based [in part] on observations made with the Spitzer Space Telescope, which was operated by the Jet Propulsion Laboratory, California Institute of Technology under a contract with NASA. Based on observations obtained at the international Gemini Observatory, which is operated by the Association of Universities for Research in
Astronomy, Inc., under a cooperative agreement with the NSF on behalf of the Gemini partnership: the National Science Foundation (United States), National Research Council (Canada), Agencia Nacional de Investigaci\'{o}n y Desarrollo (Chile), Ministerio de Ciencia, Tecnolog\'{i}a e Innovaci\'{o}n (Argentina), Minist\'{e}rio da Ci\^{e}ncia, Tecnologia, Inova\c{c}\~{o}es e Comunica\c{c}\~{o}es (Brazil), and Korea Astronomy and Space Science Institute (Republic of Korea).
This work has made use of data from the European Space Agency (ESA) mission
{\it Gaia} (\url{https://www.cosmos.esa.int/gaia}), processed by the {\it Gaia}
Data Processing and Analysis Consortium (DPAC,
\url{https://www.cosmos.esa.int/web/gaia/dpac/consortium}). Funding for the DPAC
has been provided by national institutions, in particular the institutions
participating in the {\it Gaia} Multilateral Agreement.
The Pan-STARRS1 Surveys and the PS1 public science archive have been made possible through contributions by the Institute for Astronomy, the University of Hawaii, the Pan-STARRS Project Office, the Max-Planck Society and its participating institutes, the Max Planck Institute for Astronomy, Heidelberg and the Max Planck Institute for Extraterrestrial Physics, Garching, The Johns Hopkins University, Durham University, the University of Edinburgh, the Queen's University Belfast, the Harvard-Smithsonian Center for Astrophysics, the Las Cumbres Observatory Global Telescope Network Incorporated, the National Central University of Taiwan, the Space Telescope Science Institute, the National Aeronautics and Space Administration under Grant No. NNX08AR22G issued through the Planetary Science Division of the NASA Science Mission Directorate, the National Science Foundation Grant No. AST-1238877, the University of Maryland, Eotvos Lorand University (ELTE), the Los Alamos National Laboratory, and the Gordon and Betty Moore Foundation. This research has made use of the NASA Exoplanet Archive, which is operated by the California Institute of Technology, under contract with the National Aeronautics and Space Administration under the Exoplanet Exploration Program. This research made use of Astropy,\footnote{http://www.astropy.org} a community-developed core Python package for Astronomy \citep{astropy:2013, astropy:2018}.

\section*{Data availability}

The IR data are available through \url{https://irsa.ipac.caltech.edu/cgi-bin/Gator/nph-dd} and from \url{https://vizier.u-strasbg.fr/viz-bin/VizieR} \citet[][]{vizier:J/ApJS/184/18,vizier:J/AJ/144/31}. The PS1 DR1 and DR2 data are available from \url{https://catalogs.mast.stsci.edu/panstarrs/}. The Gaia EDR3 data are available from 
\url{https://gea.esac.esa.int/archive/}. The distance data are available from \textsc{VizieR} \citet[][]{vizier:I/352}. All the above-mentioned data are available through the \textsc{Aladin} software (\url{https://aladin.u-strasbg.fr/}). The isochrones tables are available through \url{https://phoenix.ens-lyon.fr/Grids/BT-Settl/}. Gemini spectra will be available at the Argentinian Virtual Observatory, NOVA, at \url{https://nova.conicet.gov.ar/}. The light curves obtained with LCOGT on which this article is based will be available as online material.




\bibliographystyle{mnras}
\bibliography{MonR2} 




\appendix

\section{Period dependence}
\label{Ap:Period}

\begin{figure}
    \includegraphics[width=1\columnwidth]{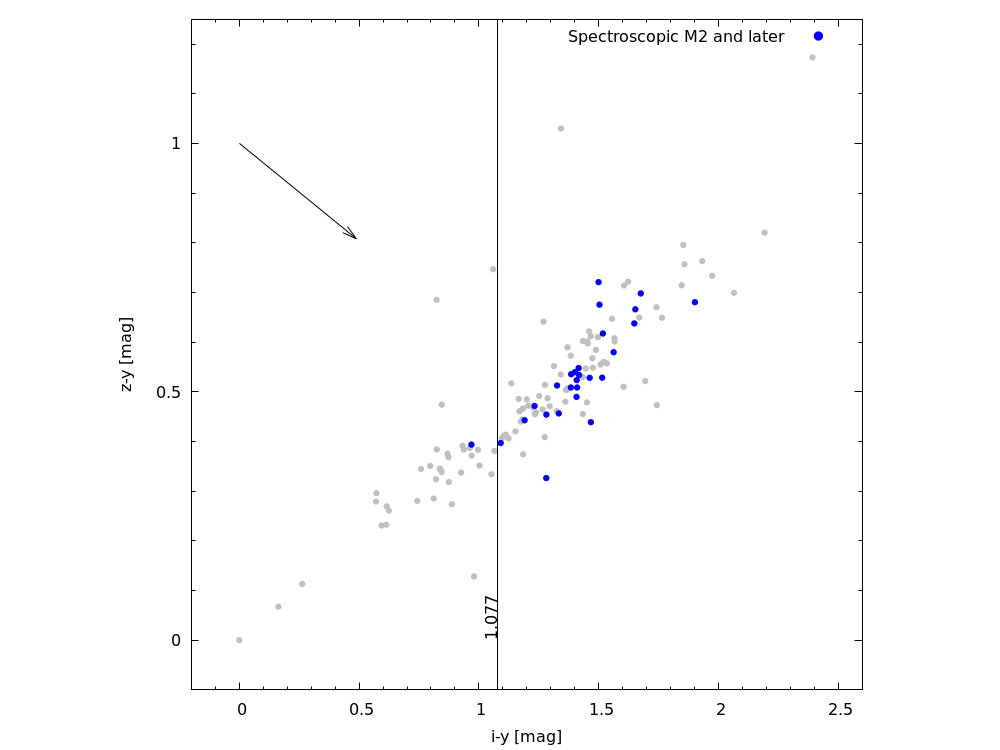}
    \caption{Colour-colour diagram for rotation-modulated variable sources with PS1 photometry (133 stars) in grey circles. In blue circles are the spectroscopic M2 and later stars. At $(i-y)$=1.077~mag is the photometric limit between M1 and earlier and M2 and later stars. The arrow at the top left is the extinction vector, corresponding to mean A$_{V}$=2.09~mag.} 
    \label{CCDfot}
\end{figure}

For the disc braking study, as our sample of spectral types is small, we decided to analyse our periodic sources by separating them on the basis of mass determination by photometry. For this, we use PS1 photometry and make the $(i-y)$ vs. $(z-y)$ colour-colour diagram shown in Fig.~\ref{CCDfot}. At $(i-y)$=1.077 we have placed the boundary between M1 and earlier and M2 and later stars. This colour limit is calculated from the values obtained by \citet[][table 4 \& 5]{2018ApJS..234....1B} and corrected by the mean $A_{V}$ obtained in this work. We have superimposed the spectroscopic sources of M2 and later to contrast this limit. Note that, as the reddening is unidirectional, the sample of M2 sample and the later stars can be contaminated by the reddened earlier stars. In contrast, the sample of M1 and earlier stars will be much less contaminated. Finally, we obtained 36 early and 97 late stars.

\begin{figure}
    \includegraphics[width=1\columnwidth]{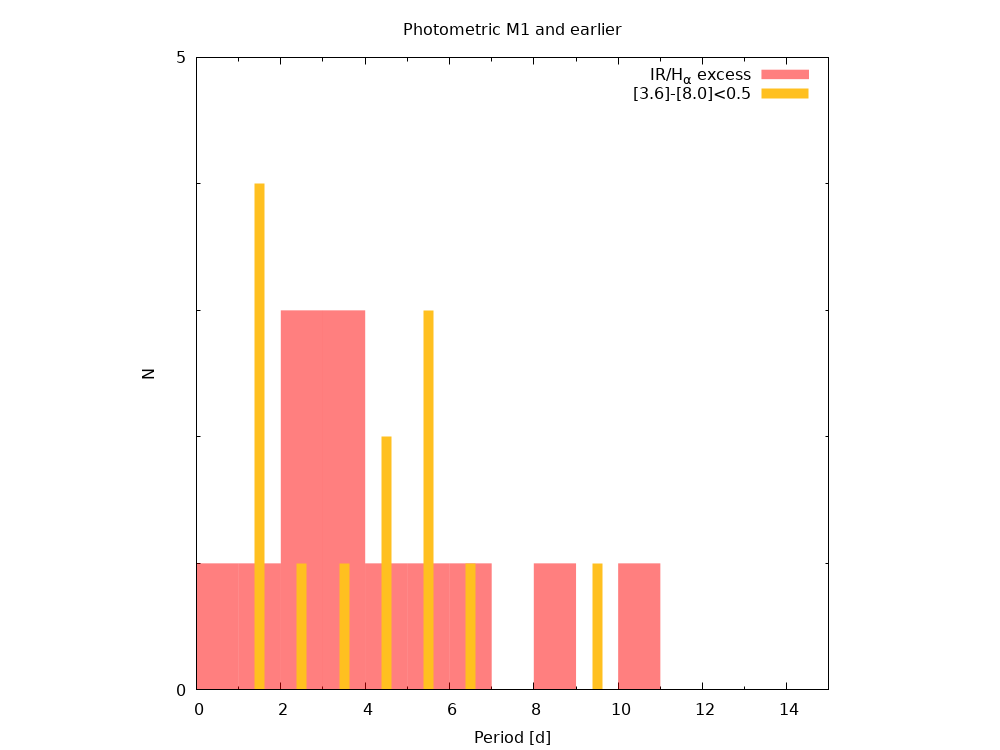}
    \includegraphics[width=1\columnwidth]{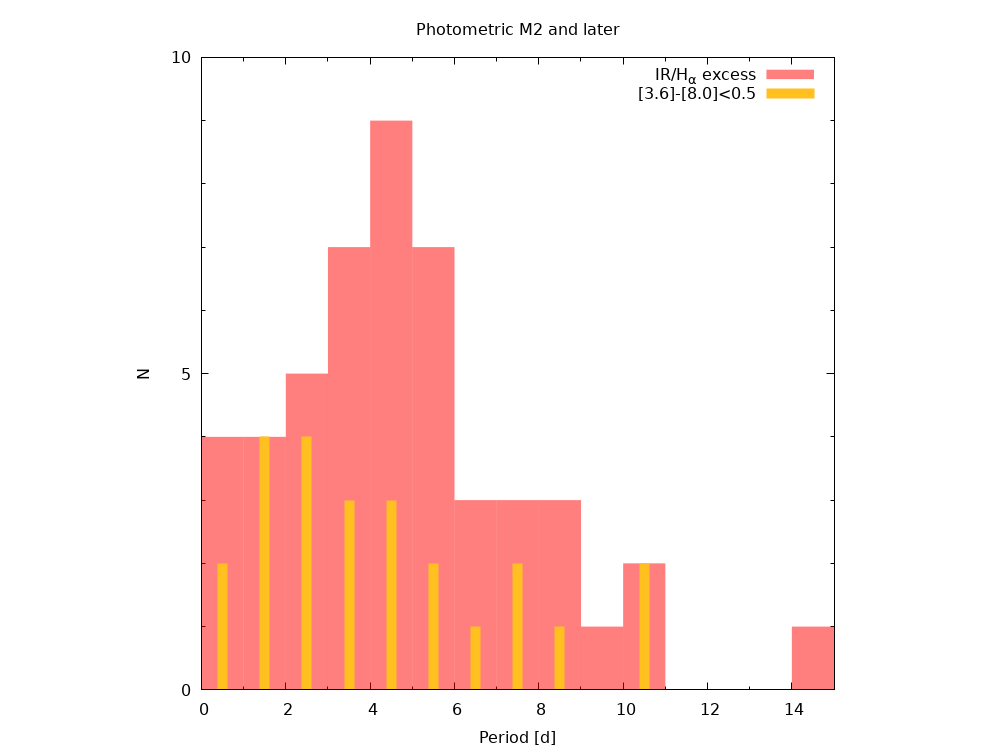}
    \caption{The period distributions of stars with (solid histogram) and without (spikes histogram) a disc. \textbf{Top:} M1 and earlier stars according to the cut shown in Fig.~\ref{CCDfot}. \textbf{Bottom:} M2 and later stars according to the same cut.}
    \label{M1M2fot}
\end{figure}

Once both samples were selected, we split them between with and non-disc sources to make the histograms of the period distribution shown in Fig.~\ref{M1M2fot} (we removed sources with periods shorter than 0.35~d). We apply the KS statistic for two samples to obtain the results shown in Tab.~\ref{tab:MfotKS}. The statistical results are inconclusive due to the small sample size. However, in the case of M2 and later stars, there is a tendency for those with a disc to rotate more slowly.

\begin{table}
\caption{KS-test for photometric samples.}
    \resizebox{\columnwidth}{!}{
    \centering
    \begin{tabular}{c|c|c|c|c|c|c}
         \hline
        Sample & Disc & No disc & No disc data & $p-value$ & D & $statistic$ \\
        \hline
        M1 \& earlier & 13 & 13 & 6 & 0.999 & 0.533 & 0.154 \\
     M2 \& later & 49 & 24 & 24 & 0.629 & 0.411 & 0.175 \\
    \hline
    \end{tabular}
    }
    \label{tab:MfotKS}
\end{table}

\section{Rotation-modulated stars and other variables}
\label{sec:other}

In this section we show the rest of the rotation-modulated variables of the Mon~R2, from Fig.~\ref{Amp101} to Fig.~\ref{quiza}.

Moreover, in our sample, we found LC with abrupt drops in magnitude which we interpret as eclipsing or occultation (by circumstellar material) events. Ten of them show multiple events during the observation time, and four stars, just an unique dimming (see Fig. \ref{binariesmulti} and Fig. \ref{binaries1ec} respectively). 
Other targets in our sample show irregular photometric  variations (see Fig. \ref{other}) that might be due to variable accretion and/or stellar activity.
However, the aim of this work is to characterise the Mon~R2 cluster, so the objects presented in this section have not been studied in depth but might be interesting for follow-up studies.  
Nevertheless, the morphologies found in our sample are varied and similar to those described by other authors, such as \citet[][]{2014AJ....147...82C,2018AJ....156...71C,2017A&A...599A..23V,2022AJ....163..263H}. They classified their variable stars according to two parameters: the stochasticity (or alternatively, periodicity) which they called Quasi-periodicity (Q), and the degree of flux asymmetry named Flux asymmetry (M). This statistical study yielded different types of variability. Certainly this classification starts with purely periodic stars due to the presence of star spots. But there are multi-periodic objects possibly due to the combination of star spots in binary pairs. In addition, there are long timescale objects that present a variability on timescales longer than 25 days. Other types of variability that exhibit asymmetry could be due to a combination of periodicity of spots with non-periodic changes of longer duration or due to unstable cycle-to-cycle variability, are the Quasi-periodic symmetric (QPS). The Quasi-periodic dipper (QPD) variable exhibits variability due to stellar occultation by circumstellar material. There are also Aperiodic dipper variables (APD). On the other hand, Burster (B) variables correspond to objects with erratic but discrete accretion bursts. Finally, the Stochastic (S) variables present non-repetitive variability patterns with relatively symmetric excursions around a mean brightness. We recommend reading the work of \citet[][]{2022AJ....163..263H} for more details.
Finally, in Fig.~\ref{nonvari2} we compared non-variables stars with periodic stars with similar magnitude for three different magnitudes bright, medium and faint magnitude. This study presents a high sensitivity that allows us to find variability in faint sources. 

\begin{figure*}
	\includegraphics[width=2\columnwidth]{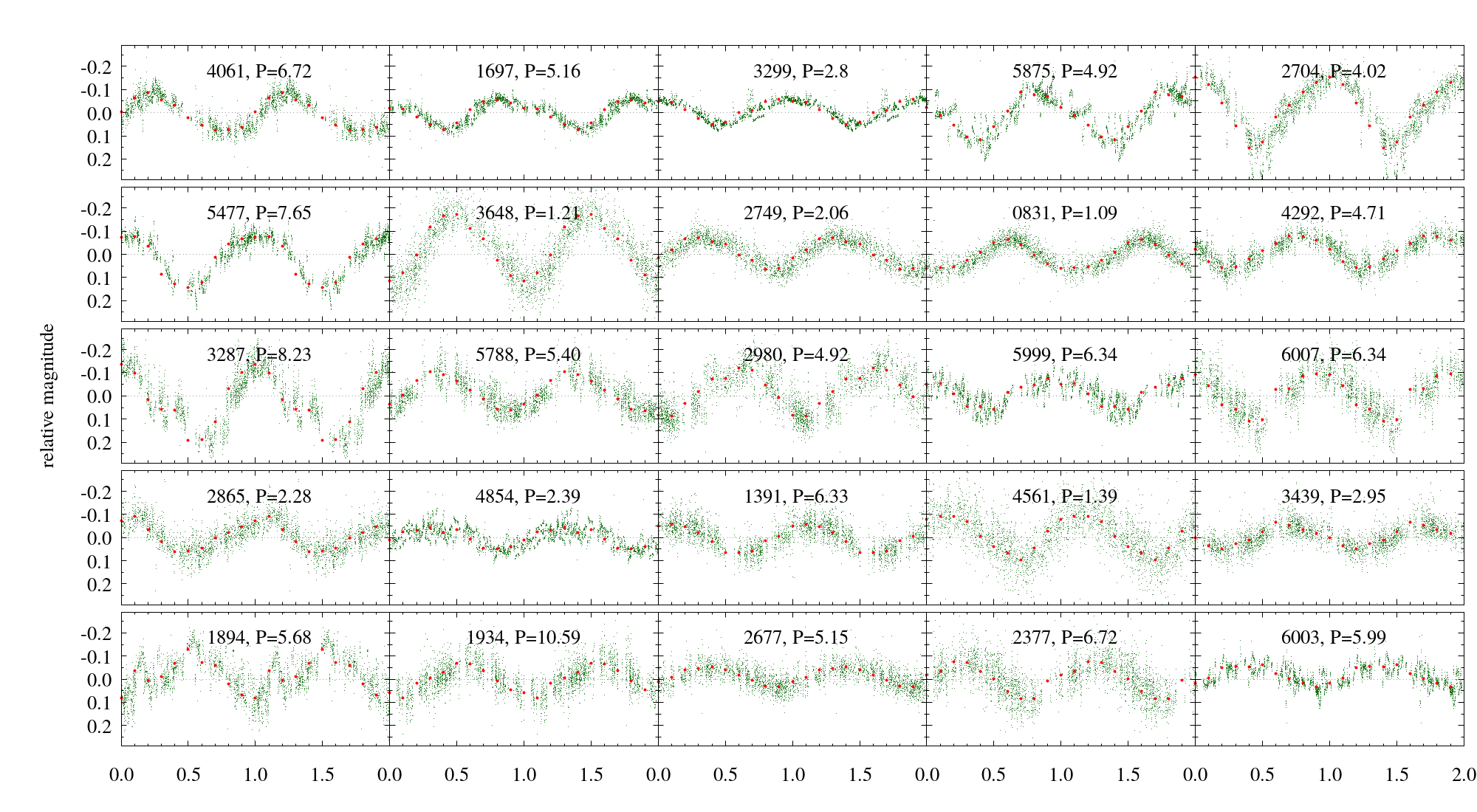}
	\caption{Phased light curves with amplitude $\geq$ 0.1 mag, ordered by periodogram power. The green dots are the $I$-band data from LCOGT. The red circles are the averages of the differential magnitude every tenth of a phase.}
\label{Amp101}
\end{figure*} 
\begin{figure*}
	\includegraphics[width=2\columnwidth]{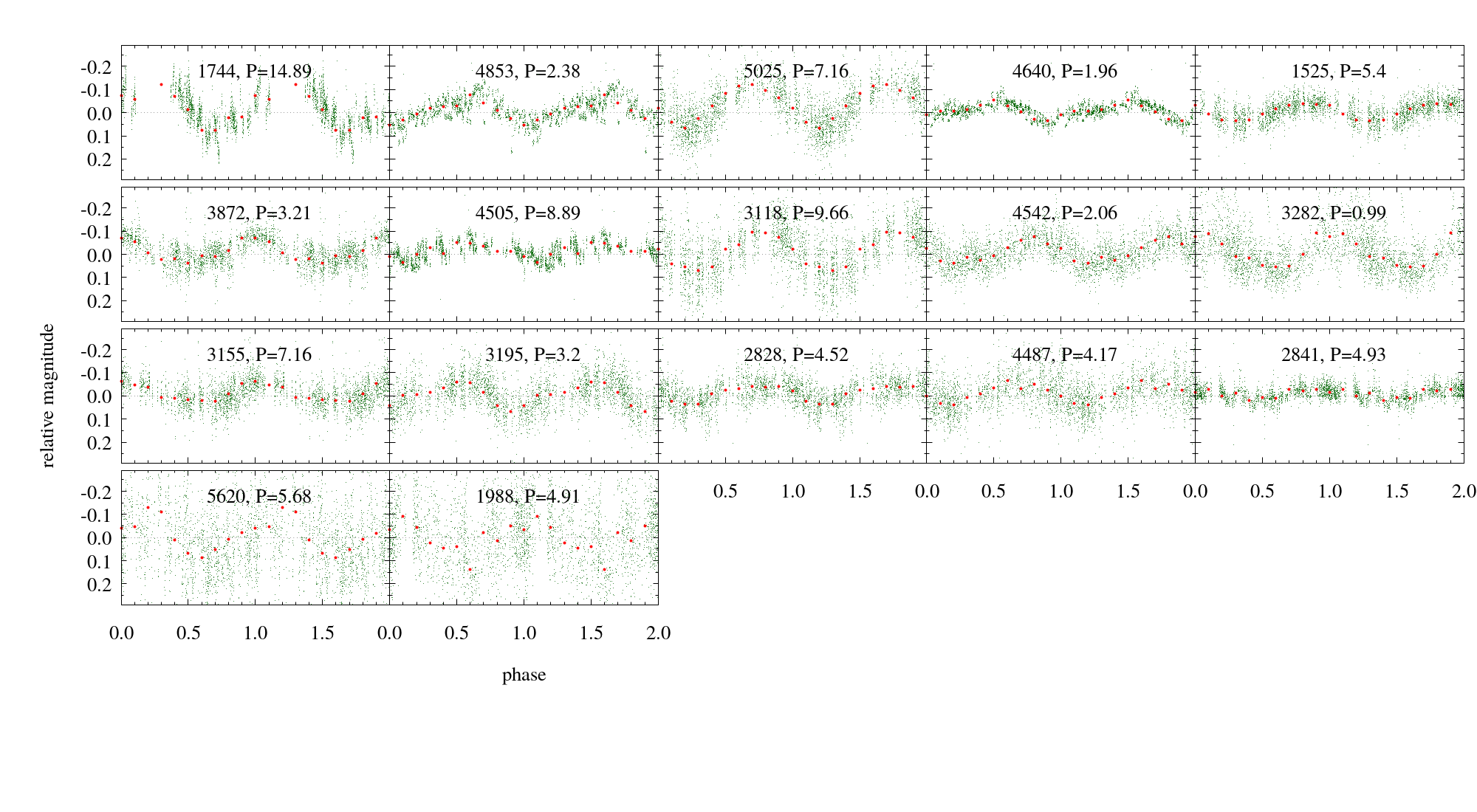}
	\caption{Continued Fig.~\ref{Amp101}.}
\label{Amp102}
\end{figure*} 
\begin{figure*}
	\includegraphics[width=2\columnwidth]{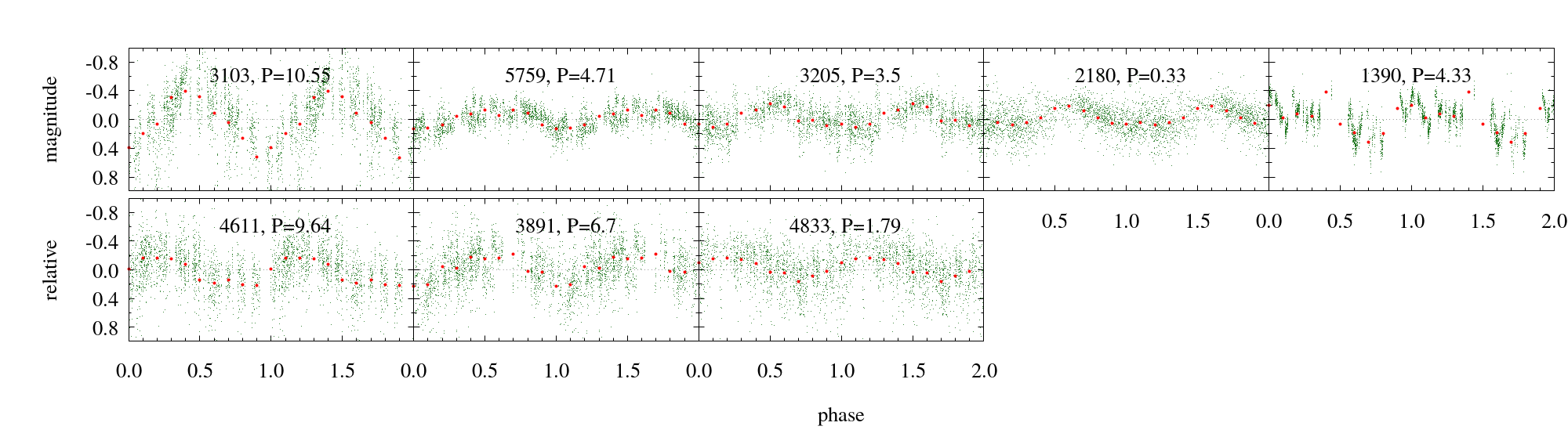}
	\caption{Continued Fig.~\ref{Amp101}.}
\label{Amp103}
\end{figure*} 

\begin{figure*}
	\includegraphics[width=2\columnwidth]{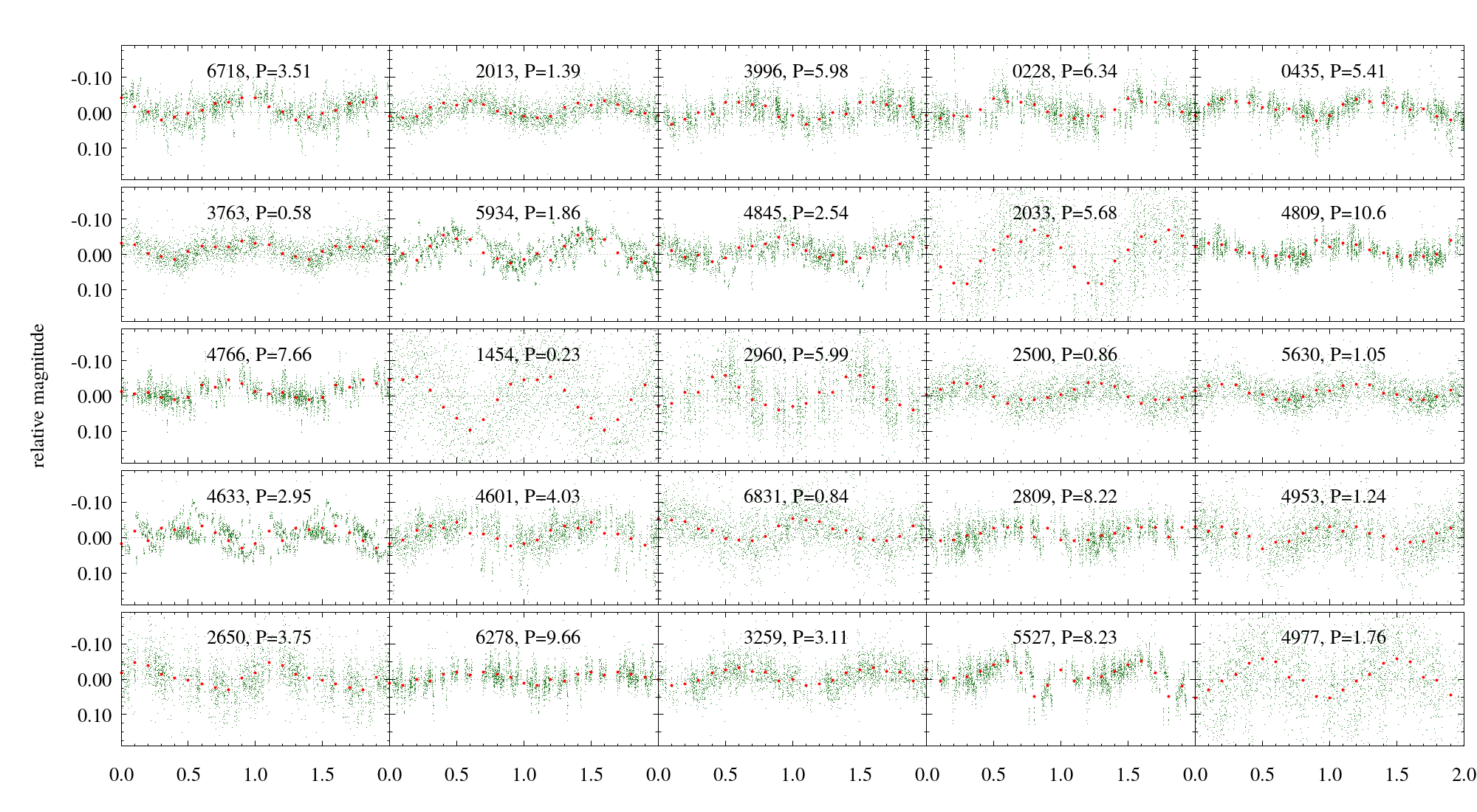}
	\caption{Phased light curves with amplitude between 0.05 to 0.1 mag, ordered for periodogram power. The green dots are the $I$-band data from LCOGT. The red circles are the averages of the differential magnitude every tenth of a phase.}
\label{Amp5101}
\end{figure*} 
\begin{figure*}
	\includegraphics[width=2\columnwidth]{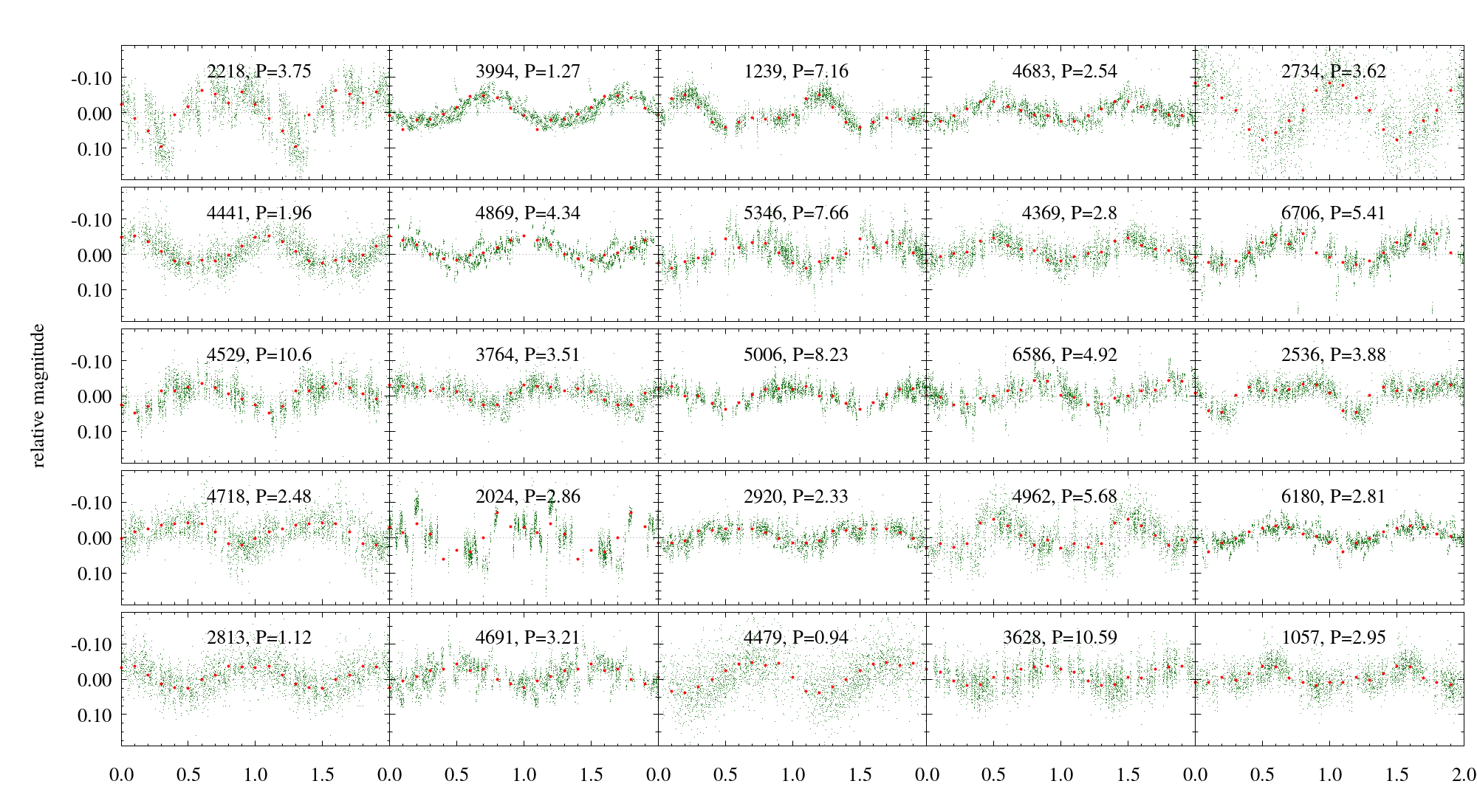}
	\caption{Continued Fig.~\ref{Amp5101}.}
\label{Amp5102}
\end{figure*} 
\begin{figure*}
	\includegraphics[width=2\columnwidth]{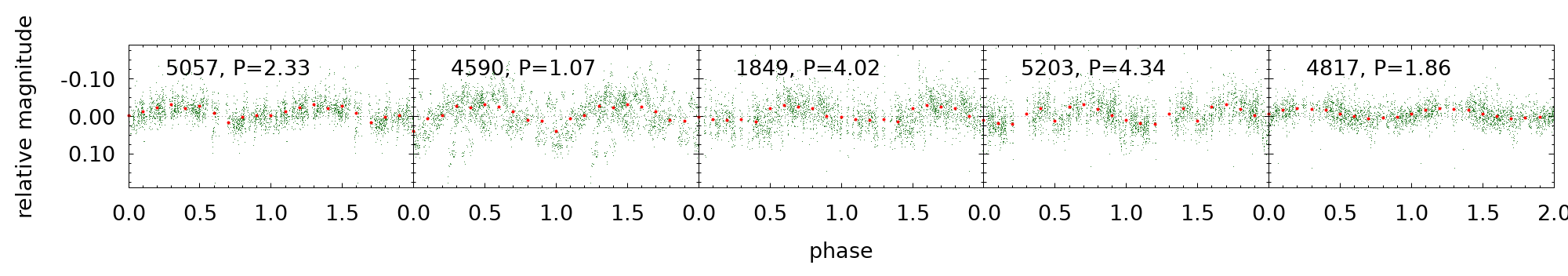}
	\caption{Continued Fig.~\ref{Amp5101}.}
\label{Amp5103}
\end{figure*} 
\begin{figure*}
	\includegraphics[width=2\columnwidth]{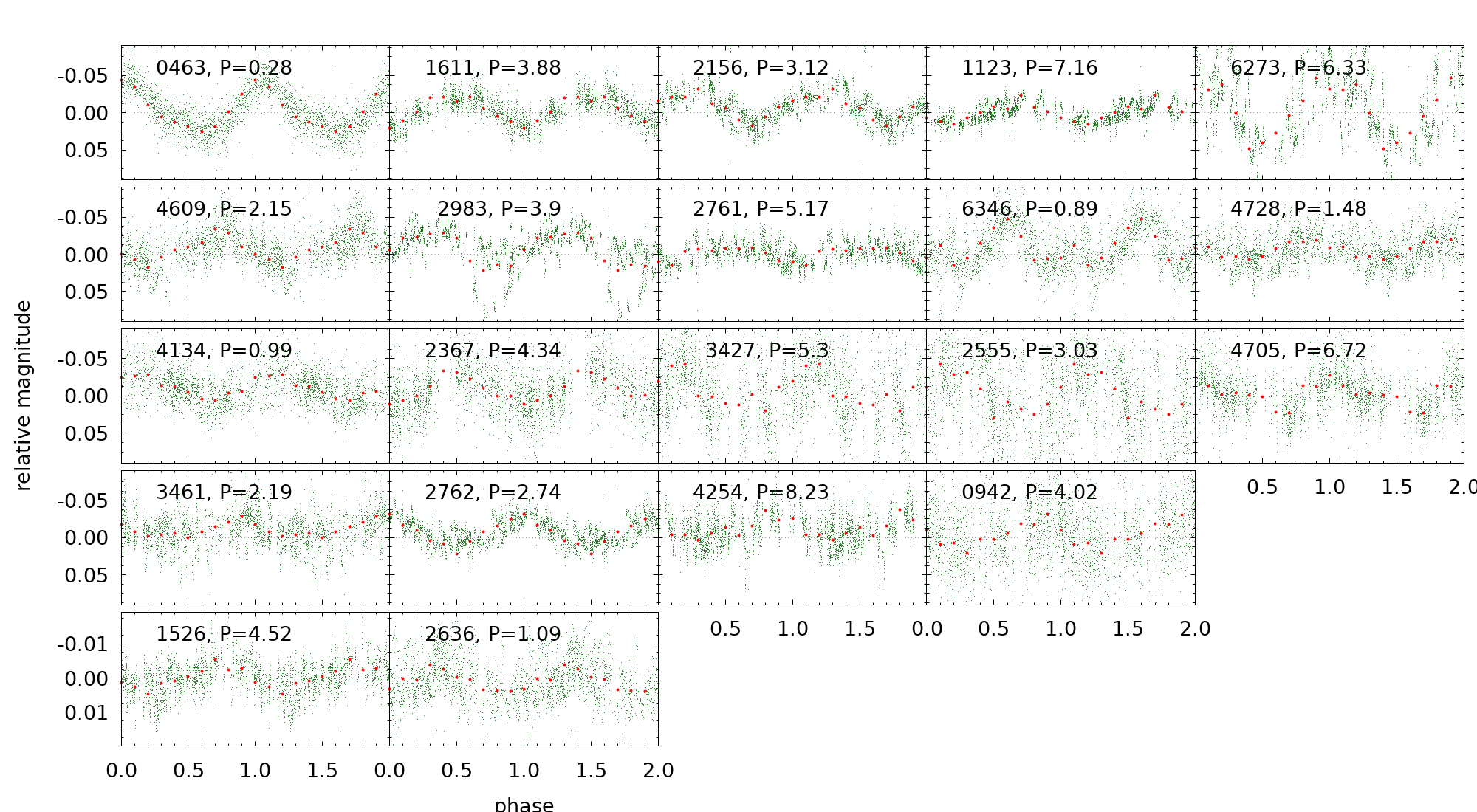}
	\caption{Phased light curves with amplitude $\leq$ 0.05 mag, ordered for periodogram power. The green dots are the $I$-band data from LCOGT. The red circles are the averages of the differential magnitude every tenth of a phase.}
\label{amp5}
\end{figure*} 

\begin{figure*}
	\includegraphics[width=2\columnwidth]{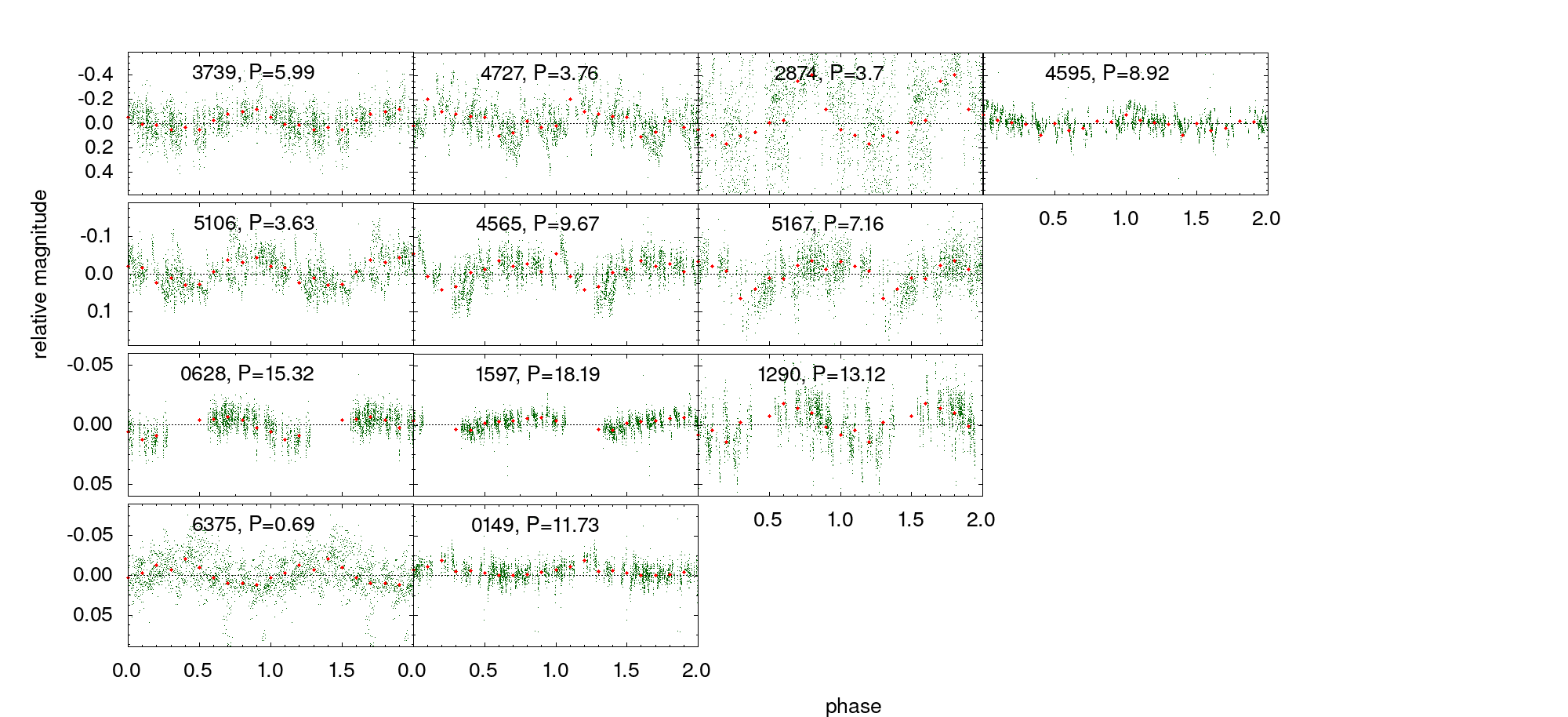}
	\caption{Phased light curves of 'possible' periodic stars, ordered for periodogram power and amplitude. The green dots are the $I$-band data from LCOGT. The red circles are the averages of the differential magnitude every tenth of a phase.}
\label{quiza}
\end{figure*}

\begin{figure*}
	\includegraphics[width=2\columnwidth]{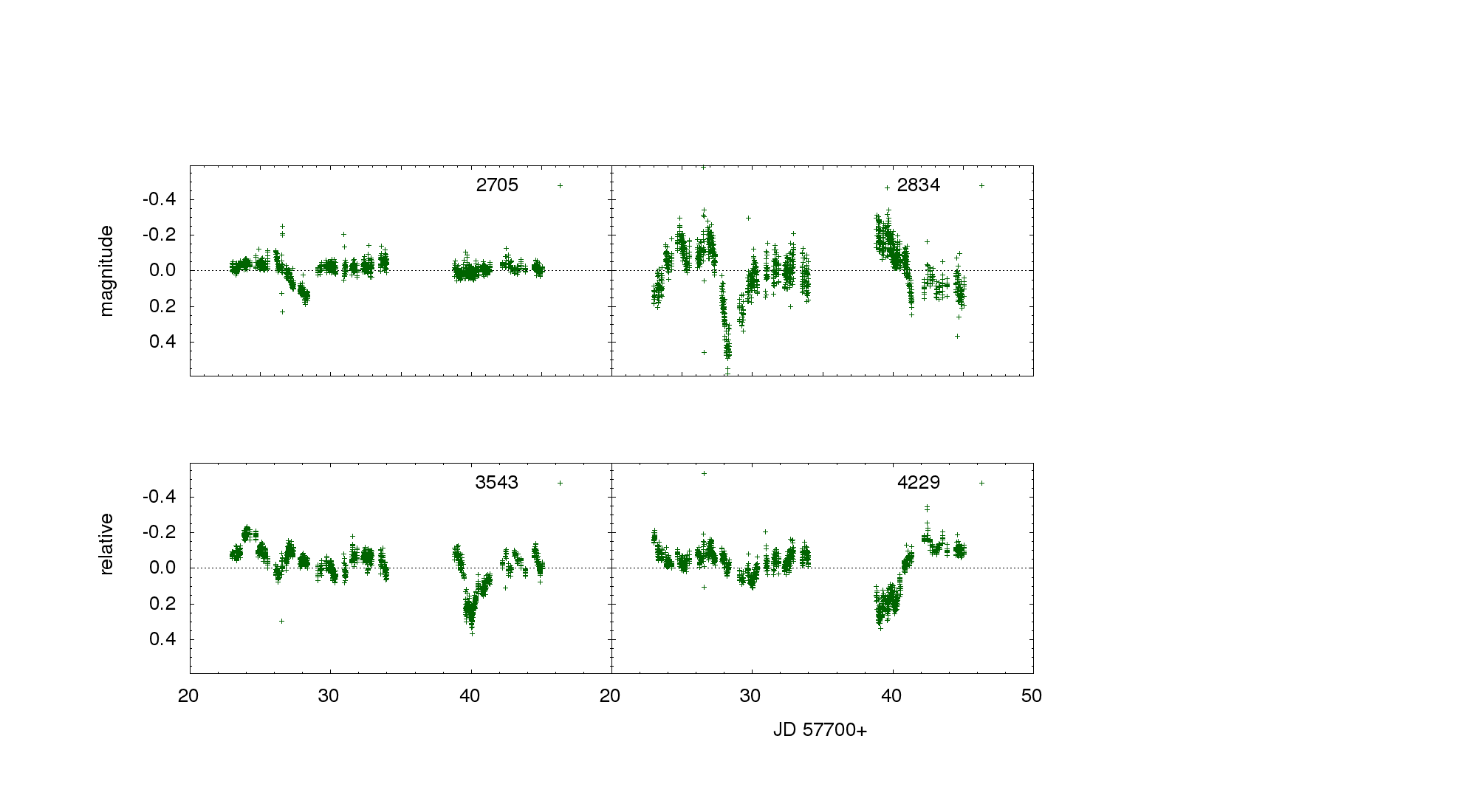}
	\caption{Light curves of stars with one eclipse during our observation time.}
\label{binaries1ec}
\end{figure*} 
\begin{figure*}
	\includegraphics[width=2\columnwidth]{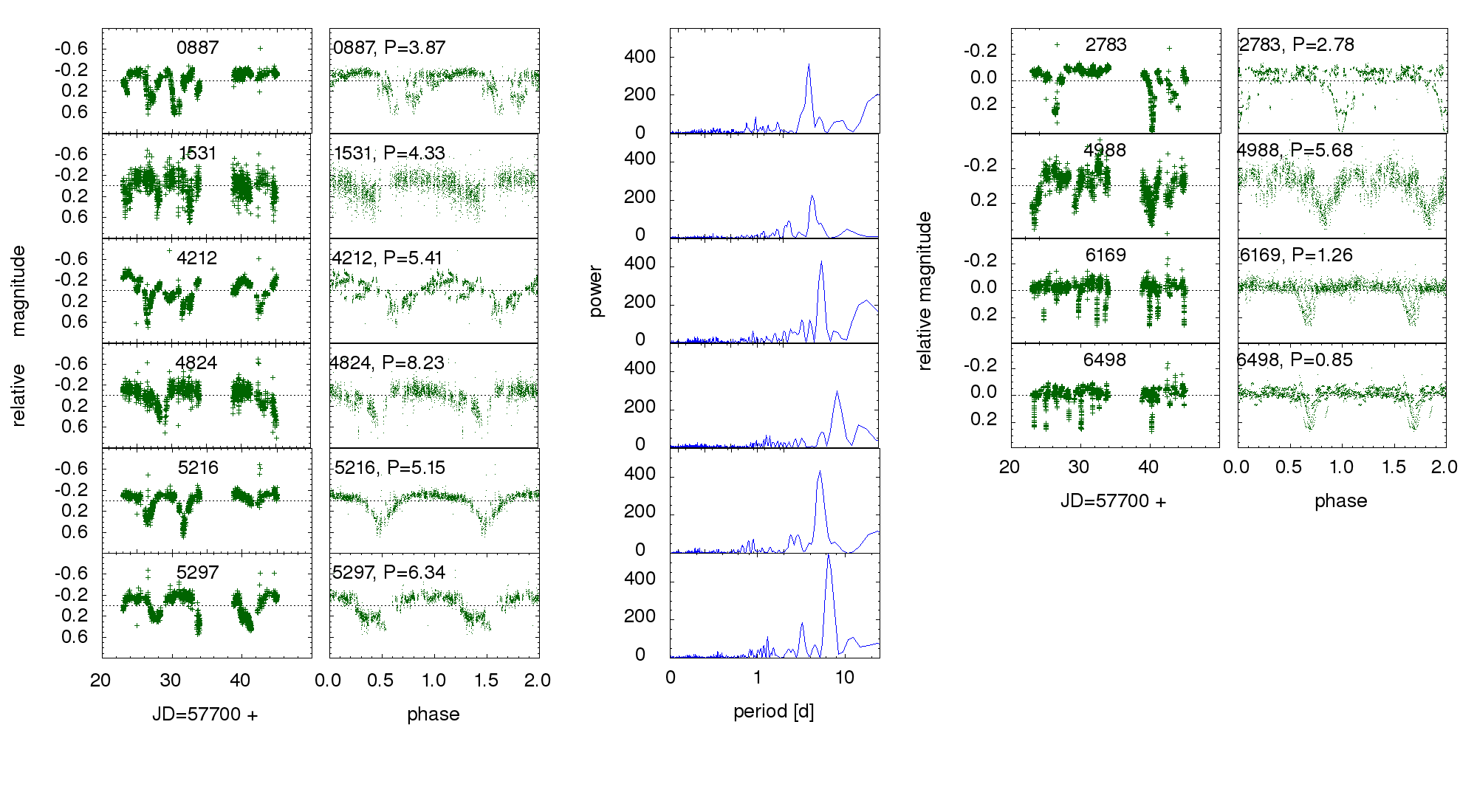}
	\caption{Light curves of stars with multi eclipses during our observation time. We presented the periodogram of best power peak, the other stars not have a clear visual power peak in periodogram.}
\label{binariesmulti}
\end{figure*} 

\begin{figure*}
	\includegraphics[width=1.6\columnwidth]{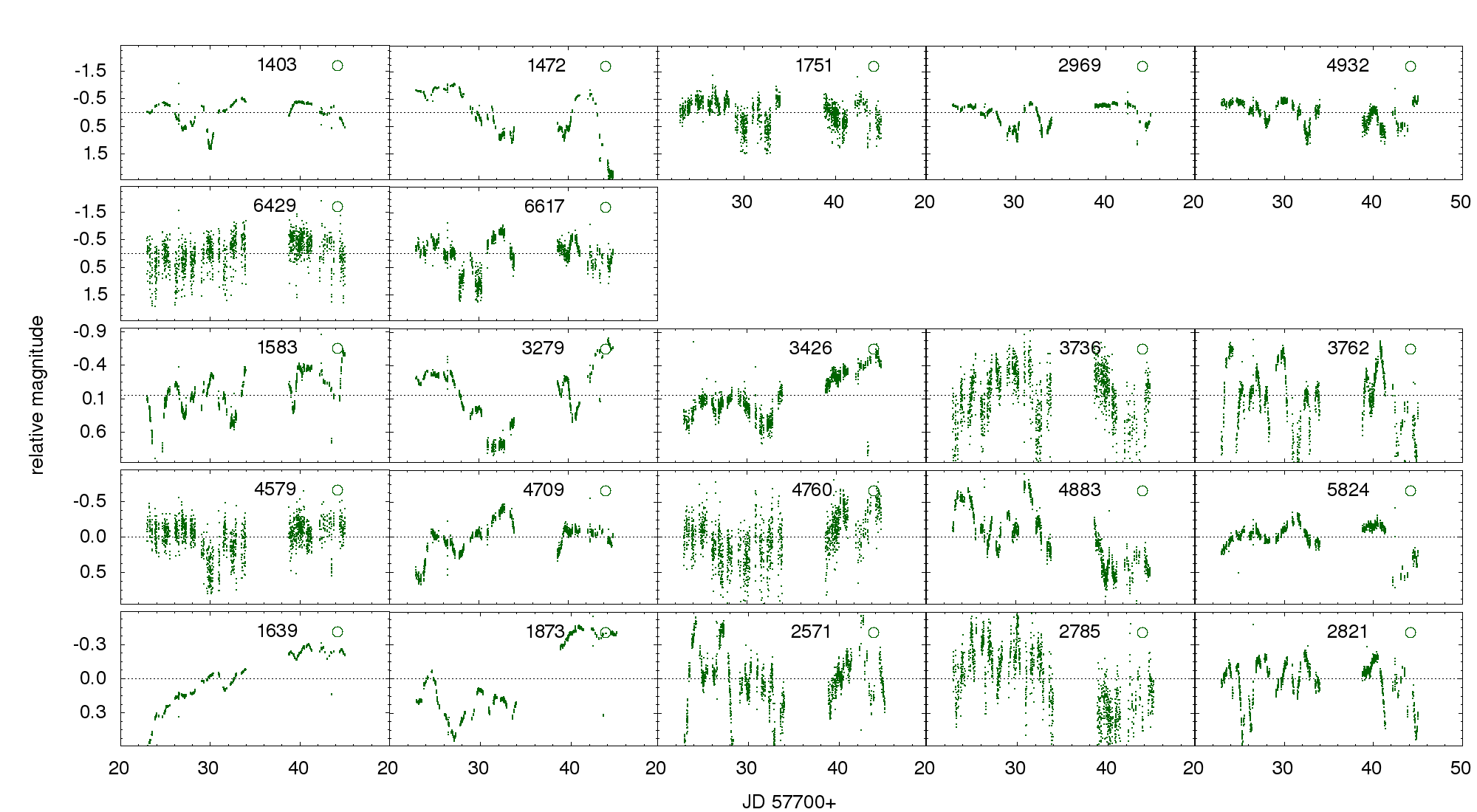}
		\includegraphics[width=1.6\columnwidth]{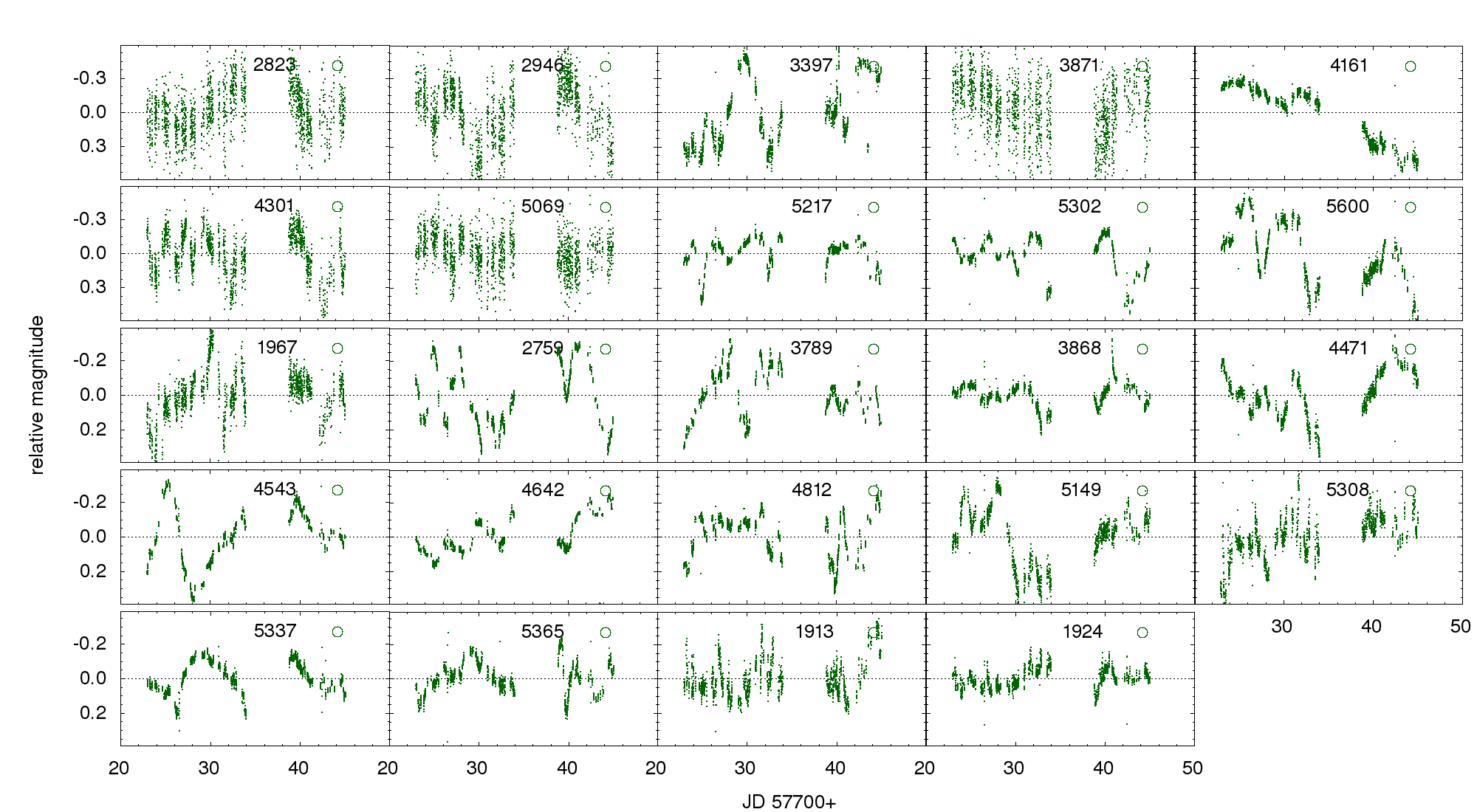}
		\includegraphics[width=1.6\columnwidth]{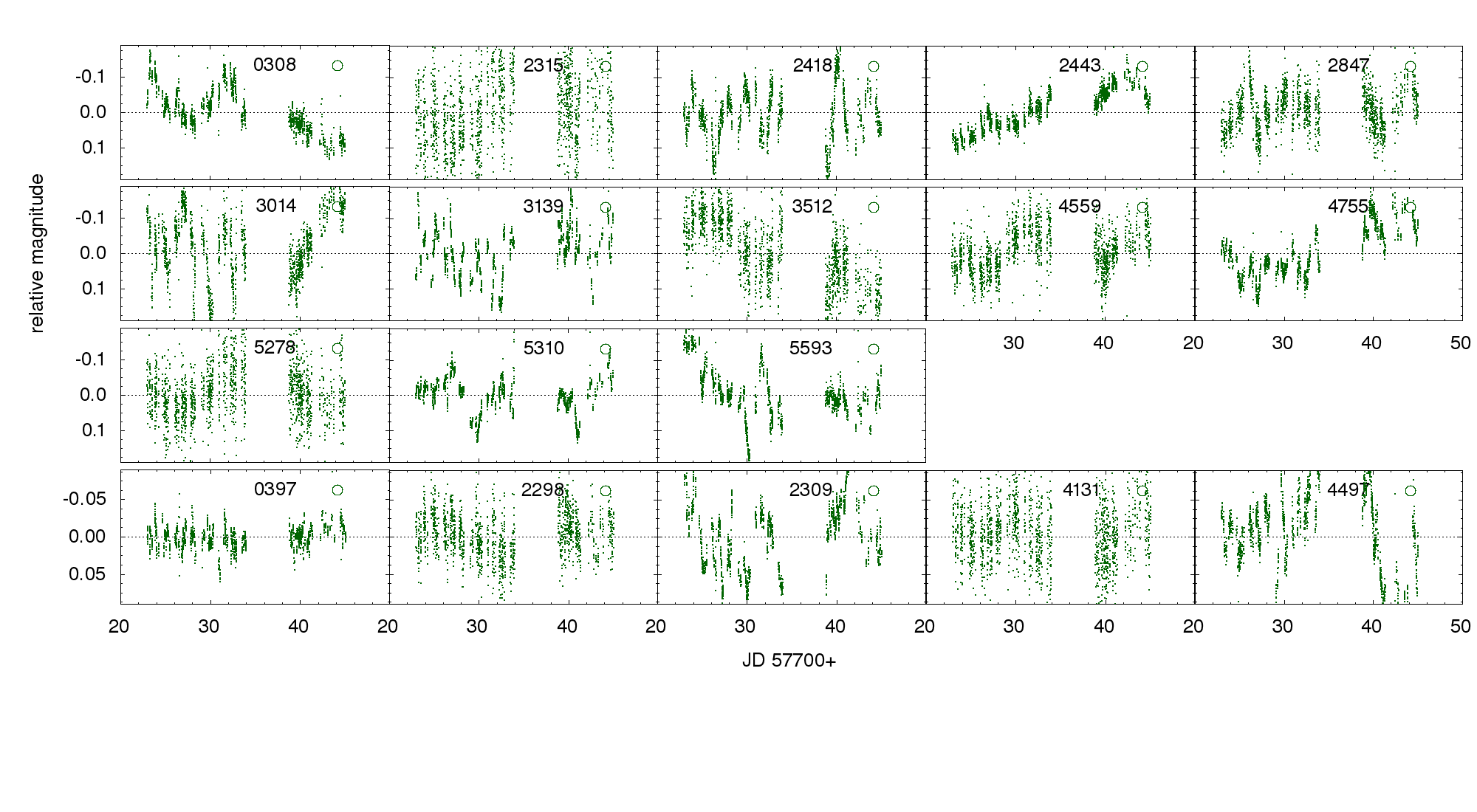}
	\caption{Light curves of no periodic variability.}
\label{other}
\end{figure*} 

\begin{figure*}
\includegraphics[width=2\columnwidth]{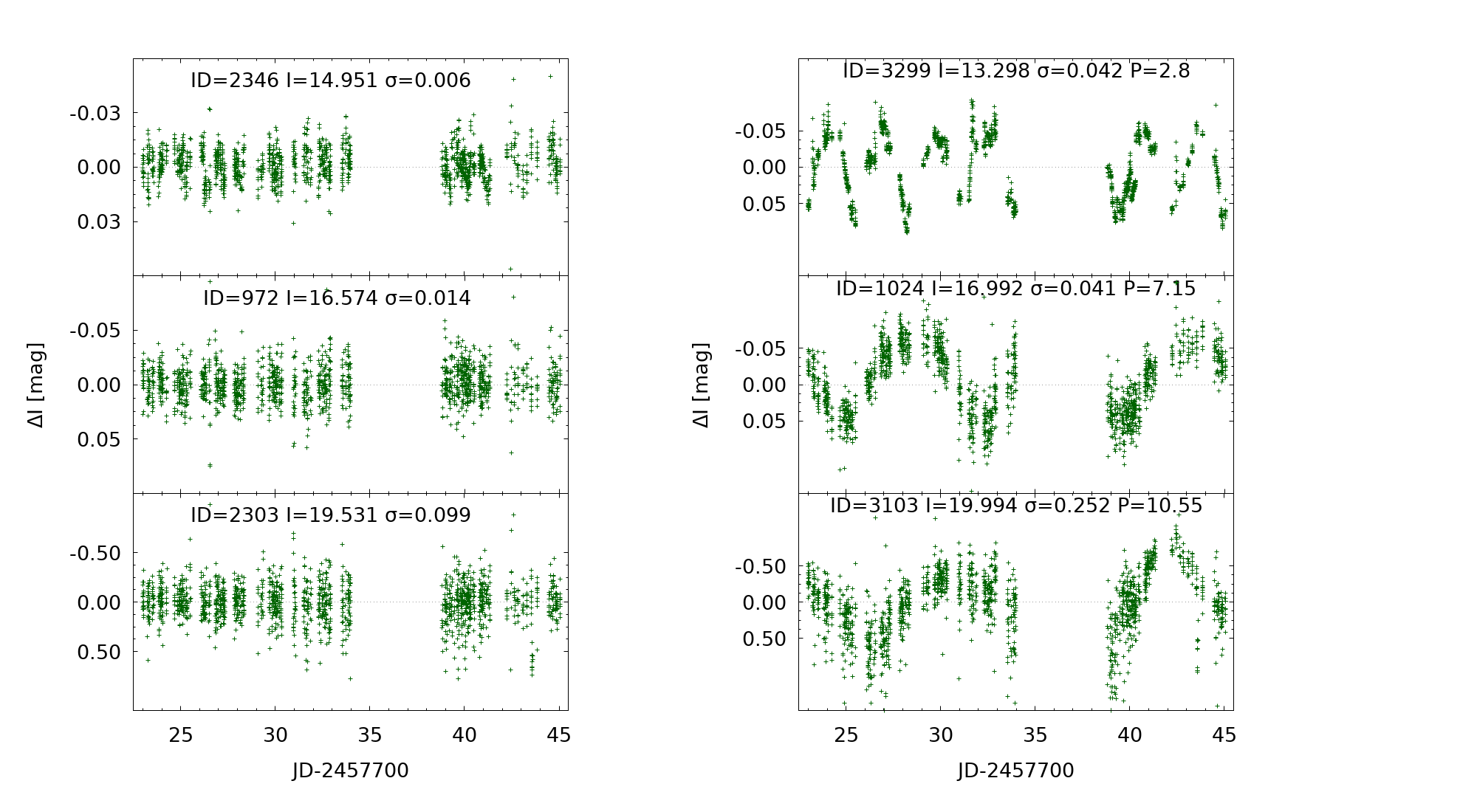}
\caption{Non-variables vs. periodic stars with similar magnitude for three different magnitudes bright, medium and faint magnitude. The LC of 3299, 1024 and 3103 are in Fig.~\ref{Amp101}, Fig.~\ref{panelPrin} and Fig.~\ref{Amp103} respectively. }
\label{nonvari2}
\end{figure*} 

\section{Gemini's fields}

\begin{figure}
	\includegraphics[width=1\columnwidth]{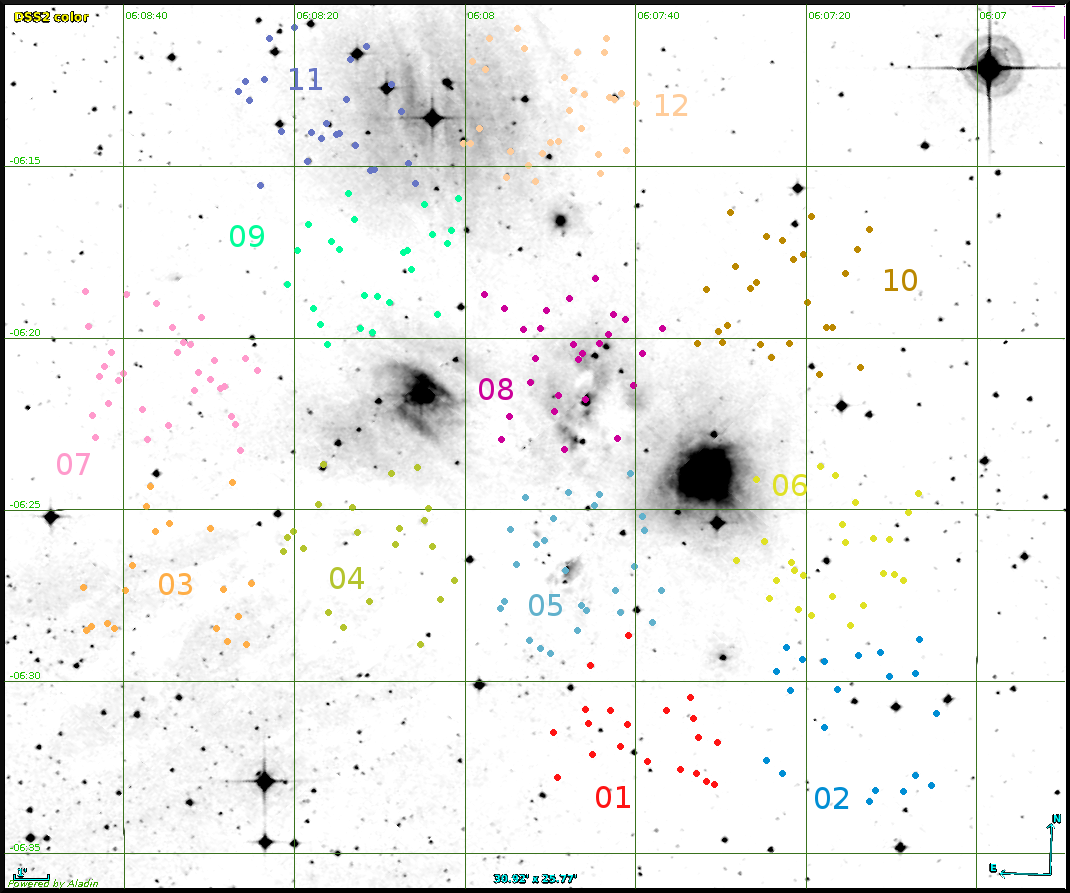}
	\caption{Sub-fields of Mon~R2 where GMOS masks were placed. Spectral targets are represented with coloured circles according to their sub-field. The background images correspond on the Digitalized Sky Survey 2 from Aladin sky atlas.}
\label{subfieldim}
\end{figure}

In Fig.~\ref{subfieldim} we show the fields of each mask used with Gemini, and the objects studied in each of them.

\section{ZAMS criterion}
\label{Ap:zams}

\begin{figure}
\includegraphics[width=1\columnwidth]{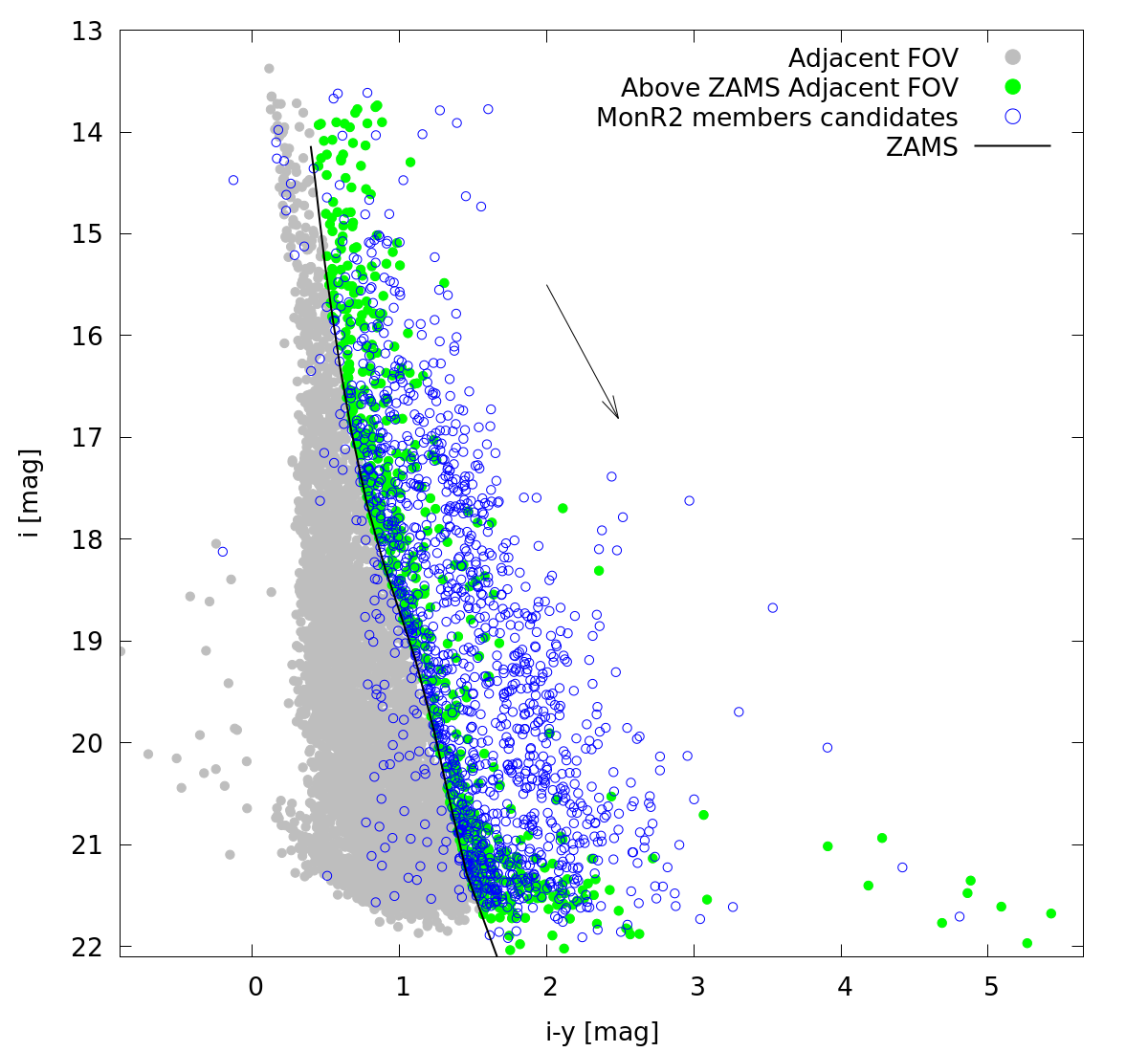}
\caption{Colour-magnitude diagram in $i$-$y$vs$i$ (PS1 DR2 photometry) of sources outside the studied Mon~R2 field (grey and green), overlapped with the candidate members (blue). In addition, we mark the ZAMS with a black line. The contamination is clearly concentrated over the ZAMS. }

\label{FOVOut}
\end{figure}

Sources above the ZAMS may be contaminated by field stars, so establishing this feature as a membership criterion leads to the largest uncertainties among the criteria. Then, to estimate this uncertainty, we performed the same procedure detailed in Sec.~\ref{sec:PS1} using the same LCOGT FOV at the same galactic latitude but shifted in longitude (coordinates in ICRS: RA=06$^{h}$09$^{m}$40$^{s}$.08 DEC=-05°57'17''.5). 
We obtained 641 objects from PS1 photometry (in the $i$ and $y$-bands) above ZAMS, compared to the 1267 objects in the Mon~R2 field. Thus, assuming that this sample contains no Mon~R2 members and that they are only field stars, we estimate the uncertainty for this criterion could be  as high as $\approx$~50~per cent (due to contamination).  In Fig.~\ref{FOVOut} we show the HR diagram with the sources of this new field that we called ``Adjacent FOV'' and the members of Mon~R2. A also performed a similar analysis by taking another field at the same galactic latitude and obtained a very consistent result. 

However, while the comparison between the number of stars above the ZAMS in both fields (control field and Mon~R2 field) is based on the assumption that the stellar density of both fields is the same, this is not necessarily true since the Mon~R2 field contains the young cluster. An alternative way to obtain the contaminant rate is to calculate the ratio of stars above the ZAMS to those below it in the control field. Then, to obtain the expected number of contaminants in the Mon~R2 field, we multiply this ratio by the number of sources below the ZAMS in this field. This method also suffers from some complications because of the differential extinction, which is not present in the control field. To alleviate this effect, we restrict the comparison to objects brighter than 21 mag in the PS1 magnitude to reach a reasonable completeness level in Mon~R2. Using this cut, we find 558 and 2690 sources above and below the ZAMS in the control field, and 1345 objects below the ZAMS in Mon~R2.  This give an estimate o contamination of (558/2690)*1345 = 279  stars. These 279 stars correspond to 26~per cent of the 1074 member candidates observed above the ZAMS in Mon~R2 (including the $i$-mag restriction mentioned above). We therefore estimate that the contamination of Mon~R2 members selected from their position in the H-R diagram to be between ¼ and 1/2 .

\section{Obtaining the distance with DR2 and EDR3}
\label{sec:distEdr3}

\begin{figure*}
\includegraphics[width=2\columnwidth]{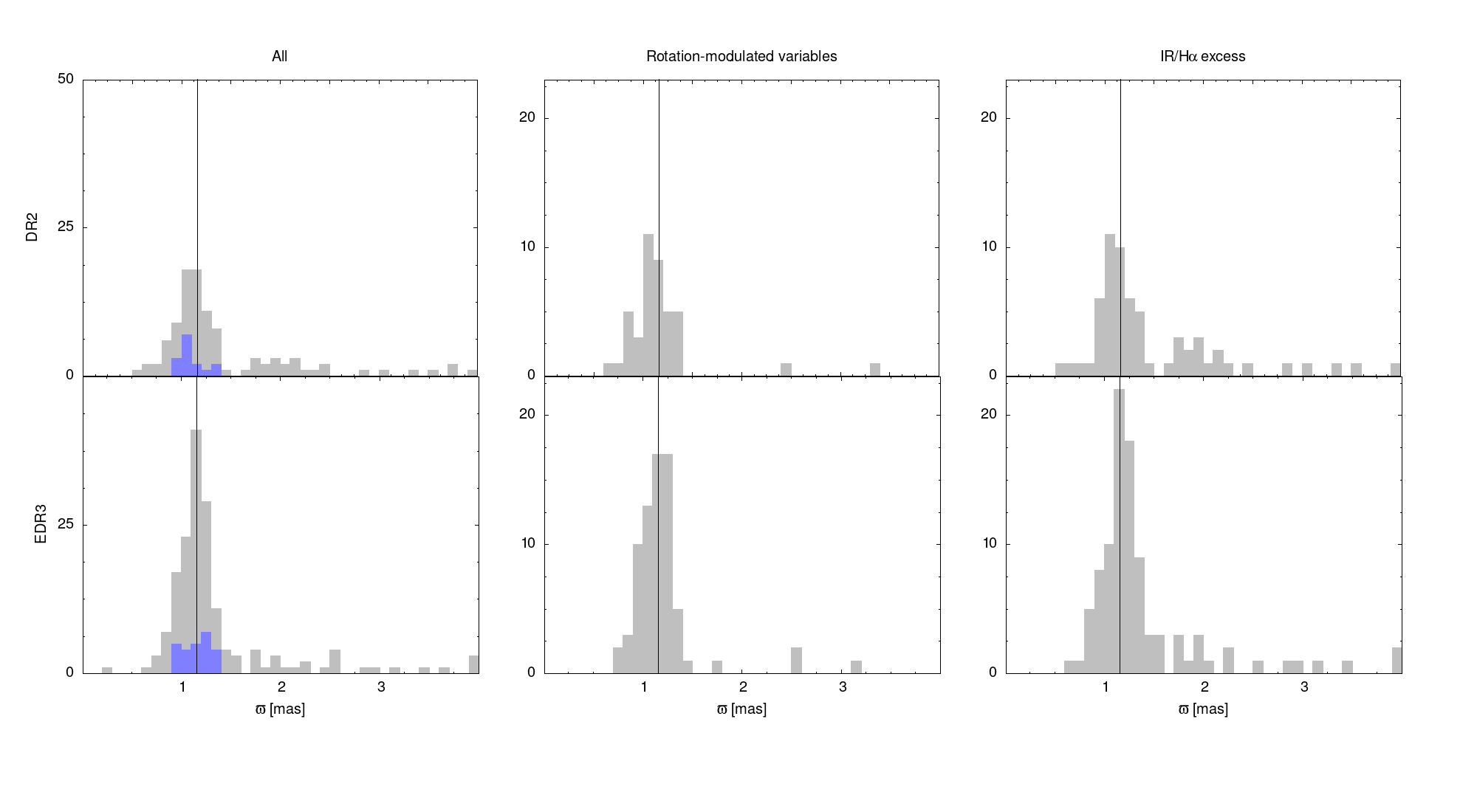}
\caption{Parallax histograms with \textit{Gaia} DR2 and EDR3 data for sources that fulfil at least two membership criteria and that have a parallax error of less than 20~per cent, avoiding negative parallaxes. The vertical line is located at the $\varpi$=1.15~mas. The blue boxes are the sources with 4 membership flags simultaneously. }

\label{plxH}
\end{figure*}

For the study of the mean distance to the cluster we have used the \textit{Gaia} DR2 and EDR3 data. Some sources have negative parallax values, so we have also performed this study with the \citet[][, BJ18]{2018AJ....156...58B} catalogue, associated to DR2 data and BJ21 associated to EDR3 data. For the parallax study, we have taken the sources which fulfilled at least two membership criteria and that have a parallax error of less than 20~per cent, avoiding negative parallaxes. In addition, we have analysed the distributions of rotation-modulated variables and/or IR/H$\alpha$-excess members. The histograms in Fig.~\ref{plxH} show that the distribution for the DR2 data has an closer peak than the distribution for the EDR3 data, 1.05~mas and 1.15~mas respectively, so DR2 locates the cluster at a greater distance.

\begin{figure*}
\includegraphics[width=2\columnwidth]{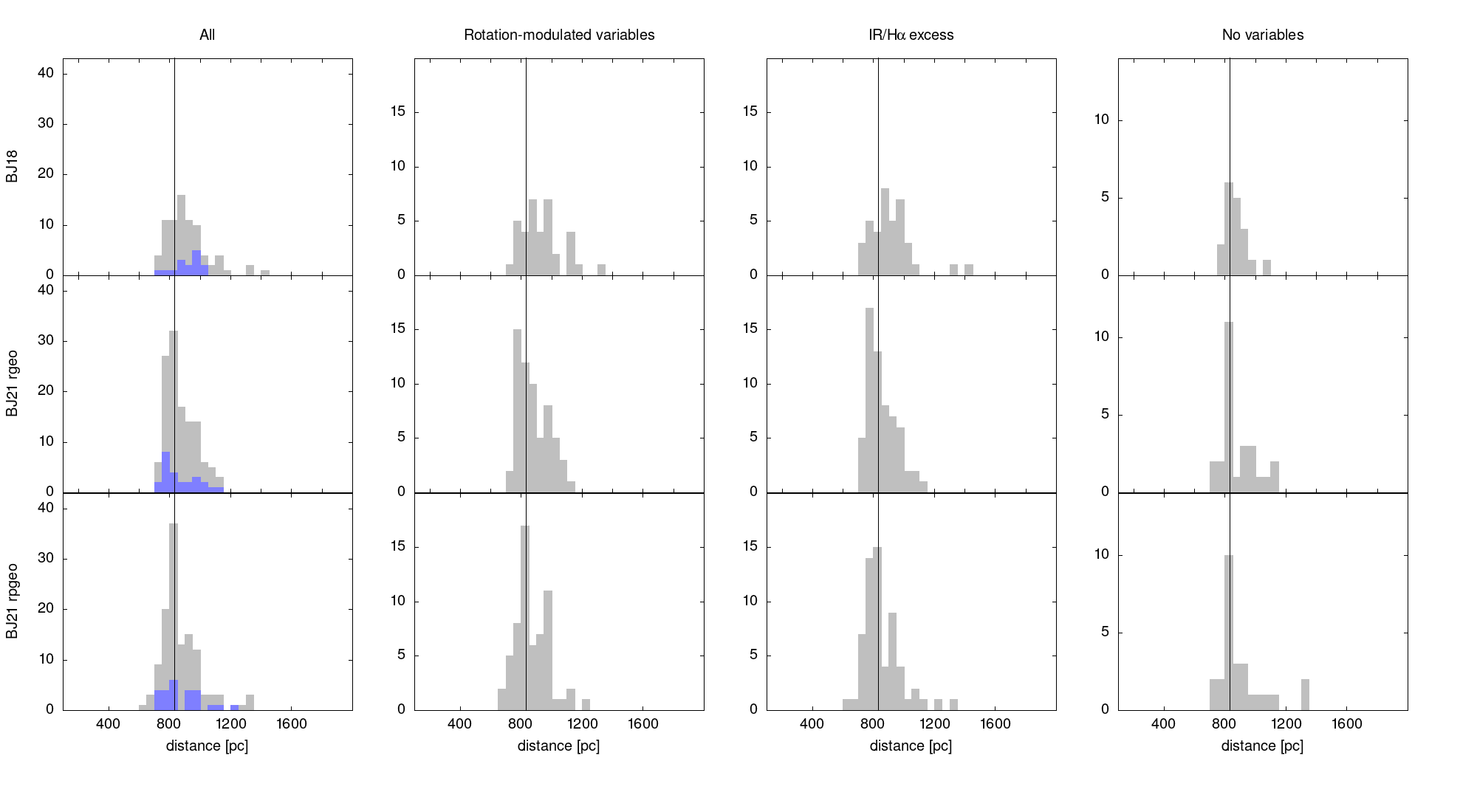}
\caption{Distance histograms with BJ18 and BJ21 data with \textit{GaiaFlag}=1 sources and parallax errors less than 20~per cent. The vertical line is located at the $d$=825~pc. The blue boxes are the sources with 4 membership flags simultaneously.}
\label{G20H}
\end{figure*}

 For the distance (in pc) study, we have taken the sources with \textit{GaiaFlag}=1 (see Tab.~\ref{MemberCrit}) that have a parallax error of less than 20~per cent. Also, we analyse these values but taking rotation-modulated variables, IR/H$\alpha$-excess or non-variable sources, this histograms are in Fig.~\ref{G20H}. Again, we find that the BJ18 data show larger distances than the BJ21 data. In addition, in the first column of these two panels, we overlay the sources that meet 4 membership flags (\textit{ZamsFlag}=1, \textit{GaiaFlag}=1, \textit{V\_type}=1 and \textit{Disc}=1, Tab.~\ref{MemberCrit}). The black vertical line in each distribution indicates the peak of the distributions associated with EDR3 ($\varpi$=1.15~mas and $d$=825~pc).

On the other hand, the two most recent catalogues (EDR3 and BJ21) show higher precision in the data because, with the parallax constraint, more sources are obtained than in the previous ones (DR2 and BJ18). 
Finally, for the BJ21 photogeometric data, we observe that the peak of the non-variable star distribution coincides with the peaks of the rest of the distributions. For the geometrical data, the peak of the non-variable sources is only 50~pc away from the peak of the periodic ones.

\section{Tables}

\subsection{Complete list of \textit{Spitzer} data.}

In Table \ref{tab:Spitzer} we presented the \textit{Spitzer} sample mentioned in subsection \ref{sec:Spz}. 

\begin{table*}
\caption{\textit{Spitzer} data. The full table is in electronic table.}
   \centering
   \resizebox{17.5cm}{!} {
    \begin{tabular}{|l|l|l|c|l|l|l|l|l|l|l|l|c|c|}
    \hline
Id     & RA          & DEC          & {[}3.6{]}                   & e{[}3.6{]} & {[}4.5{]} & e{[}4.5{]} & {[}5.8{]} & e{[}5.8{]} & {[}8.0{]} & e{[}8.0{]} & Disc & ref                   & Spz\_id                                 \\
\hline
 (1) &  (2) &  (3) & (4) & (5) & & & & & & & (6) & (7) & (8) \\
        \hline
6831.0 & 6:7:11.8152 & -6:34:7.788  & 14.1327                     & 5.90904    & 14.0449   & 5.39073    & 14.2479   & 3.65909    & 13.531    & 3.29265    & 1.0  & 1                     & J060711.83-063407.1                     \\
6832.0 & 6:7:46.608  & -6:34:3.324  & 13.7221                     & 6.17297    & 13.6658   & 5.68556    & 13.1706   & 4.67928    & 12.8134   & 4.0175     & 1.0  & 1                     & J060746.61-063402.9                     \\
6751.0 & 6:7:24.6336 & -6:34:2.532  & 17.128                      & 3.8929     & 16.7865   & 3.33417    & 999       & 999        & 14.5155   & 2.58389    & 1.0  & 1                     & J060724.42-063403.3                     \\
6706.0 & 6:7:23.3496 & -6:34:0.516  & 11.8484                     & 7.14405    & 11.7985   & 7.00212    & 11.7245   & 5.93091    & 11.7072   & 5.34458    & 0.0  & 1                     & J060723.36-063400.0                     \\
6818.0 & 6:8:7.1856  & -6:33:59.796 & 13.2748                     & 6.43814    & 13.2323   & 5.92525    & 13.117    & 4.91047    & 13.2122   & 3.70904    & 0.0  & 1                     & J060807.18-063359.3                     \\

     \hline
    \multicolumn{14}{@{}p{190mm}}{(1)~Internal identification object. (2)~Right ascension and (3)~declination in Equinox J2000. (4)~\textit{Spitzer} fluxes in magnitudes from four band [3.6], [4.5], [5.8], [8.0] in 3.6 aperture and (5)~error. (6)~Disc presence (disc=1 and no disc=0). (7)~Reference IRSA=1, G9=2 and K12=3. (10)~\textit{Spitzer} id.}
    \end{tabular}
    }
    \label{tab:Spitzer}
 \end{table*}    

\subsection{Colour excess}
\label{ColourExc}

In Table~\ref{tab:Colour} we present the colour excess obtained for the study of $A_{V}$. We show the values of the different colours analysed E($g-r$), E($r-i$), E($z-y$), E($r-z$), E($i-z$), E($i-y$).
In Table~\ref{tab:Av} we present the A$_{\lambda}$/A$_{V}$ coefficients obtained from the data in \citet[table~3]{2019ApJ...877..116W}. Negative Av values are unphysical and the result of uncertainties; therefore, Av=0.0 should be adopted in these cases.

\begin{table}
\centering
    \begin{tabular}{|c|c c|}
         \hline
      A$_{g}$/A$_{V}$   & 1.155 & $\pm$0.009 \\
      A$_{r}$/A$_{V}$   & 0.843 & $\pm$0.006 \\
      A$_{i}$/A$_{V}$   & 0.628 & $\pm$0.004 \\
      A$_{z}$/A$_{V}$   & 0.487 & $\pm$0.003 \\
      A$_{y}$/A$_{V}$   & 0.395 & $\pm$0.003 \\
         \hline
    \end{tabular}
    \caption{Coefficients obtained from the data in \citet[table 3]{2019ApJ...877..116W}}
    \label{tab:Av}
\end{table}

\begin{table*}
    \label{tab:Colour}
    \caption{Colour excess values for 124 sources. See columns notes in Table \ref{MS}, the last 12 columns are the colour excess in E($g-r$) E($r-i$), E($z-y$), E($r-z$), E($i-z$), E($i-y$) and its errors. The full table is in electronic table.}
    \resizebox{\textwidth}{!}{
    \centering
    
    \begin{tabular}{|l|l|l|l|l|l|l|l|l|c|c|c|c|c|c|c|l|l|l|l|l|l|l|l|l|l|l|}
    \hline
    Id & RA & DEC  & V\_type & P(days) & Power & V\_hs  & H$\alpha$ & SpT &  Disc & ZamsFlag & GaiaFlag & Gaia\_id & Spz\_id & PS1\_id & E($g-r$) & e\_E($g-r$) & E($r-i$) & e\_E($r-i$) & E($z-y$) & e\_E($z-y$) & E($r-z$) & e\_E($r-z$) & E($i-z$) & e\_E($i-r$) & E($i-y$) & e\_E($i-y$) \\ \hline
6346.0 & 6:7:31.716 & -6:32:55.104 & 1.0 & 0.887 & 159 & 1 & NH & M3 & 0 & 1 & 0 & 3018446360716527360 & J060731.72-063254.7 & 100140918821792158 & -2.9069 & 0.0614 & 1.787 & 0.1181 & 0.2015 & 0.0415 & 1.9979 & 0.1618 & 0.2109 & 0.0527 & 0.4124 & 0.0678 \\
6273.0 & 6:7:32.8608 & -6:32:40.38 & 1.0 & 6.337 & 312 & 1 & Ha & M0 & 1 & 1 & 0 & 3018446360716735232 & J060732.87-063239.9 & 100140918869497060 & -2.7637 & 0.0571 & 1.8433 & 0.099 & 0.217 & 0.0355 & 2.109 & 0.1397 & 0.2657 & 0.0611 & 0.4827 & 0.0648 \\
6260.0 & 6:7:22.7232 & -6:32:39.516 & 0.0 & 0 & 55 & 1 & NH & M3 & 0 & 1 & 1 & 3018447043615149056 & J060722.73-063239.1 & 100140918447237354 & -3.0902 & 0.0859 & 2.3088 & 0.1367 & 0.234 & 0.0402 & 2.5816 & 0.1706 & 0.2728 & 0.0585 & 0.5068 & 0.0763 \\
6231.0 & 6:7:34.7328 & -6:32:33.396 & 0.0 & 0 & 14 & 0 & Ha & M6 & 1 & 1 & 0 & 3018447833889325952 & J060734.74-063233.0 & 100140918947459385 & -2.4972 & 0.2678 & 1.1182 & 0.3355 & 0.108 & 0.0575 & 1.2406 & 0.3885 & 0.1224 & 0.0858 & 0.2404 & 0.1245 \\
6169.0 & 6:7:38.544 & -6:32:20.472 & 3.0 & 1.263 & 143 & 1 & NH & M6 & 1 & 1 & 1 & 3018447868248871680 & J060738.55-063220.1 & 100150919106303676 & -2.7172 & 0.145 & 1.1913 & 0.2084 & -0.028 & 0.0513 & 1.1146 & 0.2664 & -0.0767 & 0.0702 & -0.0947 & 0.1133 \\
6106.0 & 6:7:45.0384 & -6:32:7.44 & 0.0 & 0 & 12 & 0 & NH & M7 & 1 & 1 & 0 & 3018447902609799168 & J060745.04-063207.1 & 100150919376728008 & -1.4 & 0.3626 & -0.557 & 0.342 & 0.0128 & 0.0663 & -0.5996 & 0.4365 & -0.0426 & 0.1259 & -0.0198 & 0.2218 \\
6007.0 & 6:7:30.4128 & -6:31:45.912 & 1.0 & 6.337 & 436 & 0 & Ha & M4 & 1 & 1 & 0 & 3018448662818095104 & J060730.45-063145.2 & 100160918767995251 & -3.1279 & 0.14 & 1.928 & 0.1979 & 0.2035 & 0.097 & 2.1431 & 0.2872 & 0.2151 & 0.1173 & 0.4186 & 0.1077 \\
5998.0 & 6:7:30.1464 & -6:31:43.536 & 0.0 & 0 & 81 & 0 & Ha & M5 & 1 & 1 & 0 & 3018448662819492864 & J060730.15-063143.2 & 100160918756285992 & -2.5439 & 0.3004 & 1.3456 & 0.3112 & 0.2708 & 0.0523 & 1.5507 & 0.3509 & 0.2051 & 0.0621 & 0.3759 & 0.0926 \\
5875.0 & 6:7:17.8368 & -6:31:19.884 & 1.0 & 4.922 & 576 & 1 & Ha & M0 & 1 & 1 & 1 & 3018459176898943488 & J060717.84-063119.4 & 100170918243633888 & -2.8139 & 0.0723 & 1.864 & 0.1078 & 0.1675 & 0.0444 & 2.1424 & 0.1461 & 0.2784 & 0.0613 & 0.4459 & 0.0761 \\
5859.0 & 6:7:45.48 & -6:31:12.828 & 0.0 & 0 & 22 & 0 & Ha & M8 & 0 & 1 & 0 & 3018447936968356224 & J060745.49-063112.6 & 100170919395126215 & -1.53 & 0.2238 & -0.539 & 0.2114 & -0.1533 & 0.0818 & -0.7903 & 0.2905 & -0.2513 & 0.1081 & -0.4046 & 0.1827 \\
    \hline
    \end{tabular}
    }
\end{table*}

\subsection{No candidates member}
\label{sec:NoCandidates}

Some target with spectral type assigned could not be selected as members. Also, in Table \ref{NoCandSp} we show the stars with spectral type but no member indicators.

\begin{table*}
\caption{Full table of objects with spectral type but not fulfilling the membership criteria. See columns notes in Table \ref{MS}}
    \centering
    \begin{tabular}{|l|l|l|l|l|l|c|l|l|c|c|c|}
    \hline
        Id & RA & DEC  & V\_type & P(days) & Power & V\_hs & H$\alpha$ & SpT &  Disc & ZamsFlag & GaiaFlag \\ \hline
        \hline
5376 & 06:07:23.38 & -06:29:41.02 & 0 & - & 12 & 0 & Ha- & nonM & - & 0 & 0 \\ 
5184 & 06:07:52.47 & -06:28:48.36 & 0 & - & 51 & 0 & Ha- & nonM & 0 & 0 & 0 \\ 
5121 & 06:08:29.11 & -06:28:27.44 & 0 & - & 13 & 0 & Ha- & nonM & - & 0 & 0 \\ 
5095 & 06:08:41.90 & -06:28:17.40 & 0 & - & 28 & 0 & Ha- & nonM & - & 0 & 0 \\ 
4941 & 06:08:02.83 & -06:27:35.92 & 0 & - & 10 & 0 & - & M3 & - & 0 & 0 \\ 
4732 & 06:07:53.92 & -06:26:34.44 & 0 & - & 16 & 0 & NH & M3 & - & 0 & 0 \\
4152 & 06:07:25.79 & -06:24:06.37 & 0 & - & 8  & 0 & Ha & M4 & - & 0 & 0 \\
4100 & 06:07:16.51 & -06:23:59.10 & 0 & - & 16 & 0 & Ha- & nonM & - & 0 & 0 \\ 
3893 & 06:08:26.33 & -06:23:16.11 & 0 & - & 17 & 0 & Ha- & nonM & - & 0 & 0 \\ 
3624 & 06:08:34.71 & -06:22:32.80 & 0 & - & 10 & 0 & Ha- & nonM & - & 0 & 0 \\ 
3616 & 06:08:26.94 & -06:22:31.87 & 0 & - & 13 & 0 & Ha- & nonM & - & 0 & 0 \\ 
3117 & 06:08:42.77 & -06:21:07.34 & 0 & - & 16 & 0 & NH & M3 & - & - & 0 \\ 
3090 & 06:08:31.20 & -06:21:00.54 & 0 & - & 16 & 0 & Ha- & nonM & - & 0 & 0 \\
2943 & 06:08:41.38 & -06:20:25.80 & 0 & - & 17 & 0 & Ha- & nonM & - & 0 & 0 \\
2625 & 06:08:36.18 & -06:18:59.68 & 0 & - & 22 & 0 & NH & nonM & - & 0 & 0 \\ 
2622 & 06:08:36.14 & -06:18:59.18 & 0 & - & 15 & 0 & NH & nonM & - & 0 & 0 \\ 
2543 & 06:08:44.48 & -06:18:38.84 & 0 & - & 72 & 0 & Ha- & nonM & - & - & 0 \\ 
2503 & 06:07:31.73 & -06:18:34.20 & 0 & - & 10 & 0 & Ha- & nonM & - & 0 & 0 \\ 
2317 & 06:08:14.75 & -06:17:25.76 & 0 & - & 9  & 0 & NH & M5 & - & 0 & 0 \\ 
1927 & 06:08:13.61 & -06:15:47.41 & 0 & - & 7  & 0 & Ha- & nonM & - & 0 & 0 \\ 
1458 & 06:08:14.69 & -06:14:02.72 & 0 & - & 26 & 0 & Ha- & nonM & - & 0 & 0 \\ 
1365 & 06:08:16.14 & -06:13:44.65 & 0 & - & 130 & 0 & NH & M5 & - & - & 0 \\ 
1145 & 06:08:13.95 & -06:13:02.67 & 0 & - & 11 & 0 & Ha- & nonM & - & 0 & 0 \\ 
1081 & 06:08:26.52 & -06:12:49.42 & 0 & - & 16 & 0 & Ha- & nonM & - & 0 & 0 \\ 
950  & 06:08:25.67 & -06:12:30.67 & 0 & - & 10 & 0 & Ha & M3 & - & 0 & 0 \\
        \hline
    \end{tabular}
    \label{NoCandSp}
\end{table*}

\subsection{Brown dwarf candidates.}
\label{sec:BDcand}

In Table \ref{tab:BDspec} we presented the 11 spectroscopic brown dwarf stars. Moreover, in Table \ref{tab:BDcand} we presented the photometrics brown dwarf candidates. However, this list should be taken with a lot of caution as it is likely to be contaminated by reddened low-mass stars. 

\begin{table*}
    \label{tab:BDspec}
    \caption{Spectroscopic Brown Dwarf table. See columns notes in Table \ref{MS}}
    \resizebox{\textwidth}{!}{
    \centering
    \begin{tabular}{|l|l|l|l|l|l|c|l|c|c|c|c|c|c|c|}
    \hline
    Id & RA & DEC  & V\_type & P(days) & Power & V\_hs & H$\alpha$ & SpT &  Disc & ZamsFlag & GaiaFlag & Gaia\_id & Spz\_id & PS1\_id \\ \hline
6106  &  06:07:45.03  &  -06:32:07.44  &  0  &  -  &  12  & 0 & NH  &  M7  &  1  & 1  &  0 & 3018447902609799168 & J060745.04-063207.1 & 100150919376728008 \\
5859  &  06:07:45.48  &  -06:31:12.82  &  0  &  -  &  22  & 0 & Ha  &  M8  &  0  & 1  &  0 & 3018447936968356224 & J060745.49-063112.6 & 100170919395126215 \\
4667  &  06:08:18.92  &  -06:26:07.58  &  0  &  -  &  46  & 0 & Ha  &  M7  &  1  & 1  &  0 & 3019950179090847744 & J060818.92-062607.3 & 100270920788647948 \\
4559  &  06:08:20.13  &  -06:25:37.23  &  4  &  12.597  &  277  & 0 & Ha  &  M8  &  1  &  1  &  1 & 3019950179090848128 & J060820.13-062537.0 & 100280920838988066 \\
4319  &  06:07:52.89  &  -06:24:38.44  &  0  &  -  &  31  & 0 & Ha &  M9  &  1  &  1  &  0 & 3019951008020170880 & J060752.91-062437.9 & 100300919704297661 \\
3501  &  06:07:54.76  &  -06:22:16.46  &  0  &  -  &  30  & 0 & Ha  &  M9  &  1  &  1  &  0 & 3019963270150782720 & - & 100350919782144985 \\
2915  &  06:07:39.24  &  -06:20:27.52  &  0  &  -  &  97  & 0 & Ha  &  M9  &  1  &  1  &  0 & 3019963957346220288 & J060739.41-062025.3 & 100390919135431294 \\
2668  &  06:07:42.57  &  -06:19:18.12  &  0  &  -  &  10  & 0 & NH  &  M8  &  1  &  1  &  0 & - & J060742.57-061917.9 & 100410919273694448 \\
2044  &  06:07:28.86  &  -06:16:20.20  &  0  &  -  &  12  & 0 & Ha  &  M7  &  1  &  1  &  0 & - & J060728.86-061619.9 & 100470918702823749 \\
1692  &  06:07:52.62  &  -06:14:57.62  &  0  &  -  &  27  & 0 & Ha  &  M7  &  1  &  1  &  1 & 3019966435541404672 & J060752.61-061457.4 & 100500919692871279 \\
1531  &  06:07:49.99  &  -06:14:18.13  &  3  &  4.332  &  233  & 0 & Ha  &  M7  &  1  &  1  &  0 & 3019969454903697792 & J060749.99-061417.8 & 100510919583154445 \\
    \hline
    \end{tabular}
    }
\end{table*}

\begin{table*}
    
    \caption{Brown Dwarf candidates. The full table is in electronic table. See columns notes in Table \ref{MS}}
    \resizebox{\textwidth}{!}{
    \begin{tabular}{l l l l l l l l l c c c c c c}
    \hline
    Id & RA & DEC & V\_type & P(days) & Power & V\_hs & H$\alpha$ & SpT & Disc & ZamsFlag & GaiaFlag & GaiaFlag Gaia\_id & Spz\_id & PS1\_id \\
    \hline
    \hline
        6429 & 06:07:49.86 & -06:33:07.92 & 4 & 25 & 176 & 0 & - & - & - & 1 & 0 & 3018447627731892736 & J060749.86-063307.5 & 100130919577597860 \\ 
5075 & 06:07:47.88 & -06:28:16.68 & 0 & - & 10 & 0 & - & - & 1 & 1 & 0 & 3018449345717877120 & J060747.88-062816.4 & 100230919495224918 \\ 
4487 & 06:08:05.23 & -06:25:20.92 & 1 & 4.174 & 125 & 0 & - & - & 1 & 1 & 0 & 3019951076738096128 & J060805.23-062520.6 & 100290920218123505 \\ 
4308 & 06:07:45.61 & -06:24:35.20 & 0 & - & 16 & 0 & - & - & 1 & 1 & 0 & 3019962651675489024 & - & 100300919400888732 \\ 
4292 & 06:07:48.03 & -06:24:30.85 & 1 & 4.712 & 482 & 1 & - & - & 1 & 1 & 0 & 3019962655971478144 & J060748.02-062430.6 & 100310919501410198 \\ 
3898 & 06:07:51.44 & -06:23:21.40 & 0 & - & 69 & 0 & - & - & 1 & 1 & 0 & - & J060751.44-062321.0 & 100330919643433304 \\ 
3739 & 06:07:47.69 & -06:22:55.48 & 2 & 5.986 & 154 & 0 & - & - & - & 1 & 0 & 3019962862128625024 & - & 100340919487391974 \\ 
3708 & 06:07:39.99 & -06:22:52.60 & 0 & - & 10 & 0 & - & - & 1 & 1 & 0 & 3019962995272608640 & J060739.99-062252.2 & 100340919166712936 \\ 
3648 & 06:08:22.14 & -06:22:38.31 & 1 & 1.207 & 536 & 1 & - & - & 1 & 1 & 0 & 3019952000156332800 & J060822.14-062238.1 & 100340920922917700 \\ 
2277 & 06:07:39.92 & -06:17:19.75 & 0 & - & 96 & 0 & - & - & 1 & 1 & 1 & 3019967702557748224 & J060739.91-061719.5 & 100450919163533901 \\ 
1694 & 06:08:11.19 & -06:14:56.25 & 0 & - & 51 & 0 & - & - & 1 & 1 & 1 & 3019966603045411712 & J060811.19-061455.7 & 100500920466361735 \\ 
    \hline
  \end{tabular} 
  }
  \label{tab:BDcand} 
\end{table*}


\bsp	
\label{lastpage}
\end{document}